\documentclass[aps,twocolumn,prd,showpacs,nofootinbib]{revtex4}

\usepackage{bm}
\usepackage{latexsym}
\usepackage{amsfonts,amssymb,amsmath}
\usepackage{graphicx,epsfig}
\usepackage{psfrag}
\usepackage{subfigure}
\usepackage{hyperref}
\usepackage{color}
\usepackage{appendix}
\usepackage{tabularx}
\usepackage{multirow}
\usepackage{amssymb}
\usepackage{pifont}
\usepackage{lipsum}

\newcommand{\ie}{{i.e.~}}

\newcommand{\apriori}{{a priori~}}

\newcommand{\tr}{\mathrm{Tr}}

\newcommand{\dd}{\mathrm{d}}
\newcommand{\ee}{e}

\newcommand{\sss}[1]{{\scriptscriptstyle{#1}}}

\newcommand{\uPl}{\mathrm{Pl}}

\newcommand{\ucl}{\mathrm{cl}}

\newcommand{\usssPl}{\sss{\uPl}}

\newcommand{\calP}{\mathcal{P}}

\newcommand{\setR}{\mathbb{R}}

\newcommand{\Mp}{M_\usssPl}

\newcommand{\efolds}{$e$-folds~}

\newcommand{\beq}{\begin{equation}}
\newcommand{\eeq}{\end{equation}}
\newcommand{\bea}{\begin{eqnarray}}
\newcommand{\eea}{\end{eqnarray}}

\newlength{\wsingfig}
\newlength{\wdblefig}
\newlength{\wquadfig}
\newlength{\wtriplefig}
\newcommand{\Eq}[1]{Eq.~(\ref{#1})}
\newcommand{\Eqs}[1]{Eqs.~(\ref{#1})}
\newcommand{\Fig}[1]{Fig.~{\ref{#1}}}

\newcommand{\Refc}[1]{Ref.~{\cite{#1}}}
\newcommand{\Refs}[1]{Refs.~{\cite{#1}}}

\hypersetup{
    bookmarks=true,         
    unicode=false,          
    pdftoolbar=true,        
    pdfmenubar=true,        
    pdffitwindow=false,     
    pdfstartview={FitH},    
    pdftitle={My title},    
    pdfauthor={Author},     
    pdfsubject={Subject},   
    pdfcreator={Creator},   
    pdfproducer={Producer}, 
    pdfkeywords={keyword1} {key2} {key3}, 
    pdfnewwindow=true,      
    colorlinks=true,       
    linkcolor=red,          
    citecolor=cyan,        
    filecolor=magenta,      
    urlcolor=green,           
    linktocpage=true
}

\begin{document}

\title{Quantum discord of cosmic inflation: \\Can we show 
that CMB anisotropies are of quantum-mechanical origin?}

\author{J\'er\^ome Martin} \email{jmartin@iap.fr}
\affiliation{Institut d'Astrophysique de Paris, UMR 7095-CNRS,
Universit\'e Pierre et Marie Curie, 98 bis boulevard Arago, 75014
Paris, France}

\author{Vincent Vennin} \email{vincent.vennin@port.ac.uk}
\affiliation{Institute of Cosmology and Gravitation, University of
  Portsmouth, Dennis Sciama Building, Burnaby Road, Portsmouth, PO1
  3FX, United Kingdom}

\date{\today}

\begin{abstract}
  We investigate the quantumness of primordial cosmological
  fluctuations and its detectability. The quantum
  discord of inflationary perturbations is calculated for an arbitrary
  splitting of the system, and shown to be very large on
  super-Hubble scales. This entails the
  presence of large quantum correlations, due to the entangled
  production of particles with opposite momentums during inflation. 
  To determine how this is reflected at the observational level, we study
  whether quantum correlators can be reproduced by a non-discordant state, 
  \ie a state with vanishing discord that contains classical correlations only. 
  We demonstrate that this can
  be done for the power spectrum, the price to pay being twofold: first, large errors
  in other two-point correlation functions and second, the presence of intrinsic non-Gaussianity. The detectability of these two features remains to be determined but could possibly rule out a non-discordant description of the cosmic microwave background. 
  If one abandons the idea that perturbations should be modelled by quantum mechanics and wants to use a classical stochastic
  formalism instead, we show that any two-point correlators on
  super-Hubble scales can be exactly reproduced regardless of the
  squeezing of the system. The latter becomes important only for higher-order correlation functions that can be accurately reproduced only in the strong squeezing regime.
\end{abstract}

\pacs{98.80.Cq, 98.80.Qc, 03.67.-a, 03.67.Mn}
\maketitle

\section{Introduction}
\label{sec:intro}

According to the most recent cosmological
observations~\cite{Ade:2015xua,Ade:2015lrj,Ade:2015ava},
inflation~\cite{Starobinsky:1980te, Sato:1980yn, Guth:1980zm,
  Linde:1981mu, Albrecht:1982wi, Linde:1983gd} provides a mechanism
for generating large scale fluctuations~\cite{Starobinsky:1979ty,
  Mukhanov:1981xt, Hawking:1982cz, Starobinsky:1982ee, Guth:1982ec,
  Bardeen:1983qw} that fits the
data very well~\cite{Martin:2010hh,Martin:2013tda,Martin:2013nzq,Martin:2014nya,Martin:2015dha}. The
scope of this mechanism is not limited to broad phenomenological
predictions but is also deeply connected to the realm of quantum
gravity, which makes it particularly interesting from a theoretical
point of view. Indeed, according to inflation, the large scale
structure in our Universe and the cosmic microwave background (CMB)
anisotropies are nothing but quantum fluctuations of the gravitational
and inflaton fields, stretched over cosmological distances.

\par

However, the above claim is often taken with a grain of salt probably
because, in practice, astronomers analyze the data with purely
classical techniques and apparently never need to rely on the quantum
formalism to understand them. This situation raises the question of
the observational signature of the quantum nature of inflationary
perturbations~\cite{Starobinsky:1986fx,Polarski:1995jg,Lesgourgues:1996jc,Egusquiza:1997ez,Kiefer:1998qe,Perez:2005gh,Campo:2005sv,Ellis:2006fy,Kiefer:2006je,Kiefer:2008ku,Valentini:2008dq,Koksma:2010dt,Bassi:2010ss,PintoNeto:2011ui,Martin:2012pea,Canate:2012ua,Lochan:2012di,Bassi:2012bg,Das:2013qwa,Oriti:2013jga,Markkanen:2014dba,Das:2014ada,Maldacena:2015bha,Banerjee:2015zua,Singh:2015sua,Vitenti:2015lpa,Goldstein:2015mha,Leon:2015hwa,Leon:2015ija,Colin:2015tla,Valentini:2015sna}. Is quantum mechanics really necessary in order to explain the properties
of the CMB anisotropies or can everything be described in a classical
context (in which case doubts could be cast on the quantum origin
of the perturbations)? To put it differently, what would be wrong and
which properties would be missed if the analysis were performed in a
fully classical framework? The goal of this article is to address this
question both at the theoretical and observational levels. The
question is particularly timely since, in quantum information theory,
new tools that are directly relevant to the problems discussed in this article have recently been introduced, such as the quantum discord~\cite{Henderson:2001,Zurek:2001}.

\par

This paper is organized as follows. In the next section,
Sec.~\ref{sec:infpert}, we briefly review the theory of inflationary
cosmological perturbations of quantum mechanical origin. We pay
special attention to the quantum state in which the fluctuations are
placed, namely a two-mode squeezed state. In Sec.~\ref{sec:discord},
we introduce the quantum discord~\cite{Henderson:2001,Zurek:2001}
which is a quantity that has recently received a lot of attention in
the quantum information community and which is designed to measure the
quantumness of correlations present in a quantum state. It was first
applied to cosmological fluctuations in \Refc{Lim:2014uea}, where, in
the case where decoherence of perturbations~\cite{Barvinsky:1998cq,Bellini:2001hn,Lombardo:2005iz,Martineau:2006ki,Burgess:2006jn,Prokopec:2006fc,Calzetta:2008iqa,Franco:2011fg,Weenink:2011dd} occurs due to environment
degrees of freedom, the quantum discord of the joint primordial
perturbations-environment system was calculated. In this paper,
we show that, even in the absence of an environment, particles with
wavenumbers $\bm k$ and $-\bm k $ are created by the same quantum
event, and as such, are highly entangled. This gives rise to a large
quantum discord. In practice, we find that it is proportional to the
squeezing parameter which is itself proportional to
the number of \efolds spent between the Hubble radius exit during
inflation and the end of inflation (typically $\sim 50$ \efolds for
the scales observed in the CMB). We conclude that the CMB is a highly
non-classical system that is placed in an entangled quantum state. We
also consider the discord for an arbitrary splitting of the system,
showing that it is always non-vanishing except for the case where the
two sub-systems correspond to the real and imaginary parts of the
Mukhanov-Sasaki variable. This shows that a purely classical
description of cosmological perturbations necessarily fails to
reproduce all its quantum correlations. To make this statement more explicit,
we then calculate these correlators explicitly.
In Sec.~\ref{sec:quantumclassical}, we first derive the predictions
obtained from the quantum formalism when the system is placed in a two-mode squeezed state. We formulate the problem in
the language of quantum information theory in
sub-Sec.~\ref{subsec:quantumcorrel}, a new result already interesting
in itself, and then, in sub-Sec.~\ref{subsec:twoqm}, we establish the
expression of two-point correlation functions. Then we ask whether
these correlations can be reproduced by other means. In
Sec.~\ref{sec:quantumclassical}, we derive the correlators that
would be produced by a quantum state with ``classical correlations''
only (\ie a state with vanishing quantum discord). For two-point
correlation functions, see sub-Sec.~\ref{subsec:twopointclass}, we
find that, if the power spectrum is matched, the other non-discordant two-point correlation functions necessarily differ from their discordant counterparts.
Moreover, in
sub-Sec.~\ref{subsec:ng}, we calculate the four-point correlation functions and show that, in this situation, one
expects non-Gaussianities, at a level that remains to be
established. These two modifications could possibly lead to the indirect detection of 
the large CMB quantumness. In Sec.~\ref{sec:stocha}, we address the
same question but for a classical stochastic distribution. We first
present some general considerations in sub-Sec.~\ref{subsec:gene} and
then, in sub-Sec.~\ref{subsec:gaussianstocha}, we study the correlation
functions in the case of a Gaussian distribution. In this case, we
find that all observable quantities are correctly reproduced on large
scales, and we highlight the role that squeezing plays in concealing
quantum correlations from observable quantities. In particular, we
argue that it plays no role in our ability to reproduce the two-point
correlation functions. Finally, in Sec~\ref{sec:discussion}, we
present a few concluding remarks, and we end the paper with several
appendixes containing various technical aspects.
\section{Inflationary Cosmological Perturbations}
\label{sec:infpert}
Inflation is an epoch of accelerated expansion, driven by a scalar
field (the inflaton field) and taking place before the standard hot
big bang phase. Introduced for the first time nearly $35$ years ago,
it has gradually become the standard paradigm for the early Universe
because it offers simple and elegant solutions to the puzzles of the
big bang theory. 

As already mentioned in the introduction, one of the most interesting
aspects of inflation is that it gives an explanation for the origin of
the large scale structure and CMB anisotropies. According to
inflation, they are vacuum quantum fluctuations amplified by
gravitational instability and stretched over cosmological
distances. At the technical level, two types of perturbations are
produced: gravitational waves and scalar fluctuations. In the rest of
this article, we consider scalar fluctuations only but the
present analysis could straightforwardly be applied to tensor modes
(since, at leading order in slow roll, they have the same dynamics as
scalar perturbations).  The scalar sector can be described by a single
quantity, the so-called Mukhanov-Sasaki variable
\cite{Mukhanov:1981xt,Kodama:1985bj}, related to the curvature
perturbation $\zeta(\eta,{\bm x})$ through
\begin{equation}
\label{eq:defzeta}
v(\eta, {\bm x})=a\Mp \sqrt{2\epsilon_1}\zeta(\eta,{\bm x})\, .
\end{equation}
Here, $a(\eta)$ is the Friedmann-Lema\^{\i}tre-Roberston-Walker scale
factor, $\eta$ is the conformal time (related to the cosmic time $t$ by $
\dd t=a{\dd}\eta$), $\Mp$ denotes the Planck mass and
$\epsilon_1\equiv 1-(a^\prime/a)^\prime/(a^\prime/a)^2$ is the first
slow-roll parameter \cite{Schwarz:2001vv,Leach:2002ar}.
This last quantity controls whether inflation takes place or
not. Indeed, since $\ddot{a}/a=(\dot{a}/a)^2(1-\epsilon_1)$, where a
dot means derivative with respect to cosmic time, inflation is
equivalent to having $\epsilon_1<1$. Moreover, the Planck data, in
particular the measurement~\cite{Ade:2015lrj} of the scalar spectral
index $n_{_{\rm S}}=0.968\pm 0.006$ at $68$\% confidence level, and
the tensor-to-scalar ratio $95$\% constraint $r<0.11$, indicate that
we not only have $\epsilon_1<1$ but in fact $\epsilon_1\ll 1$, a
regime known as slow-roll inflation
\cite{Stewart:1993bc,Martin:1999wa,Martin:2000ak,Leach:2002ar,Schwarz:2001vv}. 

\par

The next step consists in deriving an equation of motion for
$v(\eta,{\bm x})$. Expanding the action of the system (\ie
Einstein-Hilbert action plus the action of a scalar field) up to
second order in the perturbations, one obtains the following
expression~\cite{Mukhanov:1990me}:
\begin{equation}
\label{eq:action}
{}^{\left(2\right)}\delta S=\frac{1}{2}
\int{\mathrm{d}^4x
\left[\left(v^\prime\right)^2
-\delta^{ij}\partial_iv\partial_jv
+\frac{\left(a\sqrt{\epsilon_1}\right)^{\prime\prime}}{a\sqrt{\epsilon_1}}
v^2\right]}\, .
\end{equation}
Instead of working in real space, it turns out to be convenient to go
to Fourier space. We define the Fourier transform of $v(\eta,{\bm x})$
by
\begin{equation}
\label{eq:tfv}
v\left(\eta,\bm{x}\right)=
\frac{1}{\left(2\pi\right)^{3/2}}
\int_{\mathbb{R}^3}{\dd^3\bm{k}\,
v_{\bm{k}}\left(\eta\right)}
\, \ee^{i\bm{k}\cdot \bm{x}}\, ,
\end{equation}
where the vector ${\bm k}$ denotes the comoving wavevector. Here,
because $v(\eta,{\bm x})$ is real, one has $v_{-\bm{k}}=v_{\bm{k}}^*$
(where a star denotes the complex conjugation). As will be discussed
in the following, this relation is important since it shows that, in
Fourier space, all degrees of freedoms are not independent. This is of
course expected since we now have a description of curvature
perturbations in terms of a complex function while we started from a
real one. Inserting \Eq{eq:tfv} into \Eq{eq:action}, one obtains
\begin{equation}
\label{eq:actionfourier}
{}^{\left(2\right)}\delta S=\frac{1}{2}
\int_{\mathbb{R}^3} \mathrm{d}\eta \,  \dd ^3{\bm k}
\left[v_{\bm k}'v_{\bm k}^{*}{}'
-\left(k^2-\frac{z''}{z}\right)v_{\bm k}v_{\bm k}^*\right],
\end{equation}
where the integration is performed in the full Fourier space and we
have defined the background quantity $z(\eta)$ by $z(\eta)\equiv
a\Mp\sqrt{2\epsilon_1}$. Then, under the subtraction of the total
derivative $(v_{\bm k}v_{\bm k}^*z^\prime/z)^\prime/2$ in the
Lagrangian, which leaves the action unchanged, \Eq{eq:actionfourier}
becomes
\begin{eqnarray}
\label{eq:actionfouriertotal}
{}^{\left(2\right)}\delta S &=& \frac{1}{2}
\int_{\mathbb{R}^3} \mathrm{d}\eta \,  \dd ^3{\bm k}
\Biggl[v_{\bm k}'v_{\bm k}^{*}{}'
-\frac{z'}{z}\left(v_{\bm k}'v_{\bm k}^*+v_{\bm k}v_{\bm k}^*{}'\right)
\nonumber \\ & &
+\left(\frac{z'^2}{z^2}-k^2\right)v_{\bm k}v_{\bm k}^*\Biggr].
\end{eqnarray}
From this action, the Euler-Lagrange equation reads
\begin{equation}
\label{eq:eomv}
\frac{ \dd }{ \dd \eta}\left(
\frac{\delta {\cal L}}{\delta v_{\bm k}^*{}'}\right)
-\frac{\delta {\cal L}}{\delta v_{\bm k}^*{}}
=\frac12\left[v_{\bm k}''+\left(k^2-\frac{z''}{z}\right)v_{\bm k}\right]
=0,
\end{equation}
or $v_{\bm k}''+\omega^2(k,\eta)v_{\bm k}=0$ with
$\omega^2(k,\eta)\equiv k^2-z''/z$. One recognizes the equation of
motion of a parametric oscillator, that is to say an oscillator with
time-dependent frequency. Here, the time dependence is due to the
interaction with a classical source, the background gravitational
field described by the scale factor $a(\eta)$. As discussed in
\Refc{Martin:2007bw}, this is very similar to the Schwinger effect, the
only difference being the nature of the source, an Abelian gauge field
for the Schwinger effect and, here, as already mentioned, the
gravitational field of the expanding Universe. As was noticed in
\Refs{Grishchuk:1990bj,Grishchuk:1992tw}, the cosmological
perturbations form a system which is also very similar to what one
finds in quantum optics. In particular, we can already anticipate
that, upon quantization, one will be led to the concept of
squeezing~\cite{BarnettRadmore,Caves:1985zz,Schumaker:1985zz}. Indeed, it is well known that, if quantization of
harmonic oscillators naturally leads to coherent states,
quantization of parametric oscillators leads to
squeezed states.

\par

Let us now apply the Hamiltonian formalism. Here, in order to deal
with independent variables only, we split the integral in the action
into two parts and change ${\bm k}$ into $-{\bm k}$ in the second
part. This leads to the following expression,
\begin{eqnarray}
\label{eq:actionfourierr3+}
{}^{\left(2\right)}\delta S &=& \frac{1}{2}
\int_{\mathbb{R}^{3+}} \mathrm{d}\eta \,  \dd ^3{\bm k}
\Biggl[v_{\bm k}'v_{\bm k}^{*}{}'+v_{\bm k}^*{}'v_{\bm k}'
-2\frac{z'}{z}\bigl(v_{\bm k}'v_{\bm k}^*
\nonumber \\ & &
+v_{\bm k}v_{\bm k}^*{}'\bigr)
+\left(\frac{z'^2}{z^2}-k^2\right)\left(v_{\bm k}v_{\bm k}^*
+v_{\bm k}^*v_{\bm k}\right)
\Biggr]\, ,
\end{eqnarray}
where the integral over $\bm k$ is now performed in half the Fourier
space, $\bm k \in \mathbb{R}^{3+}$. Then our next move is to define
the conjugate momentum,
\begin{equation}
p_{\bm k} =\frac{\delta {\cal L}}{\delta v_{\bm k}^*{}'} = v_{\bm k}'
-\frac{z'}{z}v_{\bm k}.
\end{equation}
It is easy to see that if $v_{\bm k}$ is a measure of the curvature
perturbation $\zeta_{\bm k}=v_{\bm k}/z$, the conjugate momentum
$p_{\bm k}$ is a measure of its time derivative since
\begin{equation}
\label{eq:defp}
\zeta_{\bm k}'(\eta )=\frac{p_{\bm k}(\eta)}{z(\eta)}.
\end{equation}
On large scales (\ie scales larger than the Hubble radius during
inflation), the solution to \Eq{eq:eomv} implies that $ \zeta_{\bm
  k}(\eta) \sim A_{\bm k}\left(1+\#_1k^2\eta^2+\#_2k^4\eta^4+\cdots
\right)+B_{\bm k}\int^{\eta} \dd \tau/z^2(\tau)\left(1+\cdots\right)$
where $A_{\bm k}$ and $B_{\bm k}$ are two scale dependent constants
and $\#_1$ and $\#_2$ two scale independent constants. During
inflation, the first branch (the ``growing mode'') is constant while
the second one (the ``decaying mode'') is decaying (hence the claim
that curvature perturbation are conserved). From the above expansion,
one has $\zeta_{\bm k}'(\eta)\sim 2 A_{\bm k}\#_1k^2\eta+B_{\bm
  k}/z^2(\eta)$ and we see that, in the super-Hubble limit where
$k\eta \rightarrow 0^{-}$, the second term of the growing mode
dominates over the leading term in the decaying mode. This means that
$\zeta_{\bm k}'(\eta)\propto 1/z(\eta)$ and, as a consequence, $p_{\bm
  k}$ tends to a constant.

\par

Let us now calculate the Hamiltonian of the system, defined as
\begin{equation}
H=\int_{\mathbb{R}^{3+}}  \dd ^3{\bm k}
\left(p_{\bm k}v_{\bm k}^*{}'+p_{\bm k}^*v_{\bm k}'-{\cal L}\right)\, .
\end{equation}
Notice that the terms $p_{\bm k}v_{\bm k}^*{}'$ and $p_{\bm k}^*v_{\bm
  k}'$ are summed up only in $\mathbb{R}^{3+}$~\cite{1989pait.book.....C}. Plugging in the
Lagrangian $\mathcal{L}$ obtained in \Eq{eq:actionfourierr3+}, one has
\begin{equation}
\label{eq:Hamiltonian:vandp}
H=\int_{\mathbb{R}^{3+}}  \dd ^3{\bm k}
  \left[p_{\bm k}p_{\bm k}^*+\frac{z'}{z}\left(p_{\bm k}^*v_{\bm k}
      +p_{\bm k}v_{\bm k}^*\right)+k^2v_{\bm k}v_{\bm k}^*\right].
\end{equation}
Quantization consists in introducing the creation and annihilation
operators $\hat{c}_{\bm k}$ and $\hat{c}_{\bm p}^{\dagger}$ obeying
the commutation relations $\left[\hat{c}_{\bm k},\hat{c}_{\bm
    p}^{\dagger}\right]=\delta \left({\bm k}-{\bm p}\right)$. From now
on, quantum operators are denoted with a hat. The Mukhanov-Sasaki
variable and its conjugate momentum are related to these operators
through the following formulas
\begin{eqnarray}
\label{eq:defc1}
\hat{v}_{\bm k} &=& \frac{1}{\sqrt{2k}}\left(\hat{c}_{\bm k}
+\hat{c}_{-{\bm k}}^{\dagger}\right), \\
\label{eq:defc2}
\hat{p}_{\bm k} &=& -i\sqrt{\frac{k}{2}}
\left(\hat{c}_{\bm k}-\hat{c}_{-{\bm k}}^{\dagger}\right).
\end{eqnarray}
Notice that \Eqs{eq:defc1} and~(\ref{eq:defc2}) mix creation and
annihilation operators associated to the modes ${\bm k}$ and $-{\bm
  k}$. This is different from what is usually done since the common
practice is to work ``mode by mode.'' These definitions are necessary
since they ensure that $\hat{v}_{-{\bm k}}=\hat{v}_{\bm k}^{\dagger}$.

\par

The final step is to express the Hamiltonian in terms of the creation
and annihilation operators. We obtain
\begin{eqnarray}
\label{eq:hami}
\hat{H} &=& \int_{\mathbb{R}^{3}} \dd ^3{\bm k}
\biggl[\frac{k}{2}\left(\hat{c}_{\bm k}\hat{c}_{\bm k}^{\dagger}
+\hat{c}_{-{\bm k}}\hat{c}_{-{\bm k}}^{\dagger}
\right)
\nonumber \\ & &
-\frac{i}{2}\frac{z'}{z}
\left(\hat{c}_{\bm k}\hat{c}_{-{\bm k}}
-\hat{c}_{-{\bm k}}^{\dagger}\hat{c}_{\bm k}^{\dagger}\right)
\biggr].
\end{eqnarray}
In this expression, the first term represents a collection of free
oscillators with energy $\omega =k$ (compatible with the fact that we
deal with massless excitations) while the second one describes the
interaction between the quantized perturbations and the classical
source. Of course, if the scale factor is constant (Minkowski
spacetime), then $z$ is a constant and this term disappears. This term
is responsible for the creation of pairs of particles. We observe that
the structure $\hat{c}_{-{\bm k}}^{\dagger}\hat{c}_{\bm k}^{\dagger}$ implies that
the particles are created with opposite momenta, thus ensuring momentum
conservation.

\par

Let us now discuss the equation of motion. It is given by the
Heisenberg equation, namely $\dd \hat{c}_{\bm k}/\dd\eta=-i\left[\hat{c}_{\bm
    k},\hat{H}\right]$. Using \Eq{eq:hami}, this leads to
 \begin{eqnarray}
i\frac{ \dd \hat{c}_{\bm k}}{ \dd \eta} &=& k\hat{c}_{\bm k}
+i\frac{z'}{z}\hat{c}_{-{\bm k}}^{\dagger}\, .
\end{eqnarray}
This can be solved by mean of the Bogoliubov transformation
\begin{eqnarray}
\label{eq:bogo1}
\hat{c}_{\bm k}(\eta)&=& u_k(\eta)\hat{c}_{\bm k}(\eta_{\rm ini})+v_k(\eta)
\hat{c}_{-{\bm k}}^{\dagger}(\eta_{\rm ini})\, ,
\end{eqnarray}
where $u_k(\eta)$ and $v_{k}(\eta)$ (not to be confused with the
Mukhanov-Sasaki variable) are two functions depending on the
wavevector modulus only\footnote{Indeed, one can see that the
  equations of motion~(\ref{eq:equ}) and~(\ref{eq:eqv}) for
  $u_k(\eta)$ and $v_{k}(\eta)$ only depend on the modulus of $\bm
  k$. So if the initial state is chosen to be the vacuum, rotationally
  invariant and hence dependent on $k$ only, $u_k(\eta)$ and $v_{k}(\eta)$
  remain independent of the orientation of $\bm k$ at any time.} and
obeying
\begin{eqnarray}
\label{eq:equ}
i\frac{ \dd u_k(\eta)}{ \dd \eta} &=& ku_k(\eta)
+i\frac{z'}{z}v_k^*(\eta), \\
\label{eq:eqv}
i\frac{ \dd v_k(\eta)}{ \dd \eta} &=& kv_k(\eta)
+i\frac{z'}{z}u_k^*(\eta).
\end{eqnarray}
They must moreover satisfy $\left \vert u_k(\eta)\right
\vert^2-\left \vert v_k(\eta)\right \vert^2 =1$ in order for the
commutation relation between the creation and annihilation operators
to be properly normalized. Let us also notice that the combination
$u_k+v_k^*$ obeys the same equation of motion as the Mukhanov-Sasaki
variable, namely $(u_k+v_k^*)''+\omega^2(u_k+v_k^*)=0$.

\par

From the above considerations, we see that solving the time dependence
of the system is equivalent to solving the Bogoliubov
system~(\ref{eq:equ})-(\ref{eq:eqv}). For this purpose, let us now
introduce two operators. The first one is the two-mode squeezing
operator $\hat{S}(r_k,\varphi_k)$ defined by
$\hat{S}(r_k,\varphi_k)={\rm e}^{\hat{B}_k} $ with
\begin{equation}
\hat{B}_k\equiv r_k{\rm e}^{-2i\varphi_k}
\hat{c}_{-{\bm k}}(\eta_{\rm ini})\hat{c}_{\bm k}(\eta_{\rm ini})
-r_k{\rm e}^{2i\varphi_k}\hat{c}_{-{\bm k}}^{\dagger}(\eta_{\rm ini})
\hat{c}_{\bm k}^{\dagger}(\eta_{\rm ini}).
\end{equation}
Clearly, from the above definition, one has
$\hat{B}_k^{\dagger}=-\hat{B}_k$. It is characterized by two
parameters, the squeezing parameter $r_k$, and the squeezing angle
$\varphi_k$. The second operator is the rotation operator
$\hat{R}_k(\theta_{k,1},\theta_{k,2})$ that can be expressed as
$\hat{R}_k(\theta_{k,1},\theta_{k,2})={\rm e}^{\hat{D}_k} $ with
\begin{equation}
\hat{D}_k\equiv -i\theta_{k,1}\hat{c}_{\bm k}^{\dagger}(\eta_{\rm ini})
\hat{c}_{\bm k}(\eta_{\rm ini})
-i\theta_{k,2}\hat{c}_{-{\bm k}}^{\dagger}(\eta_{\rm ini})
\hat{c}_{-{\bm k}}(\eta_{\rm ini})\, .
\end{equation}
We notice that we also have $\hat{D}^{\dagger}_k=-\hat{D}_k$. The rotation
operator is a priori characterized by two parameters, the rotation
angles $\theta_{k,1}$ and $\theta_{k,2}$.

\par

Let us now calculate the quantity
$\hat{R}_k^{\dagger}\hat{S}_k^{\dagger}\hat{c}_{\bm k}(\eta_{\rm
  ini})\hat{S}_k\hat{R}_k$. Using the formula ${\rm
  e}^{\hat{O}}\hat{A}{\rm
  e}^{-\hat{O}}=\hat{A}+[\hat{O},\hat{A}]+[\hat{O},[\hat{O},\hat{A}]]/2+\cdots$,
it is easy to show that
\begin{eqnarray}
\label{eq:ck}
\hat{R}_k^{\dagger}\hat{S}_k^{\dagger}\hat{c}_{\bm k}(\eta_{\rm ini})\hat{S}_k\hat{R}_k 
&=& {\rm
  e}^{-i\theta_{k,1}}\cosh r_k\hat{c}_{\bm k}(\eta_{\rm ini}) \nonumber \\ &
&\!\!\!\!\!
-{\rm e}^{i\theta_{k,2}+2i\varphi_k}\sinh r_k 
\hat{c}_{-{\bm k}}^{\dagger}(\eta_{\rm ini})\, ,\qquad\\
\label{eq:cminusk}
\hat{R}_k^{\dagger}\hat{S}_k^{\dagger}
\hat{c}_{-{\bm k}}(\eta_{\rm ini})\hat{S}_k\hat{R}_k &=& {\rm
  e}^{-i\theta_{k,2}}\cosh r_k \hat{c}_{-{\bm k}}(\eta_{\rm ini}) \nonumber
\\ & &\!\!\!\!\!  -{\rm e}^{i\theta_{1,k}+2i\varphi_k}\sinh r_k
\hat{c}_{{\bm k}}^{\dagger}(\eta_{\rm ini})\, .  
\end{eqnarray}
This relation coincides with the transport equation~(\ref{eq:bogo1})
for $\hat{c}_{\bm k}$ provided that
\begin{equation}
\hat{c}_{\bm k}(\eta)=\hat{R}_k^{\dagger}\hat{S}_k^{\dagger}
\hat{c}_{\bm k}(\eta_{\rm ini})\hat{S}_k\hat{R}_k,
\end{equation}
with $\theta_{k,1}=\theta_{k,2}\equiv \theta_k$, and 
\begin{eqnarray}
\label{eq:defu}
u_k(\eta) &=& {\rm e}^{-i\theta_k}\cosh r_k, \\
\label{eq:defv}
v_k(\eta) &=& - {\rm e}^{i\theta_k+2i\varphi_k}\sinh r_k.
\end{eqnarray}
One checks that, indeed, $\left \vert u_k(\eta)\right \vert^2-\left
  \vert v_k(\eta)\right \vert^2 =1$.

\par

We conclude from the previous analysis that the full Hilbert space of
the system ${\cal E}$ can be factorized into independent products of
Hilbert spaces for modes ${\bm k}$ and $-{\bm k}$: ${\cal
  E}=\Pi_{k\in\mathbb{R}^{3+}}{\cal E}_{\bm k}\otimes {\cal E}_{-{\bm
    k}}$. Each of these terms is a composite or a bipartite system,
and evolves from the initial vacuum state $\vert 0_{\bm k},0_{-{\bm
    k}}\rangle $ into a two-mode squeezed state
\begin{equation}
\label{eq:qstatevacuum}
\hat{S}(r_k,\varphi_k)\hat{R}(\theta_k)\vert 0_{\bm k},0_{-{\bm k}}\rangle .
\end{equation}
One can explicitly check that the vacuum state is rotationally
invariant, $\hat{R}(\theta_k)\vert 0_{\bm k},0_{-{\bm k}}\rangle
=\vert 0_{\bm k},0_{-{\bm k}}\rangle $, which implies that if the
initial state is the vacuum state, $\theta_k$ cancels out in all
physical quantities, as will be checked in what follows. More
explicitly, using operator ordering theorems~\cite{BarnettRadmore},
one has
\begin{align}
& \hat{S}(r_k,\varphi_k) =\exp\left[-
{\rm e}^{2i\varphi_k}\tanh r_k \hat{c}_{-{\bm k}}^{\dagger}(\eta_{\rm ini})
\hat{c}_{{\bm k}}^{\dagger}(\eta_{\rm ini})\right]
\nonumber \\ & \times
\exp\biggl\{-\ln \left(\cosh r_k\right)
\bigl[\hat{c}_{\bm k}^{\dagger}(\eta_{\rm ini})\hat{c}_{\bm k}(\eta_{\rm ini})
\nonumber \\ &
+\hat{c}_{-{\bm k}}(\eta_{\rm ini})
\hat{c}_{-{\bm k}}^{\dagger}(\eta_{\rm ini})\bigr]\biggr\}
\nonumber \\ & \times
\exp\left[{\rm e}^{-2i\varphi_k}\tanh r_k \hat{c}_{-{\bm k}}(\eta_{\rm ini})
\hat{c}_{{\bm k}}(\eta_{\rm ini})\right]\, ,
\end{align}
and straightforward calculations lead to
\begin{eqnarray}
\hat{S}(r_k, \varphi_k)  \hat{R}(\theta_k)\vert 0_{\bm k},0_{-{\bm k}}\rangle
\nonumber
\qquad\qquad\qquad\qquad\qquad \\
 = \frac{1}{\cosh r_k} \sum _{n=0}^{\infty}
{\rm e}^{2in\varphi_k}(-1)^n\tanh ^n r_k \vert n_{\bm k},n_{-{\bm k}}\rangle .
\label{eq:qstate}
\end{eqnarray}
This is a two-mode squeezed state, well known in the context of
quantum optics as an entangled state. Indeed, in a very sketchy
way (restricting the sum to $n=0$ and $n=1$ and ignoring the
coefficients of the expansion, just for the purpose of illustration),
it is of the form
\begin{equation}
\label{eq:entangle}
\vert \Psi \rangle\sim \frac{1}{\sqrt{2}}
\left(\vert 0_{\bm k}\rangle \vert 0_{-{\bm k}}\rangle 
+\vert 1_{\bm k}\rangle \vert 1_{-{\bm k}}\rangle \right),
\end{equation}
which, indeed, is in an inseparable form since it cannot be written as
$\vert \Psi _{\bm k}\rangle \otimes \vert \Psi_{-{\bm k}}\rangle
$. Entangled states are considered as the most non-classical states
and the above discussion shows that the CMB is placed in such a
state. Moreover, it is known that in the strong squeezing limit, the state~(\ref{eq:qstate}) exactly tends towards an Einstein-Podolsky-Rosen (EPR)~\cite{PhysRev.47.777} quantum state. We are therefore led to conclude that the CMB is a highly non-classical system.

\par

Before concluding this section, it is interesting to derive the
equations satisfied by the parameters $r_k$, $\varphi_k$ and
$\theta_k$. From \Eq{eq:qstate}, we see that their time evolution
controls the time evolution of the quantum
state~(\ref{eq:qstate}). Using \Eqs{eq:equ}-(\ref{eq:eqv})
and~(\ref{eq:defu})-(\ref{eq:defv}), one obtains
\begin{eqnarray}
\label{eq:r}
\frac{ \dd r_k}{ \dd \eta} &=& 
-\frac{z'}{z}\cos\left(2\varphi_k\right), \\
\label{eq:phi}
\frac{ \dd \varphi_k}{ \dd \eta} &=& -k+
\frac{z'}{z}
\coth \left(2 r_k\right)
\sin \left(2\varphi_k\right),\\
\label{eq:theta}
\frac{ \dd \theta_k}{ \dd \eta} &=& k-
\frac{z'}{z}
\tanh r_k\sin \left(2\varphi_k\right).
\end{eqnarray}
The two first equations form a closed system, and $\theta_k$ can be
derived once $r_k$ and $\varphi_k$ are known. This is expected since,
as already said, $\theta_k$ is a spurious parameter that does not play
any physical role. Exact solutions are available only in the exact de
Sitter case, for which we have
\begin{eqnarray}
\label{eq:r:deSitter}
r_k(\eta) &=& -\arg \sinh \left(\frac{1}{2k\eta}\right), \\
\label{eq:varphi:deSitter}
\varphi_k(\eta) &=& \frac{\pi}{4}
-\frac12 \arctan \left(\frac{1}{2k\eta}\right), \\
\theta_k (\eta)&=& k\eta +\arctan \left(\frac{1}{2k\eta}\right).
\end{eqnarray}
On super-Hubble scales, at the end of inflation, we have $ r_k
\rightarrow +\infty$ (notice that $r_k$ is positive because
conformal time is negative during inflation), $\varphi_k \rightarrow
\pi/2$ and $\theta_k \rightarrow -\pi/2$, which remains true for quasi
de Sitter expansions.
\section{Inflationary Discord}
\label{sec:discord}
The discussion of the last section indicates that the quantumness of the CMB is \apriori large. Therefore, it
seems important to better quantify this property. Recently, in the
context of quantum information theory, Henderson and
Vedral~\cite{Henderson:2001} and Ollivier and Zurek~\cite{Zurek:2001}
have introduced the quantum discord, which aims to quantitatively
measure the quantumness of a system. The goal of this section is to
calculate the quantum discord of inflationary perturbations.

\par

Quantum discord is defined as follows. Let us consider a bipartite
system ${\cal E}={\cal E}_{\bm k}\otimes {\cal E}_{-{\bm k}}$ (we
write the two systems ``${\bm k}$" and ``$-{\bm k}$'' for obvious
reasons but it should be clear that the following considerations are
in fact valid for any bipartite system). The idea is to introduce two
ways of calculating the mutual information between the two subsystems
that coincide for classical correlations but not necessarily in
quantum systems. The difference between the two results will then
define the quantum discord. The first measure of mutual information is
provided by the von Neumann entropy,
\begin{equation}
{\cal I}({\bm k},-{\bm k})=S\left[\hat{\rho}({\bm k})\right]
+S\left[\hat{\rho}(-{\bm k})\right]-S\left[\hat{\rho}({\bm k},-{\bm k})\right],
\end{equation}
where $S$ is the entropy defined by $S=-\tr \left(\hat{\rho} \log_2
  \hat{\rho}\right)$, $\hat{\rho} $ being the density matrix of the
system under consideration. Let us recall that the density matrix of
the sub-system ``${\bm k}$'' (respectively ``$-{\bm k}$'') is obtained
from the full density matrix $\hat{\rho}({\bm k},-{\bm k})$ by tracing
over the degrees of freedom associated with ``$-{\bm k}$''
(respectively ``${\bm k}$''), that is to say $\hat{\rho}({\bm
  k})=\tr_{-{\bm k}}\left[\hat{\rho}({\bm k},-{\bm k})\right]$
(respectively $\hat{\rho}(-{\bm k})=\tr_{{\bm k}}\left[\hat{\rho}({\bm
    k},-{\bm k})\right]$).

\par

Then let us imagine that a measurement of an observable of the
sub-system ``$-\bm k$'' is performed through a projector $\hat{\Pi}_j$
(defined on $\mathcal{E}_{-{\bm k}}$). This transforms the state into
$\hat{\rho}\rightarrow \hat{\rho}\hat{\Pi}_j/p_j$, with probability
$p_j\equiv \tr \left[\hat{\rho}\left({\bm k},-{\bm
      k}\right)\hat{\Pi}_j\right]$. If we only access
$\mathcal{E}_{\bm k}$, tracing over $-\bm k$, the state becomes
\begin{equation}
\label{eq:projectedState}
\hat{\rho} ({\bm k};\hat{\Pi}_j)=\tr _{-{\bm k}}
\left[\frac{\hat{\rho}({\bm k},-{\bm k})\hat{\Pi}_j}
{p_j}\right]\, .
\end{equation}
Now, if one performs a set of all possible measurements through a
complete set\footnote{More precisely, $\lbrace \hat{\Pi}_j\rbrace$
  forms a positive operator valued measure, obeying the partition of
  unity $\Sigma_j \hat{\Pi}_j = 1$. They generalize complete sets in
  the sense that they do not need to be orthogonal.} of projectors
$\lbrace \hat{\Pi}_j\rbrace$, an alternative definition of mutual
information is given by
\begin{equation}
{\cal J}({\bm k},-{\bm k})=S\left[\hat{\rho}({\bm k})\right]
-\sum _j p_j S\left[\hat{\rho} ({\bm k};\hat{\Pi}_j)\right]\, .
\end{equation}
Classically, thanks to Bayes theorem, we have ${\cal J}({\bm k},-{\bm
  k})={\cal I}({\bm k},-{\bm k})$. Quantum mechanically however, this
need not be true and the deviation from the previous equality
therefore represents the quantumness of the correlations contained in
a given state. This is why quantum discord is defined by
\begin{equation}
\label{eq:defdiscord}
\delta \left({\bm k},-{\bm
      k}\right)\equiv \min_{\lbrace \hat{\Pi}_j \rbrace } 
\left[{\cal I}({\bm k},-{\bm
  k})-{\cal J}({\bm k},-{\bm
  k})\right],
\end{equation}
where we minimize over all possible sets of measurements in order to
avoid dependence on the projectors. More detailed discussions on the
physical interpretation and meaning of quantum discord and entanglement can be found in
\Refs{Huang:1994,Henderson:2001,Zurek:2001}.

\par

Let us now calculate the quantum discord of inflationary
perturbations. For the two-mode squeezed state~(\ref{eq:qstate}), the
density matrix is given by
\begin{eqnarray}
\hat{\rho}({\bm k},-{\bm k})&=& \frac{1}{\cosh^2 r_k}
\sum_{n,n^\prime=0}^\infty\ee^{2i(n-n^\prime)\varphi_k}
(-1)^{n+n^\prime}
\nonumber\\ &  &
\tanh^{n+n^\prime} r_k\vert n_{\bm k} , n_{-\bm k}\rangle
\langle n_{\bm k}^\prime , n_{-\bm k}^\prime\vert\, .
\end{eqnarray}
The reduced density matrix $\hat{\rho}({\bm k})$ is obtained from the
full density matrix by tracing out degrees of freedom associated to
``$-{\bm k}$''. One has
\begin{eqnarray}
\hat{\rho}({\bm k}) &=& \sum_{n=0}^{\infty}\langle n_{-{\bm k}}
\vert \hat{\rho}({\bm k},-{\bm k})\vert n_{-{\bm k}} \rangle \\
&=& \frac{1}{\cosh ^2r_k}
\sum_{n=0}^{\infty}
\tanh ^{2n}r_k\vert n_{\bm k}\rangle \langle n_{\bm k}\vert,
\end{eqnarray}
which is a thermal state with inverse temperature $\beta_k=-\ln
\tanh^2 r_k$. We have of course a similar equation for
$\hat{\rho}(-{\bm k})$ where $\vert n_{\bm k}\rangle \langle n_{\bm
  k}\vert$ is replaced by $\vert n_{-{\bm k}}\rangle \langle n_{-{\bm
    k}}\vert$.

\par

Our next move is to calculate the entropy of the different density
matrices appearing in the expression of the discord. As can be shown
explicitly, the entropy of $\hat{\rho}({\bm k},-{\bm k})$ vanishes
since we deal with a pure state. Since $\hat{\rho}({\bm k})$
represents a thermal state, its entropy is simply given by~\cite{Demarie:2012}
\begin{equation}
\label{eq:thermalentropy}
S\left[\hat{\rho}({\bm k})\right]=\left(1+\langle \hat{n}_k\rangle\right)
\log_2\left(1+\langle \hat{n}_k\rangle\right)-
\langle \hat{n}_k\rangle \log_2 \langle \hat{n}_k\rangle ,
\end{equation}
where $\langle \hat{n}_k \rangle =\sinh^2r_k$ is the mean occupation
number. Obviously, this formula is also valid for $\hat{\rho}(-{\bm
  k})$. Finally, the quantity $S\left[\hat{\rho} ({\bm
    k};\hat{\Pi}_j)\right]$ remains to be calculated. In
Appendix~\ref{sec:appendixpure}, we show that $\hat{\rho} ({\bm
  k};\hat{\Pi}_j)$ is in fact a pure state and, consequently, its
entropy is zero. Therefore, it follows that the discord is given by
\begin{figure}[t]
\begin{center}
\includegraphics[width=0.45\textwidth,clip=true]{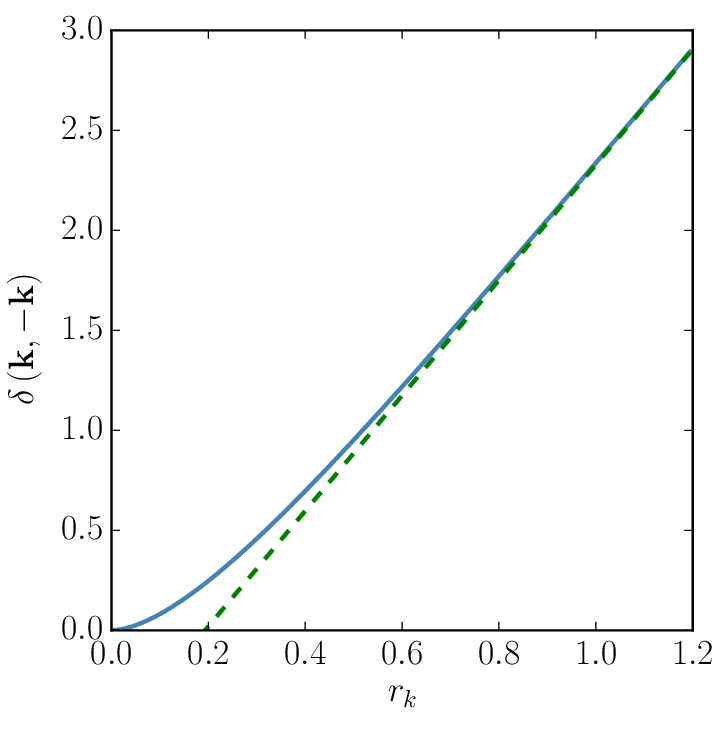}
\caption{Quantum discord $\delta \left({\bm k},-{\bm k}\right)$ of
  cosmological scalar perturbations during inflation, as a function of
  the squeezing parameter $r_k$. The solid blue line is the
  result~(\ref{eq:discord}) while the dotted green line is the large
  squeezing expansion~(\ref{eq:discord:appr}).}
\label{fig:discord}
\end{center}
\end{figure}
\begin{eqnarray}
\label{eq:discord}
\delta \left({\bm k},-{\bm k}\right)
&=& S\left[\hat{\rho}(-{\bm k})\right]
=\cosh ^2r_k\log_2\left(\cosh ^2 r_k\right)
\nonumber \\ & &
-\sinh ^2r_k\log_2\left(\sinh ^2 r_k\right)\, .
\end{eqnarray}
The corresponding function is displayed in \Fig{fig:discord}. One can
see that except for $r_k=0$, the discord is not zero, and, therefore,
the quantum state of the perturbations is not classical. In fact,
$r_k=0$ corresponds to a coherent state~\cite{1926NW.....14..664S,2012JPhA...45a0101T}. Such states are often called
``quasi classical'' and are known to be the ``most classical'' states
since they follow the classical trajectory in phase space, with
minimal spread. In the strong squeezing limit however, $r_k\rightarrow
\infty$, we have
\begin{equation}
\label{eq:discord:appr}
\delta \left({\bm k},-{\bm k}\right)= 
\frac{2}{\ln 2} r_k-2+\frac{1}{\ln 2}
+{\cal O}\left({\rm e}^{-2r_k}\right)\, .
\end{equation}
Let us recall that for the modes within the CMB window, at the end of
inflation, $r_k\sim 50$.  We conclude that the CMB is placed in a
state which is ``very quantum.'' This means that it is certainly
impossible to reproduce all the correlation functions in a classical
picture as we are going to demonstrate explicitly in the following
section.  In fact, the results of this section show that, strictly
speaking, there is no transition to a classical behavior since the
discord only grows. But we will also see that the situation is subtle
and that, nevertheless, a classical treatment of the perturbations can
partially be employed, in a sense that will be carefully discussed in
the following.

\begin{figure}[t]
\begin{center}
\includegraphics[width=0.45\textwidth,clip=true]{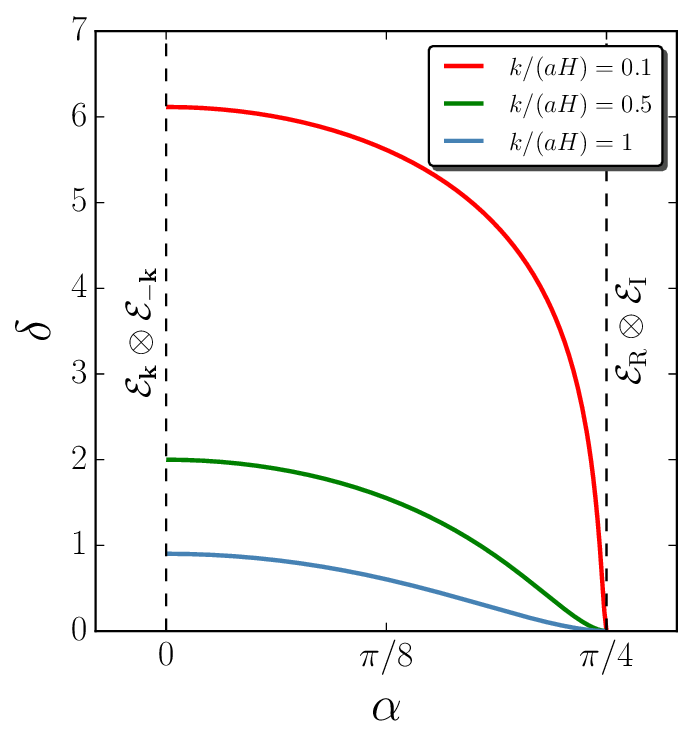}
\caption{Quantum discord for different system subdivisions (labeled by
  $\alpha$) and different times during inflation.}
\label{fig:GeneralizedDiscord}
\end{center}
\end{figure}

Let us also mention that quantum discord is always defined relatively to a given division into two subsystems. For example, in what precedes, the quantum discord was shown to be large for the bipartite system ${\cal E}={\cal E}_{\bm k}\otimes {\cal E}_{-{\bm k}}$. A priori, with a different division, one would have obtained a
different result. For example, let us write the
Hamiltonian~(\ref{eq:Hamiltonian:vandp}) in terms of the real and
imaginary (or Hermitian and anti-Hermitian) parts of $v_{ {\bm k}}$
and $p_{{\bm k}}$, denoted $v_{\bm k}^\mathrm{R}$, $v_{\bm
  k}^\mathrm{I}$, $p_{\bm k}^\mathrm{R}$, $p_{\bm k}^\mathrm{I}$, and
defined by $v_{\bm k} =(v_{\bm k}^{\mathrm{R}}
+iv_{\bm k}^{\mathrm{I}})/\sqrt{2}$ and $p_{\bm k} =(p_{\bm k}^{\mathrm{R}}
+ip_{\bm k}^{\mathrm{I}})/\sqrt{2}$.
Notice that the relation $v_{\bm k}^{\dagger}=v_{-{\bm k}}$ implies
that $v_{\bm k}^{\mathrm{R}}=v_{-{\bm k}}^{\mathrm{R}}$ and $v_{\bm
  k}^{\mathrm{I}}=-v_{-{\bm k}}^{\mathrm{I}}$. One has
\begin{align}
H= &\int_{\mathbb{R}^{3+}}  \dd ^3{\bm k}
\left[\frac12 \left(p_{\bm k}^\mathrm{R}\right)^2 
+\frac{k}{2} \left(v_{\bm k}^\mathrm{R}\right)^2 
+ \frac{z'}{z} {v_{\bm k}^\mathrm{R}}{p_{\bm k}^\mathrm{R}} \right]
\nonumber \\
& +\int_{\mathbb{R}^{3+}}  \dd ^3{\bm k}\left[
\frac12 \left(p_{\bm k}^\mathrm{I}\right)^2 
+
\frac{k}{2} \left(v_{\bm k}^\mathrm{I}\right)^2 + 
\frac{z'}{z} {v_{\bm k}^\mathrm{I}}{p_{\bm k}^\mathrm{I}} \right]\, .
\end{align}
One can see that the ``real'' and ``imaginary'' sectors are in fact
independent. This suggests that the quantum discord calculated with
respect to ${\cal E}={\cal E}_{\mathrm{R}}\otimes {\cal
  E}_{\mathrm{I}}$ should vanish. It is therefore interesting to
calculate the quantum discord calculated with respect to an arbitrary
subdivision of our system ${\cal E}={\cal E}_{1}\otimes {\cal
  E}_{2}$. Starting from $\hat{c}_{\bm k}$ and $\hat{c}_{-\bm k}$, the most generic pair of independent ladder operators can be written as
\begin{align}
\label{eq:defa1}
\hat{a}_1 & =\cos\alpha \hat{c}_{\bm k} + \sin\alpha \hat{c}_{-{\bm k}}\\
\label{eq:defa2}
\hat{a}_2 & =\ee^{2 i\alpha}\left( 
\cos \alpha \hat{c}_{-{\bm k}} - \sin \alpha \hat{c}_{\bm k}\right),
\end{align}
where $\alpha $ is an angle varying from $0$ to $\pi/4$. This defines
a one-parameter family of subdivision such that, if $\alpha =0$, then
${\cal E}={\cal E}_{\bm k}\otimes {\cal E}_{-{\bm k}}$ and, if
$\alpha=\pi/4$, then ${\cal E}={\cal E}_{\mathrm{R}}\otimes {\cal
  E}_{\mathrm{I}}$. One can then proceed and calculate the
corresponding discord along the lines already explained before. This
is what is done in detail in Appendix~\ref{sec:genediscord}. The
result is displayed in \Fig{fig:GeneralizedDiscord} for a few values
of $k/(aH)$. As expected, one can check that the discord always
vanishes for $\alpha=\pi/4$ (${\cal E}={\cal E}_{\mathrm{R}}\otimes
{\cal E}_{\mathrm{I}}$) while it is maximal when $\alpha=0$ (${\cal
  E}={\cal E}_{\bm k}\otimes {\cal E}_{-{\bm k}}$). It also increases
with the squeezing level, since more pairs of particles with
wavenumbers $\bm k$ and $-{\bm k}$ are produced as time proceeds, in
agreement with \Fig{fig:discord}.

The above analysis undoubtedly shows that the CMB is a strongly
discordant quantum system. In particular, it contains genuine quantum
correlations between the ${\bm k}$ and $-{\bm k}$ modes (and in fact,
as the previous calculation demonstrates, between any ``1'' and ``2''
modes except if $\alpha=\pi/4$). A priori, this makes a classical
description of the CMB impossible. Therefore, the fact that, in
practice, the CMB data are analyzed without any reference to the
quantum formalism appears to be rather paradoxical. The rest of this
paper is devoted to this question.

\section{Quantum Predictions for the Perturbations}
\label{sec:quantumclassical}
In practice, the quantities of observational interest are correlation functions. In this section, as a first step, we derive these correlation functions for the full quantum state~(\ref{eq:qstate}). We make use of the quantum information formalism, which sheds some new light on the standard calculation of inflationary cosmological perturbations.
\subsection{Quantum information theory and cosmological perturbations}
\label{subsec:quantumcorrel}
A two-mode squeezed state being Gaussian, its characteristic function
is Gaussian as well and all correlation functions can be calculated
from the covariance matrix. To compute this covariance matrix, we can
use the techniques of quantum information
theory~\cite{2007PhR...448....1W,Demarie:2012}. In order to do so, we
must formulate the problem such that this formalism can be
applied. This technical question actually plays an important role in
the following. The point is that the quantum information formalism is
designed mode by mode while, as already noted,
\Eqs{eq:defc1} and~(\ref{eq:defc2}) mix creation and annihilation
operator associated with two different modes, ${\bm k}$ and $-{\bm
  k}$. We must therefore reformulate the problem such that
``coordinates'' (``position'' and ``momentum'') are Hermitian
operators defined for a single mode. This leads us to introduce the
quantities $\hat{q}_{\bm k}$ and $\hat{\pi}_{\bm k}$ such that
\begin{eqnarray}
\label{eq:defq}
\hat{q}_{\bm k} &=& \frac{1}{\sqrt{2k}}\left(\hat{c}_{\bm k}
+\hat{c}_{\bm k}^{\dagger}\right)\\
\label{eq:defpi}
\hat{\pi}_{\bm k} &=& -i\sqrt{\frac{k}{2}}\left(\hat{c}_{\bm k}
-\hat{c}_{\bm k}^{\dagger}\right).
\end{eqnarray}
Here, and in the rest of this article, we work in the Schr\"odinger
picture where $\hat{q}_{\bm k}$ and $\hat{\pi}_{\bm k}$ are viewed as
constant and the state is evolving.  Contrary to \Eqs{eq:defc1}
and~(\ref{eq:defc2}) defining $v_{\bm k}$ and $p_{\bm k}$,
\Eqs{eq:defq} and~(\ref{eq:defpi}) do not mix different modes. Making
use of \Eqs{eq:defc1} and~(\ref{eq:defc2}), one can establish the
relation between the two sets of variables, which read
\begin{eqnarray}
\label{eq:vqpi}
\hat{v}_{\bm k} &=& \frac12\left[\hat{q}_{\bm k}+\hat{q}_{-{\bm k}}+\frac{i}{k}
\left(\hat{\pi}_{\bm k}-\hat{\pi}_{-{\bm k}}\right)\right] \\
\label{eq:pqpi}
\hat{p}_{\bm k} &=& \frac{1}{2i}
\left[k\left(\hat{q}_{\bm k}-\hat{q}_{-{\bm k}}\right)
+i\left(\hat{\pi}_{\bm k}+\hat{\pi}_{-{\bm k}}\right)\right]\, .
\end{eqnarray}
In particular, since $\hat{q}_{\bm k}=\hat{q}_{\bm k}^{\dagger}$ and
$\hat{\pi}_{\bm k}=\hat{\pi}_{\bm k}^{\dagger}$, one can check that we
indeed have $\hat{v}_{\bm k}^{\dagger} =\hat{v}_{-{\bm k}}$. The
quantum information formalism provides us with all correlation
functions for combinations of $\hat{q}_{\bm k}$ and $\hat{\pi}_{\bm
  k}$, and using \Eqs{eq:vqpi} and~(\ref{eq:pqpi}), correlation
functions for combinations of $\hat{v}_{\bm k}$ and $\hat{p}_{\bm k}$
can be inferred.
  
\par

Let us now introduce the characteristic function. It is a real function
defined on a four-dimensional real space, given by
\begin{equation}
\chi(\xi)=\tr\left[\hat{\rho} \hat{{\cal W}}(\xi)\right],
\label{eq:chi:def}
\end{equation}
where $\hat{{\cal W}}(\xi)$ is the Weyl operator, namely
\begin{equation}
\label{eq:WeylOperator:def}
\hat{{\cal W}}(\xi)={\rm e}^{i\xi ^{\rm T}\hat{R}},
\end{equation}
with $\hat{R}=\left(k^{1/2}\hat{q}_{\bm k},k^{-1/2}\hat{\pi}_{\bm
    k},k^{1/2}\hat{q}_{-{\bm k}},k^{-1/2}\hat{\pi}_{-{\bm
      k}}\right)^{\rm T}$. In the above expressions, $\hat{\rho}$ is
obviously the density operator.

\par

In appendix~\ref{sec:appendixcm}, we show that the two-mode squeezed
state has a Gaussian characteristic function given by
\begin{equation}
\chi (\xi)
={\rm e}^{-\xi^{\rm T}\gamma \xi/4}, 
\label{eq:chi:gamma}
\end{equation}
where $\gamma $ is the covariance matrix, related to the two-point
correlation functions by $\langle
\hat{R}_j\hat{R}_k\rangle=\gamma_{jk}/2+iJ_{jk}/2$. Here, $J$ is the
commutator matrix, $iJ_{j,k}=[\hat{R}_j,\hat{R}_k]$, given by $J=J_1
\oplus J_1$ with
\begin{equation}
\label{eq:defJ}
J_1=
\begin{pmatrix}
0 & 1 \\
-1 & 0
\end{pmatrix}.
\end{equation}
The covariance matrix of the two-mode squeezed state~(\ref{eq:qstate})
has been derived in appendix~\ref{sec:appendixcm} and this leads to
\begin{eqnarray}
\label{eq:vvminus}
\langle \hat{v}_{\bm k}\hat{v}_{{\bm p}}\rangle &=& \frac{1}{2k}
\left[\cosh(2r_k)-\sinh(2r_k)\cos(2\varphi_k)\right]\delta({\bm k}+{\bm p}),
\nonumber \\
\\
\label{eq:ppminus}
\langle \hat{p}_{\bm k}\hat{p}_{{\bm p}}\rangle &=& \frac{k}{2}
\left[\cosh(2r_k)+\sinh(2r_k)\cos(2\varphi_k)\right]\delta({\bm k}+{\bm p}),
\nonumber  \\ \\
\label{eq:vpminus}
\langle \hat{v}_{\bm k}\hat{p}_{{\bm p}}\rangle &=& \left[\frac{i}{2}-
\frac{1}{2}\sinh(2r_k)\sin(2\varphi_k)\right]\delta({\bm k}+{\bm p}).
\end{eqnarray}

\subsection{Two-point correlation functions}
\label{subsec:twoqm}

Using the above results, one can proceed and calculate the power
spectrum for the states~(\ref{eq:qstate}). Using \Eq{eq:tfv}, the
correlation function of the Mukhanov-Sasaki variable is given by
\begin{equation}
\label{eq:correlgeneral}
\langle \hat{v}(\eta,{\bm x})\hat{v}(\eta,{\bm y})\rangle 
=\frac{1}{(2\pi)^3}\int  \dd {\bm k} \dd {\bm p}
\langle \hat{v}_{\bm k}(\eta)\hat{v}_{\bm p}(\eta)\rangle 
{\rm e}^{i{\bm k}\cdot {\bm x}+i{\bm p}\cdot {\bm y}}.
\end{equation}
For the two-mode squeezed state, the only correlation function
$\langle \hat{v}_{\bm k}(\eta)\hat{v}_{\bm p}(\eta)\rangle$ that does not vanish
is when ${\bm p}=-{\bm k}$. From \Eq{eq:vvminus}, it follows that
\begin{align}
\langle \hat{v}(\eta,{\bm x})\hat{v}(\eta,{\bm y})\rangle 
=&\frac{1}{(2\pi)^3}\int  \dd {\bm k}
\frac{1}{2k}\bigl[\cosh(2r_k)
\nonumber \\  &
-\sinh(2r_k)\cos(2\varphi_k)\bigr]
{\rm e}^{i{\bm k}\cdot ({\bm x}-{\bm y})}.
\end{align}
We notice that the result only depends on $\vert {\bm x}-{\bm
  y}\vert $ as expected in a homogeneous and isotropic universe. As a
consequence, one can perform the angular integration and one obtains
\begin{align}
\langle \hat{v}(\eta,{\bm x})\hat{v}(\eta,{\bm y})\rangle 
 =\frac{4\pi }{(2\pi)^3}\int_0^{\infty} \frac{ \dd k}{k}
\frac{\sin (k\vert {\bm x}-{\bm y}\vert)}{k\vert {\bm x}-{\bm y}\vert}
\frac{k^3}{2k}
\nonumber \\  \times 
\bigl[\cosh(2r_k)
-\sinh(2r_k)\cos(2\varphi_k)\bigr],
\end{align}
or, using \Eq{eq:defzeta},
\begin{align}
\langle \hat{\zeta}(\eta,{\bm x})\hat{\zeta}(\eta,{\bm y})\rangle 
=\frac{4\pi }{(2\pi)^3}\frac{1}{a^2\Mp^22\epsilon_1}
\int_0^{\infty} \frac{ \dd k}{k}
\frac{\sin (k\vert {\bm x}-{\bm y}\vert)}{k\vert {\bm x}-{\bm y}\vert}
\nonumber \\ \times 
\frac{k^3}{2k}
\bigl[\cosh(2r_k)
-\sinh(2r_k)\cos(2\varphi_k)\bigr]\, .
\end{align}
To go further, one needs to specify $r_k$ and $\varphi_k$. In the de Sitter
limit, they are given by \Eqs{eq:r:deSitter}
and~(\ref{eq:varphi:deSitter}), and on super-Hubble scales ($-k\eta\ll
1$), this gives rise to $\cosh(2r_k)\simeq \sinh(2r_k)\simeq 1+1/(2k^2\eta^2)$ and
$\cos(2\varphi_k) \simeq
1-2k^2\eta^2$. As a consequence, the
power spectrum $P_v$ of $v(\eta,{\bm x})$, defined through $\langle\hat{v}_{\bm k} \hat{v}_{\bm p}\rangle = 2\pi^3\calP_v/k^3\delta({\bm k}-{\bm p})$, becomes ${\cal P}_v=1/(2\pi^2
2\eta^2)$ and the power spectrum of curvature perturbations can be
expressed as
\begin{equation}
\label{eq:Pzeta:standard}
{\cal P}_{\zeta}=\frac{H^2}{8\pi^2\Mp^2\epsilon_1},
\end{equation}
where we have used that in de Sitter spacetimes,
$a(\eta)=-1/(H\eta)$. This formula is the standard result~\cite{Martin:2004um,Martin:2015dha}. Of course,
strictly speaking, we have $\epsilon_1=0$ for exact de Sitter (there
are no density perturbations in de Sitter) so, in fact, the above
result is valid at leading order in slow roll only.

\par

Let us now try to evaluate other correlators, in particular the one
involving the curvature perturbation and its time derivative. For a
two-mode squeezed state, using \Eq{eq:defp}, we have
\begin{align}
\langle 
&\hat{\zeta}(\eta,{\bm x}) \hat{\zeta}'(\eta,{\bm y})+
\hat{\zeta}'(\eta,{\bm x})\hat{\zeta}(\eta,{\bm y})\rangle
\nonumber \\ 
& =\frac{1}{(2\pi)^3}\frac{1}{z^2(\eta)}
\int  \dd {\bm k}
\left(\langle \hat{v}_{\bm k}\hat{p}_{-{\bm k}}\rangle 
+\langle \hat{p}_{\bm k}\hat{v}_{-{\bm k}}\rangle\right)
{\rm e}^{i{\bm k}\cdot ({\bm x}-{\bm y})}.
\end{align}
The next step is to use \Eq{eq:vpminus} together with the fact that
$\left[\hat{v}_{\bm k},\hat{p}_{-{\bm k}}\right]=i$. This commutation relation
ensures that the term $i/2$ in \Eq{eq:vpminus} cancels out and that
the above correlation function is real. Then, one arrives at
\begin{align}
\langle 
\hat{\zeta}(\eta,{\bm x})& \hat{\zeta}'(\eta,{\bm y})+
\hat{\zeta}'(\eta,{\bm x})\hat{\zeta}(\eta,{\bm y})\rangle
\nonumber \\ 
&=-\frac{4\pi }{(2\pi)^3}\frac{1}{z^2(\eta)}
\int_0^{\infty} \frac{ \dd k}{k}
\frac{\sin (k\vert {\bm x}-{\bm y}\vert)}{k\vert {\bm x}-{\bm y}\vert}
\nonumber \\ & \times 
k^3
\sinh(2r_k)\sin(2\varphi_k).
\label{eq:TMSS:zetazetaprime}
\end{align}
In the de Sitter super-Hubble limit, this gives rise to $\langle \hat{\zeta}
\hat{\zeta}'+\hat{\zeta}'\hat{\zeta} \rangle \sim 1/z\sim \eta$ which is consistent with
the fact that $\zeta_{\bm k}$ and $p_{\bm k}=z\zeta_{\bm k}'$ are
constant on large scales.

\par

Finally, it is also interesting to evaluate the two-point correlation
function $\langle \hat{\zeta}'(\eta,{\bm x}) \hat{\zeta}'(\eta,{\bm
  y})\rangle$. Straightforward manipulations lead to
\begin{align}
\langle
\hat{\zeta}'(\eta,{\bm x}) \hat{\zeta}'(\eta,{\bm y})\rangle
& =\frac{1}{(2\pi)^3}\frac{1}{z^2(\eta)}
\int  \dd {\bm k} \langle \hat{p}_{\bm k}\hat{p}_{-{\bm k}}\rangle 
{\rm e}^{i{\bm k}\cdot ({\bm x}-{\bm y})}
\\ 
\label{eq:zetapzetap}
&=\frac{4\pi }{(2\pi)^3}\frac{1}{z^2(\eta)}
\int_0^{\infty} \frac{ \dd k}{k}
\frac{\sin (k\vert {\bm x}-{\bm y}\vert)}{k\vert {\bm x}-{\bm y}\vert}
\nonumber \\ & \times
k^3\frac{k}{2}\bigl[
\cosh(2r_k)
+\sinh(2r_k)\cos(2\varphi_k)\bigr].
\end{align}
For the exact de Sitter case, we have $\cosh(2r_k)
+\sinh(2r_k)\cos(2\varphi_k)=1$ and this correlator is proportional to
$\eta^2$ as expected. 

\section{Can the Quantum Correlation Functions Be Obtained with a non-Discordant State?}
\label{sec:classicalcorrel}

In the previous section, we have established the
correlation functions of the full quantum state. Given that, as already mentioned,
the CMB is usually analyzed in a classical framework, we now study to which
extent a ``classical state,'' \ie a non-discordant
quantum state, can reproduce these results. Notice that, the discord
being non-vanishing, we already know for sure that it cannot be the case
exactly. As a consequence, here, we rather want to study whether the difference between the exact and the classical approaches is large enough to be seen observationally, which, if so, would be considered as a proof that the perturbations are of quantum-mechanical origin.

\subsection{Two-point correlation functions}
\label{subsec:twopointclass}

The most generic expression for a bipartite system containing
classical correlations only is given by~\cite{Werner:1989zz,
  Zurek:2001, Datta:2010}
\begin{eqnarray}
\label{eq:classicalstate}
\hat{\rho}_{\rm cl} &=& \bigotimes _{{\bm k}\in \setR^{3+}}\sum_{n,m}p_{nm}({\bm k})
\hat{\rho}_n({\bm k})\otimes \hat{\rho}_m(-{\bm k})
\\
&=& \bigotimes _{{\bm k}\in \setR^{3+}}
\sum _{n,m} p_{nm}({\bm k})\vert n_{\bm k}\rangle \langle n_{\bm k}\vert 
\otimes \vert m_{-{\bm k}}\rangle \langle m_{-{\bm k}}\vert, 
\label{eq:classicalstate:2}
\end{eqnarray}
where the $p_{nm}$'s satisfy $\sum_{n,m}p_{nm}({\bm k})=1$. The $p_{nm}$'s
represent a joint probability distribution function that we do not
need to specify for the moment. Several comments are in order at this
point. Firstly, the above density matrix represents a quantum state that is \apriori anisotropic. Indeed, here, the momentum space is split into two
parts, $\setR^{3+}$ and $\setR^{3-}$, and the number of
quantas present in each corresponding mode \apriori differs. In the following,
we denote ``${\bm k}$'' a vector belonging to $\setR^{3+}$ and ``$-{\bm
  p}$'' a vector belonging to $\setR^{3-}$. Therefore, in
\Eq{eq:classicalstate:2}, $\vert m_{-{\bm k}}\rangle $ denotes
a state with $m$ quanta for $-{\bm k}\in \setR^{3-}$, the vector $-{\bm
  k}$ being of course minus the vector ${\bm k}$. Secondly, in
appendix~\ref{sec:appendixdiscord}, we explicitly show that, as stated above, the
quantum discord vanishes for the
state~(\ref{eq:classicalstate:2}). Thirdly, it is also important to
remark, as demonstrated in appendix~\ref{sec:appendixcmclassical},
that this state is not a Gaussian state. As a consequence, one cannot
define a covariance matrix. One has to proceed directly from the above
density matrix in order to calculate the different two-point (and higher)
correlation functions. Straightforward calculations give
\begin{align}
\label{eq:vvppclass}
\langle \hat{v}_{\bm k}\hat{v}_{\bm p}\rangle _{\rm cl}&=
\langle \hat{v}_{-{\bm k}}\hat{v}_{-{\bm p}}\rangle _{\rm cl}
= 0, 
\\
\label{eq:vvpmclass}
\langle \hat{v}_{\bm k}\hat{v}_{-{\bm p}}\rangle _{\rm cl}
& =\langle \hat{v}_{-{\bm k}}\hat{v}_{{\bm p}}\rangle _{\rm cl}
\nonumber \\ &=
\frac{1}{2k}\left[1+
\sum _{n,m}n\left(p_{nm}+p_{mn}\right)\right]\delta({\bm k}-{\bm p}).
\end{align}
The last expression contains a Dirac function which indicates that the
result is non-vanishing only if ${\bm k}={\bm p}$, ${\bm p} \in
\setR^{3+}$ being the vector obtained from $-{\bm p}\in \setR^{3-}$ by
multiplying by minus one. It is also interesting to notice that it can be written as the correlation function calculated in the Bunch-Davies vacuum, $1/(2k)$, multiplied by $1+\langle \hat{N} \rangle$, where $\langle \hat{N} \rangle$ is the mean of the particle number operator $\hat{N}=c_{{\bm k}}^\dagger c_{{\bm k}} + c_{-{\bm k}}^\dagger c_{-{\bm k}}$, in agreement with \Refs{2009PhRvD..80l3537D, 2011PhRvD..83f3526A}.
The other auto correlation functions read
\begin{align}
\label{eq:ppppclass}
\langle \hat{p}_{\bm k}\hat{p}_{\bm p}\rangle _{\rm cl}&=
\langle \hat{p}_{-{\bm k}}\hat{p}_{-{\bm p}}\rangle _{\rm cl}
= 0, 
\\
\label{eq:ppmclass}
\langle \hat{p}_{\bm k}\hat{p}_{-{\bm p}}\rangle _{\rm cl}
&=\langle \hat{p}_{-{\bm k}}\hat{p}_{{\bm p}}\rangle _{\rm cl}
\nonumber \\ &
=
\frac{k}{2}\left[1+
\sum _{n,m}n\left(p_{nm}+p_{mn}\right)\right]\delta({\bm k}-{\bm p}).
\end{align}
In particular, for any distribution $p_{nm}$, one has the following
property,
\begin{equation}
k^2\langle \hat{v}_{\bm k}\hat{v}_{-{\bm k}}\rangle_\ucl 
=\langle \hat{p}_{\bm k}\hat{p}_{-{\bm k}}
\rangle_\ucl,
\label{eq:classical:consistencyrelation}
\end{equation}
while one can check, see \Eqs{eq:vvminus} and~(\ref{eq:ppminus}), that
this is not the case for the two-mode squeezed state. We conclude that
a classical state~(\ref{eq:classicalstate}) cannot reproduce all the
correlation functions of the two-mode squeezed state in accordance
with the fact that they have a different quantum discord.

\par

Finally, the cross correlation functions can be written as
\begin{align}
\label{eq:vpclass}
\langle \hat{v}_{\bm k}\hat{p}_{\bm p}\rangle _{\rm cl}&= 
\langle \hat{v}_{-{\bm k}}\hat{p}_{-{\bm p}}\rangle _{\rm cl}=0, \\
\label{eq:vppmclass}
\langle \hat{v}_{\bm k}\hat{p}_{-{\bm p}}\rangle _{\rm cl}
&=\frac{i}{2}\left[1+
\sum _{n,m}n\left(p_{nm}-p_{mn}\right)\right]\delta({\bm k}-{\bm p})\\
\label{eq:vpmpclass}
\langle \hat{v}_{-{\bm k}}\hat{p}_{{\bm p}}\rangle _{\rm cl}
&=\frac{i}{2}\left[1+
\sum _{n,m}n\left(p_{mn}-p_{nm}\right)\right]\delta({\bm k}-{\bm p}).
\end{align}
The fact that $\langle \hat{v}_{\bm k}\hat{p}_{-{\bm p}}\rangle _{\rm
  cl}\neq \langle \hat{v}_{-{\bm k}}\hat{p}_{{\bm p}}\rangle _{\rm
  cl}$ is a consequence of the anisotropy of the quantum
state~(\ref{eq:classicalstate:2}). If one does not want to violate
homogeneity and isotropy, one simply has to require the
joint probability $p_{nm}$ to be symmetric, and the cosmological
principle is satisfied when
\begin{equation}
\label{eq:pnmsummetric}
p_{nm}=p_{mn}.
\end{equation}

Let us now be more accurate and repeat the calculation of the power
spectrum for the classical state. 
Using \Eqs{eq:vvppclass} and~(\ref{eq:vvpmclass}), it
is obvious that the standard power spectrum can always be
recovered provided the coefficients $p_{nm}$ are chosen such that
$\langle \hat{v}_{\bm k}(\eta)\hat{v}_{-{\bm k}}(\eta)\rangle_{\rm
  cl}=\langle \hat{v}_{\bm k}(\eta)\hat{v}_{-{\bm k}}(\eta)\rangle$,
that is to say
\begin{align}
\label{eq:pquantum}
1+\sum_{n,m} 2n p_{nm}=
\cosh(2r_k) -\sinh(2r_k)\cos(2\varphi_k)\, .
\end{align}
A possible solution is to take $p_{nm}=p_n\delta(n-m)$ with $p_n$
being given by a thermal distribution, $p_n=\left(1-{\rm
    e}^{-\beta_k}\right){\rm e}^{-\beta_k n}$, with
\begin{equation}
\label{eq:betak}
\beta_k =-\ln \left[\frac{\cosh(2r_k) -\sinh(2r_k)\cos(2\varphi_k)-1}
{\cosh(2r_k) -\sinh(2r_k)\cos(2\varphi_k)+1}\right].
\end{equation}
In the super-Hubble limit, the corresponding temperature goes to
infinity. The classical state~(\ref{eq:classicalstate}) can then be
written
\begin{equation}
\hat{\rho}_{\rm cl}=\left(1-{\rm e}^{-\beta_k}\right)
\sum_{n=0}^{+ \infty}{\rm e}^{-\beta_k n}
\left \vert n_{\bm k},n_{-{\bm k}}\rangle 
\langle n_{\bm k},n_{-{\bm k}} \right \vert\, .
\label{eq:rho:class}
\end{equation}
In other words, we have a thermal ``gas'' of pairs of particles with
momentum ${\bm k}$ and $-{\bm k}$. At the level of the power spectrum,
this ``classical'' description leads to the same result as the quantum
one.

As was done before for the two-mode squeezed state, let us now try to
evaluate other correlators, involving the curvature perturbation and
its time derivative. Using \Eqs{eq:vpclass},
(\ref{eq:vppmclass}) and~(\ref{eq:vpmpclass}) together with \Eq{eq:pnmsummetric}, we immediately see
that
\begin{align}
\langle 
\hat{\zeta}(\eta,{\bm x})& \hat{\zeta}'(\eta,{\bm y})+
\hat{\zeta}'(\eta,{\bm x})\hat{\zeta}(\eta,{\bm y})\rangle_{\rm cl}=0.
\end{align}
Several comments are in order. Firstly, we see that the classical
correlator is not equal to the quantum one; see \Eq{eq:TMSS:zetazetaprime}. It is possible to ensure
that the quantum and classical power spectra coincide, but then the
price to pay is that the other correlators differ. This is of course
conceptually very important. It means that by treating the problem
classically, rather than quantum mechanically, we are really doing
something wrong: some observables are not correctly calculated. This
is consistent with the fact that a two-mode squeezed state has a large
discord. Secondly, in practice, the difference between the quantum and
the classical results is tiny and unobservable probably forever. Indeed, the classical correlator is exactly zero and the quantum
one is proportional to $1/z\sim {\rm e}^{-N_*}$, where $N_*$ is the
number of \efolds between the Hubble radius exit during inflation and the
end of inflation. Modes probed in the CMB typically have $N_*\sim 50$
and, hence, the quantum correlator is very small for the CMB
scales. We conclude that, at the level of the two-point function, a
classical description is acceptable. Thirdly, in any case, it should be
clear that the classical description is an effective one only. The
state~(\ref{eq:classicalstate}) is introduced by hand and is not a
solution of the quantum Einstein equations.

\par

Finally, let us repeat the calculation of the correlator involving two
derivatives of the curvature perturbations for the classical
situation. We now have 
\begin{align}
\langle
\hat{\zeta}'(\eta,{\bm x}) \hat{\zeta}'(\eta,{\bm y})\rangle_{\rm cl}
& =\frac{1}{(2\pi)^3}\frac{1}{z^2(\eta)}
\int  \dd {\bm k} \langle \hat{p}_{\bm k}\hat{p}_{-{\bm k}}\rangle_{\rm cl} 
{\rm e}^{i{\bm k}\cdot ({\bm x}-{\bm y})}
\\ 
&=\frac{4\pi }{(2\pi)^3}\frac{1}{z^2(\eta)}
\int_0^{\infty} \frac{ \dd k}{k}
\frac{\sin (k\vert {\bm x}-{\bm y}\vert)}{k\vert {\bm x}-{\bm y}\vert}
\nonumber \\ & \times
k^3k \sum_{n,m}p_{nm}\left(n+\frac12\right).
\end{align}
But the coefficients $p_{nm}$ were specified when we required the
quantum and classical power spectra to be the same. This means that, as
already discussed, the above correlator is in fact explicitly
known. From \Eq{eq:pquantum} indeed, one has
\begin{align}
\langle
\hat{\zeta}'(\eta,{\bm x}) \hat{\zeta}'(\eta,{\bm y})\rangle_{\rm cl}
=\frac{4\pi }{(2\pi)^3}\frac{1}{z^2(\eta)}
\int_0^{\infty} \frac{ \dd k}{k}
\frac{\sin (k\vert {\bm x}-{\bm y}\vert)}{k\vert {\bm x}-{\bm y}\vert}
\nonumber \\  \times
k^3\frac{k}{2} \bigl[
\cosh(2r_k)
-\sinh(2r_k)\cos(2\varphi_k)\bigr],
\end{align}
which is different from \Eq{eq:zetapzetap}. In the de Sitter
super-Hubble limit, this gives that $\langle
\hat{\zeta}'\hat{\zeta}'\rangle _{\rm cl}$ is constant, in accordance with \Eq{eq:classical:consistencyrelation}. In fact, this last equation shows that $\langle
\hat{\zeta}'\hat{\zeta}'\rangle _{\rm cl}$ and $\langle
\hat{\zeta}\hat{\zeta}\rangle_{\rm cl}$ always scale in the same way. Since we have imposed $\langle
\hat{\zeta}\hat{\zeta}\rangle _{\rm cl}=\langle
\hat{\zeta}\hat{\zeta}\rangle$, and since $\langle
\hat{\zeta}\hat{\zeta}\rangle$ is constant on large scales, so is $\langle
\hat{\zeta}'\hat{\zeta}'\rangle _{\rm cl}$. Therefore, we
conclude that $\langle \hat{\zeta'}\hat{\zeta'}\rangle \sim {\rm
  e}^{-2N_*}$ and $\langle \hat{\zeta'}\hat{\zeta'}\rangle _{\rm
  cl}$ strongly differ ($\langle
\hat{\zeta}'\hat{\zeta'}\rangle/\langle
\hat{\zeta'}\hat{\zeta'}\rangle _{\rm cl}\sim {\rm e}^{-2N_*}\ll 1$)
and that, as before, the classical description cannot reproduce the
quantum results. Contrary to before however, the classical correlator is not
exponentially small, leaving thus the possibility to distinguish the two states experimentally.
\subsection{Non-Gaussianities}
\label{subsec:ng}

We have just seen that two-point correlators for a discordant and for
a non-discordant states differ but that this difference is, in
practice, very small. Another way to differentiate these two states is to look at
non-Gaussianities. All three-point correlation functions vanish in both the two-mode squeezed and classical states; hence one has to
consider four-point correlation functions. For the classical state, making use of the same techniques as above, one finds
\begin{widetext}
\begin{align}
\left\langle \hat{\zeta}_{{\bm k}_1} \hat{\zeta}_{{\bm k}_2} \hat{\zeta}_{{\bm k}_3} \hat{\zeta}_{{\bm k}_4}\right\rangle_\ucl  = &  \left\langle \hat{\zeta}_{{\bm k}_1} \hat{\zeta}_{{\bm k}_2} \right\rangle_\ucl \left\langle \hat{\zeta}_{{\bm k}_3} \hat{\zeta}_{{\bm k}_4}  \right\rangle_\ucl + \left\langle \hat{\zeta}_{{\bm k}_1} \hat{\zeta}_{{\bm k}_3} \right\rangle_\ucl \left\langle \hat{\zeta}_{{\bm k}_2} \hat{\zeta}_{{\bm k}_4}  \right\rangle_\ucl + \left\langle \hat{\zeta}_{{\bm k}_1} \hat{\zeta}_{{\bm k}_4} \right\rangle_\ucl \left\langle \hat{\zeta}_{{\bm k}_2} \hat{\zeta}_{{\bm k}_3}\right\rangle_\ucl  
 \nonumber \\ &  
+\frac{\delta_{{\bm k}_1+{\bm k}_2+{\bm k}_3+{\bm k}_4,{\bm 0}}}{2 k_1^2 z^4}   \left\lbrace 4 \sum_n n^2 p_n(k_1)-4 \left[\sum_n n p_n\left(k_1\right)\right]^2 - \sum_n n\left(n+1\right) p_n\left(k_1\right)\right\rbrace
 \nonumber \\ &  \times
 \left(\delta_{{\bm k}_1,-{\bm k}_2}\delta_{{\bm k}_1,{\bm k}_3}+\delta_{{\bm k}_1,-{\bm k}_2}\delta_{{\bm k}_1,-{\bm k}_3}+\delta_{{\bm k}_1,{\bm k}_2}\delta_{{\bm k}_1,-{\bm k}_3}\right).
 \label{eq:NG:class}
\end{align}
\end{widetext}
In this expression, we have kept the coefficients $p_n(k)$ unspecified so that the following generic comments can be made. The first line of \Eq{eq:NG:class} corresponds to the standard disconnected part of the four-point correlation function. For the two-mode squeezed state, it is the only term that is obtained,
which is in agreement with Wick's theorem given that, as already explained, the two-mode squeezed state is Gaussian. This means that the second and third lines of \Eq{eq:NG:class} correspond to a non-Gaussian contribution. This one is not of the local type and, in fact, corresponds to a scale dependent $g_{{}_\mathrm{NL}}$, notably due to the presence of the $\delta$ terms. Therefore, it remains to generate templates corresponding to this structure in order to determine whether such a four-point function is already excluded or not. Let us however notice that for the thermal state introduced above \Eq{eq:betak}, the term in braces is given by $(\cosh\beta_{k_1}-1)^{-1}$. On super-Hubble scales, $\beta_{k_1}\rightarrow 0$ and this term blows up. Let us also remark that the non-Gaussianities studied here are due to the intrinsic non-Gaussian character~\cite{Martin:1999fa, Gangui:2002qc} of the classical state and not to the non-linearities of the theory. But of course we expect those non-linearities to produce extra contributions, at the level of both the three- and four-point correlation functions~\cite{Holman:2007na,2011PhRvD..83f3526A}. We conclude that constraining the four-point correlation function would be an important test since it has the potential (depending on the level of classical non-Gaussianity) to rule out non discordant primordial states, hence indirectly proving the large quantumness of the CMB anisotropies.
\section{Can the Quantum Correlation Functions Be Obtained in a Classical Stochastic Description?}
\label{sec:stocha}
In section~\ref{sec:classicalcorrel}, we studied to what extent a
classical state can reproduce the quantum correlation
functions. We saw that, in principle, it cannot (even if, in practice, it remains to be seen whether it can be detected). Although it was called classical, the situation was still described in terms of a quantum
state. This is why here, we go one step further and even ignore the
quantum formalism, working only in terms of a stochastic probability
distribution.
\subsection{General properties}
\label{subsec:gene}
If an effective classical stochastic description is possible, quantum
averages of any operator $\hat{A}$ should be given by an integral of a
stochastic distribution over classical phase space, namely
\begin{equation}
\langle \hat{A} \rangle = \langle A\rangle _{\rm stocha},
\label{eq:Wigner:mean:superHubble}
\end{equation}
with 
\begin{equation}
\label{eq:meanstocha}
\langle A\rangle _{\rm stocha}\equiv 
\int A(R) W(R) \dd^4 R\, .
\end{equation}
As usual, a hat denotes a physical observable $\hat{A}=A(\hat{R})$,
while $A$ (no hat) is an ordinary real function. The vector $R$,
previously defined below \Eq{eq:WeylOperator:def}, represents the
classical variables in phase space and, obviously, $\hat{R}$ denotes
the corresponding operators $\hat{R}=\left(k^{1/2}\hat{q}_{\bm
    k},k^{-1/2}\hat{\pi}_{\bm k},k^{1/2}\hat{q}_{-{\bm
      k}},k^{-1/2}\hat{\pi}_{-{\bm k}}\right)^{\rm T}$. The quantity
$W(R)$ represents a probability density function. The stochastic
description is possible and meaningful if three conditions are met:
first, $W$ is positive everywhere and normalized to unity (so that it
can be interpreted as a distribution function); second, $W$ obeys the
classical equation of motion (see below); and, third,
\Eq{eq:Wigner:mean:superHubble} is valid.

\par

Let us now discuss how $W$ can be concretely found (if it can). For
this purpose, we introduce the Weyl transform $\tilde{A}$ of the
operator $\hat{A}$ through the following expression,
\begin{align}
\label{eq:Weyltransform:def}
\tilde{A}& \left(q_{\bm k},\pi_{\bm k}, q_{-\bm k}, 
\pi_{-\bm k}\right)\equiv \int\dd x\, \dd y \, 
\ee^{-i \pi_{\bm k} x -i\pi_{-\bm k} y} 
\nonumber\\ 
& \times \left\langle q_{\bm k}+\frac{x}{2},q_{-\bm k}+\frac{y}{2}   
\left \vert \hat{A} \right \vert q_{\bm k}
-\frac{x}{2},q_{-\bm k}-\frac{y}{2} \right\rangle,
\end{align}
which builds a real function in phase space, $\tilde{A}$, out of the
quantum operator $\hat{A}$. It is important to stress that $\tilde{A}$
is a function (\ie not an operator) which, for instance, means that
$\tilde{A}\tilde{B}=\tilde{B}\tilde{A}$. However, $\widetilde{AB}\neq
\widetilde{BA}$ in general. It is also important to notice that, a
priori, $\tilde{A}\neq A$. In fact, in
Appendix~\ref{sec:Wigner:meanValue}, the following exact result is
established,
\begin{equation}
\label{eq:weylR}
\widetilde{R}_i=R_i, \quad 
\widetilde{R_j R_k}=R_j R_k+\frac{i}{2}J_{jk},
\end{equation}
where the matrix $J$ is defined in \Eq{eq:defJ}. Moreover, since $R$
and $V\equiv \left(k^{1/2}v_{\bm k},k^{-1/2}p_{\bm k},k^{1/2}v_{-{\bm
      k}},k^{-1/2}p_{-{\bm k}}\right)^{\rm T}$ are linearly related
through \Eqs{eq:vqpi} and~(\ref{eq:pqpi}), this relation translates
into a similar one for $V$, namely
\begin{equation}
\widetilde{V}_i=V_i\, ,\quad \widetilde{V_j V_k}
=V_j V_k+\frac{i}{2}I_{jk}\, ,
\label{eq:Weyl:ViVj}
\end{equation}
where $I$ is the anti-diagonal matrix with coefficients $\lbrace
-1,1,-1,1 \rbrace$ on the antidiagonal and verifies
$iI_{j,k}=[\hat{V}_j,\hat{V}_k]$. Therefore, the only non-vanishing
terms of the $I$ matrix are the ones corresponding to products of the
form $v_{\bm k} p_{-\bm k}$. Even more general relations are
established in Appendix~\ref{sec:Wigner:meanValue}.

\par

The fundamental property of the Weyl transform is that
\begin{equation}
\tr \left(\hat{A}\hat{B}\right) 
=\int \tilde{A}(R)\tilde{B}(R) \frac{\dd^4R}{\left(2\pi\right)^2}.
\end{equation}
Given that $\langle \tilde{A}\rangle =\tr
\left(\hat{\rho}\hat{A}\right)$, this immediately suggests taking for
$W$ the Weyl transform of the density operator,
$W=\tilde{{\rho}}/(2\pi)^2$ [the factor $(2\pi)^2$ originates from the
fact that we consider a system whose phase space is $\mathbb{R}^4$;
had we considered a system with phase space $\mathbb{R}^{2n}$, this
factor would have been $(2\pi)^n$] namely
\begin{align}
\label{eq:Wigner:def}
W(R) \equiv &  \frac{1}{(2\pi)^2}
\int\dd x\, \dd y \, \ee^{-i \pi_{\bm k} x -i\pi_{-\bm k} y} 
\nonumber\\ 
& \times \left\langle q_{\bm k}+\frac{x}{2},q_{-\bm k}+\frac{y}{2}   
\biggl\vert \hat{\rho} \biggr \vert q_{\bm k}
-\frac{x}{2},q_{-\bm k}-\frac{y}{2} \right\rangle\, .
\end{align}
Then it follows that any expectation value of a physical observable
can be obtained through the relation
\begin{equation} \langle \hat{A} \rangle =
\int \tilde{A}\left(R\right) W\left(R\right) \dd^4 R\, ,
\label{eq:meanvalue:Wigner:Weyl}
\end{equation}
or, in other words,
\begin{equation}
\label{eq:meanqstocha}
\langle \hat{A} \rangle =\langle \tilde{A} \rangle _{\rm stocha}.
\end{equation}
It is worth emphasizing again that this equation is exact. As a
consequence, the condition given by
\Eq{eq:Wigner:mean:superHubble} is realized if the Weyl
transform equals its classical counter part, that is to say if
\begin{equation}
\label{eq:weylcondition}
\tilde{A}=A.
\end{equation}

\begin{figure*}[t]
\begin{center}
\includegraphics[width=0.45\textwidth,clip=true]{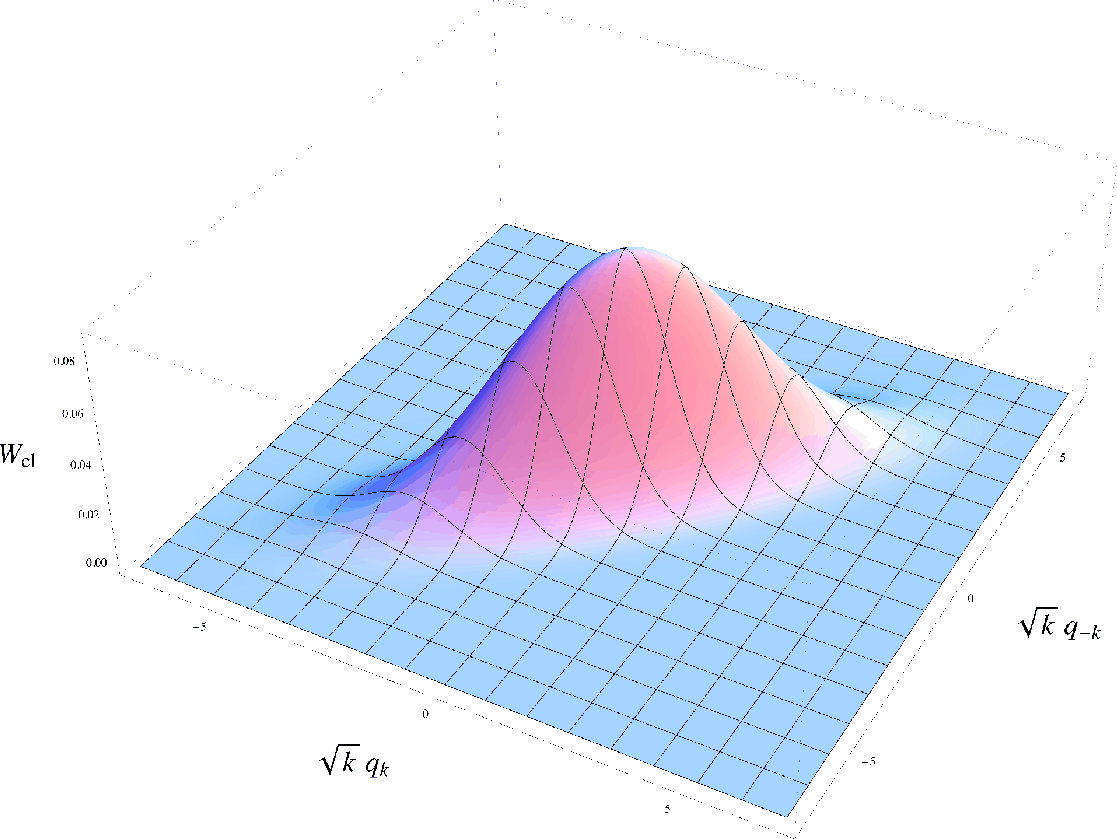}
\includegraphics[width=0.45\textwidth,clip=true]{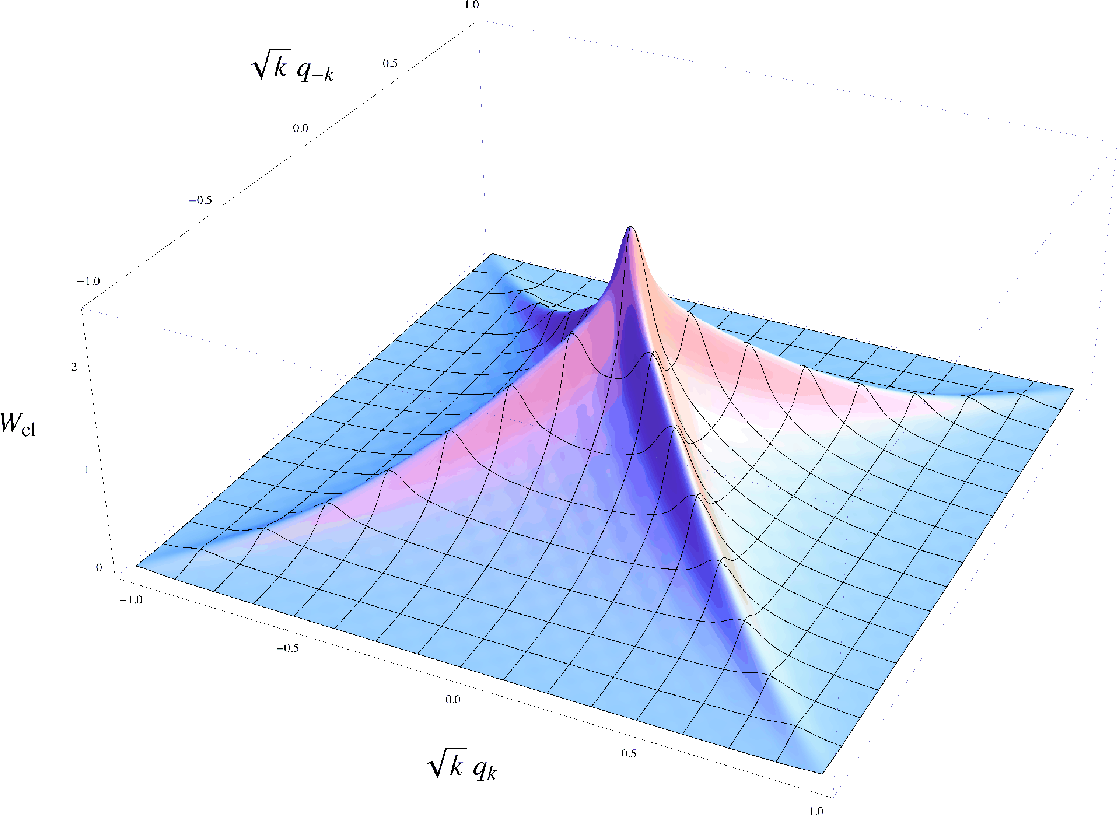}
\caption{Marginalized Wigner function $W(q_{\bm k},q_{-\bm k})=\int W
  (q_{\bm k},\pi_{\bm k},q_{-{\bm k}},\pi_{-{\bm k}})\dd\pi_{\bm k}\dd
  \pi_{-{\bm k}}$, for the super-Hubble mode $k/(aH)=0.3$. The left
  panel corresponds to the two-mode squeezed state while the right
  panel is for the classical state~(\ref{eq:rho:class}). One can see
  that as explained in the text and contrary to the two-mode squeezed
  state, the classical state is both non-Gaussian and unsqueezed.}
\label{fig:Wigner}
\end{center}
\end{figure*}

Let us conclude this sub-section by noticing
that, as is well known, $W(R)$ given by \Eq{eq:Wigner:def} is
the Wigner function~\cite{1932PhRv...40..749W,Case:2008}. For illustrative purposes, we now give the Wigner
function of the two-mode squeezed and classical states. With the
covariance matrices calculated in section~\ref{sec:quantumclassical},
one obtains an explicit expression
for the two-mode squeezed state Wigner function, namely (see appendix~\ref{appendix:Wigner:Gaussian})
\begin{align}
W = & \frac{1}{\pi^2}\exp\biggl[-\left(k q_{\bm k}^2 
+k q_{-\bm k}^2+\frac{\pi_{\bm k}^2}{k}+\frac{\pi_{-\bm k}^2}{k}\right)
\cosh(2r_k) 
\nonumber \\ &
-2\left(q_{\bm k}\pi_{-\bm k} + q_{-\bm k}\pi_{\bm k}\right)
\sin\left(2\varphi_k\right)\sinh\left(2r_k\right)
\nonumber \\ &
-2\left(k q_{\bm k} q_{-\bm k} - \frac{\pi_{\bm k}\pi_{-\bm k}}{k}\right)
\cos\left(2\varphi_k\right)\sinh\left(2r_k\right)
\biggr].
\label{eq:Wigner:qpi}
\end{align}
It is a positive, Gaussian function. Using
\Eqs{eq:vqpi} and~(\ref{eq:pqpi}), the corresponding expression in terms
of the $v_{\bm k}$ and $p_{\bm k}$ variables can be established as
well. One can notice that due to the last term, $W$ can neither be
factorized as $W=W_{\bm k}(q_{\bm k},\pi_{\bm k})W_{-{\bm k}}(q_{- \bm
  k},\pi_{- \bm k})$ nor as $W=W_{\bm k}(v_{\bm k},p_{\bm k})W_{-{\bm
    k}}(v_{- \bm k},p_{- \bm k})$, which reflects the presence of
correlations between the modes ${\bm k}$ and $-{\bm k}$. These
correlations vanish in the sub-Hubble limit [where
$\cos(2\varphi_k)\sinh(2r_k)=0$], but in the super-Hubble limit, the
Wigner function gets squeezed along the directions ``${\bm k}=-{\bm
  k}$,''
\begin{align}
W\propto \delta\left(q_{\bm k}-q_{-\bm k}\right)
\delta\left(\pi_{\bm k}+\pi_{-\bm k}\right)   \, ,
\end{align}
hence the term squeezing.

\par

Let us now calculate the Wigner function of the classical
state~(\ref{eq:classicalstate:2}). Combining \Eqs{eq:wigchi}
and~(\ref{eq:chiclassical}), one obtains after a few manipulations
\begin{align}
W_\ucl = & \frac{1}{\left(2\pi\right)^2}
 \int_0^\infty x\dd x\int_0^\infty y \dd y
 \nonumber\\  &
\exp\left[-\frac{\coth(\beta_k/2)}{4}\left(x^2+y^2\right)\right]
I_0\left[\mathrm{cosech}\left(\frac{\beta_k}{2}\right)\frac{xy}{4}\right]
\nonumber\\  &
J_0\left(x\sqrt{kq_{\bm k}^2+\frac{\pi_{\bm k}^2}{k}}\right)
J_0\left(y\sqrt{kq_{-\bm k}^2+\frac{\pi_{-\bm k}^2}{k}}\right)
\, ,
\end{align}
where $\beta_k$ has been defined in \Eq{eq:betak} and $I_0$ and $J_0$
are Bessel functions. As already pointed out in
section~\ref{sec:classicalcorrel}, one can see that the classical
state is a non-Gaussian one.\footnote{Let us notice that
  when one of the modes is integrated out, the classical Wigner
  function becomes Gaussian. For example, the marginalized Wigner
  function in the plane $(q_{\bm k},\pi_{\bm k})$, obtained by
  marginalizing over $q_{-\bm k}$ and $\pi_{-\bm k}$, can be expressed
  as the Fourier transform of the restrained characteristic function
  $\chi\left(\eta_1,\eta_2,0,0\right)$, and one obtains
\begin{align}
W_\ucl\left(q_{\bm k},\pi_{\bm k}\right) = 
\frac{1}{\pi}\tanh\left(\frac{\beta_k}{2}\right)
\ee^{-\tanh\left(\frac{\beta_k}{2}\right)
\left(kq_{\bm k}^2+\frac{\pi_{\bm k}^2}{k}\right)}\, .
\end{align}
Therefore, the non-Gaussianity of the classical Wigner function
directly reflects the (classical) correlations between the two modes
$\bm k$ and $- \bm k$.} It is also interesting to notice that, in the
sub-Hubble limit, $\beta_k\rightarrow\infty$, the argument of the
$I_0$ function vanishes and the integrals over $x$ and $y$ can be
factorized. This means that, in this limit, the Wigner function
becomes separable, $W=W_{\bm k}(q_{\bm k},\pi_{\bm k})W_{-{\bm
    k}}(q_{- \bm k},\pi_{- \bm k})$. When $\beta_k$ decreases, this is
not the case anymore and the two modes are correlated. This is similar
to what was noted for the two-mode squeezed state. However, these
correlations are of a different nature. Indeed, since the classical
Wigner function depends on the phase-space variables only through the
two combinations $kq_{\bm k}^2+\pi_{\bm k}^2/k$ and $kq_{-\bm
  k}^2+\pi_{-\bm k}^2/k$, it is rotationally invariant inside each
mode. In particular, squeezing does not take place. This is consistent
with the results obtained in the previous sections. In
\Fig{fig:Wigner}, we display the
marginalized Wigner functions $W(q_{\bm k},q_{-{\bm k}})$ for the
two-mode squeezed and classical states.

\subsection{The case of a gaussian state}
\label{subsec:gaussianstocha}
So far, the discussion was fully generic. We now discuss whether the
three conditions for the validity of a stochastic classical
description mentioned before, namely $W$ being positive everywhere, $W$
obeying the classical equation of motion and the validity of
\Eq{eq:Wigner:mean:superHubble} or, equivalently, of
\Eq{eq:weylcondition}, are satisfied for a Gaussian state.

\par

Let us start with the first condition, \ie the positivity of the
Wigner function. In appendix~\ref{appendix:Wigner:Gaussian}, it is
shown that, for Gaussian states, the characteristic function of which
is of the form given by \Eq{eq:chi:gamma}, one has
\begin{equation}
W\left(R\right)=\frac{1}{\pi^2\sqrt{\det \gamma}}
\ee^{-R^\mathrm{T}\gamma^{-1} R}\, .
\label{eq:Wigner:CovarianceMatrix}
\end{equation}
In this case, the Wigner function is therefore a (correctly
normalized) Gaussian function with covariance matrix $\gamma$ (hence
its name). This also means that it is positive at any time and, thus,
indeed satisfies our first condition. Notice that this is a general
feature of Gaussian states and is completely independent from the fact
that we have squeezing (large or not). For instance, a coherent state
also has a positive Wigner function.

\par

Let us now examine the second condition. In
appendix~\ref{appendix:Wigner:EOM}, we show that, for any quadratic
Hamiltonian, the Wigner function obeys a classical equation of
motion. Indeed, if one differentiates \Eq{eq:Wigner:def} with respect
to time and makes use of the Schr\"odinger equation, one obtains
\begin{align}
\frac{\dd W}{\dd \eta}=& 
\left\lbrace H_{\bm
    k},W \right\rbrace_{\mathrm{PB}}\, ,
\label{eq:Liouville}
\end{align}
where the right-hand side of this equation is the Poisson bracket
between the Hamiltonian $H_{\bm k}$ and the Wigner function. The above
equation therefore describes a Liouville evolution. This means that if
one starts from a collection of pairs of point-like particles [the
first one living in $(q_{\bm k},\pi_{\bm k})$, the other one in
$(q_{-\bm k},\pi_{-\bm k})$] at time $\eta$ that mimic the
distribution $W(\eta)$, and if one lets them all evolve according to
the classical Hamilton equations, the distribution calculated at later
time $\eta^\prime$ matches $W(\eta^\prime)$. We therefore conclude
that the second condition is also met. Again, this is completely
independent of whether we have large squeezing or not and is
true for any quadratic Hamiltonian.

\par

Finally, we come to the third condition, namely the equality between
quantum and stochastic correlators. Using \Eqs{eq:weylR}
and~(\ref{eq:meanqstocha}), one has
\begin{align}
\langle \hat{R}_j\hat{R}_k\rangle 
&=\langle\widetilde{R_jR_k}\rangle _{\rm stocha}
=\left \langle R_jR_k+\frac{i}{2}J_{jk}\right \rangle_{\rm stocha}
\\
&=\langle R_jR_k \rangle_{\rm stocha}+\frac{i}{2}J_{jk}.
\end{align}
This implies that for combinations of $\hat{R}_j$ and $\hat{R}_k$ such
that $J_{jk}=0$, the stochastic distribution reproduces exactly the
two-point quantum correlators. The only cases where $J_{jk}\neq 0$
correspond to mixed terms and we find
\begin{align}
\langle q_{\bm k}\pi_{\bm k}\rangle _{\rm stocha}
&=\langle q_{-{\bm k}}\pi_{-{\bm k}}\rangle _{\rm stocha}
=\langle \pi_{\bm k}q_{\bm k}\rangle _{\rm stocha}
\nonumber \\
&=\langle \pi_{-{\bm k}}q_{-{\bm k}}\rangle _{\rm stocha}
=0.
\end{align}
In terms of the Mukhanov-Sasaki variable and its conjugate momentum,
this means that
\begin{align}
\label{eq:vvminusstocha}
\langle v_{\bm k}v_{{\bm p}}\rangle _{\rm stocha}&= \frac{1}{2k}
\left[\cosh(2r_k)-\sinh(2r_k)\cos(2\varphi_k)\right]
\nonumber \\ & \times
\delta({\bm k}+{\bm p}),
\\
\label{eq:ppminusstocha}
\langle p_{\bm k}p_{{\bm p}}\rangle _{\rm stocha}&= \frac{k}{2}
\left[\cosh(2r_k)+\sinh(2r_k)\cos(2\varphi_k)\right]
\nonumber \\ & \times
\delta({\bm k}+{\bm p}),
\\
\label{eq:vpminusstocha}
\langle v_{\bm k}p_{{\bm p}}\rangle _{\rm stocha}&= 
-\frac{1}{2}\sinh(2r_k)\sin(2\varphi_k)\delta({\bm k}+{\bm p}).
\end{align}
These equations should be compared to their quantum counterparts,
namely \Eqs{eq:vvminus}, (\ref{eq:ppminus})
and~(\ref{eq:vpminus}).  We notice that the only correlator which is
not identical is the last one: $\langle \hat{v}_{\bm k}\hat{p}_{-{\bm
    k}}\rangle\neq \langle v_{\bm k}p_{-{\bm k}}\rangle _{\rm
  stocha}$; see \Eqs{eq:vpminus}
and~(\ref{eq:vpminusstocha}). There is an additional $i/2$ in the
quantum correlator which makes the difference. However, contrary
to the classical state studied in section~\ref{sec:classicalcorrel},
all the correlators of $\hat{\zeta}$ and $\hat{\zeta'}$ are correctly
reproduced. This is evident for $\langle \hat{\zeta}
\hat{\zeta}\rangle$ and $\langle \hat{\zeta'}\hat{\zeta'}\rangle $
since they do not involve the two-point function $\langle \hat{v}_{\bm
  k}\hat{p}_{-{\bm k}}\rangle$ but it is also true that
\begin{align}
\langle & 
\hat{\zeta}(\eta,{\bm x}) \hat{\zeta'}(\eta,{\bm y})+
\hat{\zeta'}(\eta,{\bm x})\hat{\zeta}(\eta,{\bm y})\rangle=
\nonumber \\
&= \langle 
\zeta(\eta,{\bm x})\zeta'(\eta,{\bm y})+
\zeta'(\eta,{\bm x})\zeta(\eta,{\bm y})\rangle_{\rm stocha}.
\end{align}
The reason is that the above operator is Hermitian and, therefore, the
complex term $i/2$ cancels out. This means that, as far as the two-point
correlators are concerned, the quantum and stochastic descriptions cannot be
observationally distinguished. Notice that this conclusion is
independent of the value of $r_k$ and is valid for any value of the
squeezing parameter.

\begin{center}
\begin{table*}[t]
\newcolumntype{L}[1]{>{\raggedright\let\newline\\\arraybackslash\hspace{0pt}}m{#1}}
\newcolumntype{C}[1]{>{\centering}m{#1}}
\newcolumntype{R}[1]{>{\raggedleft\let\newline\\\arraybackslash\hspace{0pt}}m{#1}}
\begin{tabular}{{|c||C{2.7cm}|C{2.7cm}||C{2.7cm}|C{2.7cm}|}}
\hline
\multirow{3}{*}{Correlation function} &  \multicolumn{2}{c||}{Classical state}& \multicolumn{2}{c|}{Stochastic distribution}\\
\cline{2-5}
& Generic & Super-Hubble   & Generic & Super-Hubble \newline (squeezed)  \tabularnewline
\hline
$\calP_\zeta\propto\langle \zeta_{\bm k}\zeta_{-{\bm k}}\rangle$ & \multicolumn{2}{c||}{Succeeds} & \multicolumn{2}{c|}{Succeeds} \tabularnewline
\hline
Other non-Hermitian two-point correlators & \multirow{2}{*}{Fails}  & \multirow{2}{*}{Strongly fails} & Fails & Succeeds \tabularnewline
\cline{1-1}\cline{4-5}
Other Hermitian two-point correlators &   &  & \multicolumn{2}{c|}{Succeeds} \tabularnewline
\hline
Higher order Hermitian correlators & Fails & Strongly fails & Fails & Succeeds \tabularnewline
\hline
\end{tabular}
\caption[]{Reproducibility of the two-mode squeezed state correlation functions by a classical state (see section~\ref{sec:quantumclassical}) and a classical stochastic distribution (see section~\ref{sec:stocha}).} 
\label{tab:summary} 
\end{table*}
\end{center}

Actually, squeezing only plays a role when computing higher order
observable quantities involving the momentum. For example, in
Appendix~\ref{sec:Wigner:meanValue}, it is shown that
$\widetilde{q_{\bm k}^2 \pi_{\bm k}^2}=q_{\bm k}^2 \pi_{\bm
  k}^2+2iq_{\bm k}\pi_{\bm k}-1/2$ and $\widetilde{\pi_{\bm k}^2
  q_{\bm k}^2}=\pi_{\bm k}^2 q_{\bm k}^2-2i\pi_{\bm k}q_{\bm
  k}-1/2$. Making use of \Eq{eq:meanqstocha}, this means
that \begin{align} \left\langle q_k^2\pi_k^2 + \pi_k^2 q_k^2
  \right\rangle_\mathrm{stocha} = \left\langle q_k^2\pi_k^2 + \pi_k^2
    q_k^2 \right\rangle - 1\, ,
\end{align}
so that the difference between the quantum and stochastic correlators
is still present for Hermitian operators. However, from the results of
section~\ref{subsec:quantumcorrel}, one has $\left\langle q_k^2\pi_k^2
  + \pi_k^2 q_k^2 \right\rangle = \cosh^2\left(2r_k\right)/2$ which is
exponentially large when $r_k\gg 1$. Therefore, in the large squeezing
limit, the stochastic description gives accurate results for these
correlators too.

As mentioned before, a difference between the quantum and stochastic cases arises only if the corresponding correlators contain $\hat{q}_{\bm k}$ and the momentum. In particular, we show in Appendix~\ref{sec:Wigner:meanValue} that $\widetilde{q^n_{\bm    k}}=q_{\bm k}^n$. However, this does not mean that higher order terms, \ie such as $\langle \hat{\zeta}_{\bm   k}^4\rangle$, will also be correctly and automatically reproduced since $\hat{\zeta}_{\bm k}=\hat{v_{\bm k}}/z$ is expressed in terms of $\hat{q}_{\bm k}$ but also in terms of $\hat{\pi}_{\bm k}$; see \Eq{eq:vqpi}. In other words, higher order correlators of $\hat{\zeta}_{\bm k}$ will precisely involve terms with various powers of $\hat{q}_{\bm k}$ and $\hat{\pi}_{\bm k}$ for which the squeezing plays a role.

It is usually said that a stochastic classical distribution
successfully reproduces the quantum correlators of cosmological
fluctuations in the large squeezing limit. The results of this section
shed some light on this statement. The fact that the Wigner function
is positive and evolves under the classical equations of motion is
independent of the squeezing level and is just a consequence of having
a quadratic Hamiltonian and a Gaussian state. For observable
(\ie Hermitian) operators, two-point correlators are also correctly
reproduced by the Wigner function regardless of the squeezing level,
and it is only when considering higher order correlators that
squeezing comes into play.\footnote{A remaining non-trivial question
  is whether a stochastic classical description is valid to describe (even approximately)
  correlators of non analytical functions in $v$ and $p$ (or $q$ and
  $\pi$)~\cite{2005PhRvA..71b2103R}.} Therefore, our ability to calculate
exactly the power spectrum in the stochastic approach has nothing to
do with the large squeezing limit but is in fact due to the property
that the Hamiltonian is quadratic.
\section{Discussion and Conclusion}
\label{sec:discussion}
Let us now summarize our main findings. The presence of quantum
correlations in a quantum state can be characterized by the recently
proposed quantum
discord~\cite{Henderson:2001,Zurek:2001}. Cosmological perturbations
produced during inflation are placed in a two-mode squeezed state, for
which we have calculated the quantum discord and showed that it is
very large at the end of inflation. This means that primordial
perturbations represent a highly non-classical system.

We have also computed the discord for a general splitting of the
system and shown that, except for the specific case where the two sub-systems correspond to the real and imaginary parts of the Mukhanov-Sasaki variable, the quantum discord is always very large on super-Hubble scales. In an ordinary situation, such as a Bell-type experiment where
the spin states of two remote systems are measured, the splitting is
unambiguous, and the two subsystems are obviously the two particles that
have travelled far away from each other. But the situation is less
clear when a measurement of a ``continuous'' system is performed. A
priori, a measurement corresponds to a realization which provides us
with all the values of the $q_{\bm k}$'s, \ie a map of the CMB. How
should we split this system into two subsystems? These tricky
questions are in fact connected to deeper issues related to the status
of the measurement problem in a cosmological context; see
\Refs{Sudarsky:2009za,Martin:2012pea}.

\par

In order to determine how the large quantum discord is reflected at
the observational level, we have compared the correlation
functions of the two-mode squeezed states with the ones that would be
obtained (i) from a ``classical state'', that is to say a state with
vanishing quantum discord, that contains classical correlations only;
and (ii) from a classical stochastic distribution. The results are
summarized in table~\ref{tab:summary}. One can design a classical
state that reproduces the observed primordial power spectrum. However, large
differences with the two-mode squeezed state correlators then appear
in other two-point functions, and such classical states are intrinsically
non-Gaussian. Then, if one
is ready to abandon the idea that perturbations should be described in the framework of quantum mechanics and wants to use a classical stochastic distribution
instead, we have seen that because the Hamiltonian is quadratic and
the two-mode squeezed state is Gaussian, the Wigner function provides
such a distribution, evolving under the classical equation of motion
and allowing us to reproduce all observable (\ie Hermitian) two-point
correlators. It is important to stress that the squeezing plays no
role at this level. The large squeezing is important only if one wants
to correctly reproduce non-Hermitian or higher order correlators.

Let us conclude with a few remarks. First, the fact that a classical
stochastic distribution succeeds in reproducing the two-mode squeezed
state late time correlators is due to most phase information being
contained in the decaying mode, which is not observable in the large
squeezing limit in the minimal setup we considered here. More
complicated situations, such as multi-field inflation~\cite{Wands:2007bd} where entropic
perturbations can source the decaying mode on large scales, or models
where features in the potential~\cite{Leach:2001zf,Jain:2008dw,Hazra:2010ve} give rise to transient violations of
slow roll (hence modified squeezing histories) may therefore lead to
different conclusions. Second, in light of the results of this
paper, the presence of large quantum correlations in primordial
perturbations can be seen as a remarkable feature of inflation. It
would be interesting to see whether this property is also present in
alternative scenarios~\cite{Battefeld:2014uga} to inflation, such as bouncing cosmologies~\cite{Pinto-Neto:2013npa} or
ekpyrotic setups~\cite{Battarra:2013cha}. 

The answer to the question raised in the title of this paper is yes in principle, and maybe in practice. On top of the possibles channels we exhibited in this article, other observations, such as Bell-type experiments, may also be capable of proving the existence of large quantum correlations in the CMB. We plan to investigate these issues in future publications~\cite{Martin:2016tbd, Martin:2017zxs}.\\

\section*{ACKNOWLEDGEMENTS}
We thank Rathul Nath Raveendran for pointing out a mistake in Sec.~\ref{sec:stocha}, and Kenta Ando for pointing out typos in Sec.~\ref{sec:infpert}, in a previous version of this article.
This work is supported by STFC Grants No. ST/K00090X/1 and No. ST/L005573/1.\\
%
\appendix
\onecolumngrid
\section{The Projected State is a Pure State}
\label{sec:appendixpure}
In this appendix, we show that the projected state defined
in~\Eq{eq:projectedState} is a pure state, and that, as a consequence,
its entropy vanishes. Let us start from a general bipartite state
which can always be written as
\begin{equation}
  \vert \psi \rangle=\sum _{n,m}c_{nm}\vert n_{\bm k},m_{-{\bm k}}\rangle.
\end{equation}
Then, the corresponding density matrix reads
\begin{equation}
  \hat{\rho}({\bm k},-{\bm k})=\sum_{n,m}\sum_{p,q}c_{nm}c_{pq}^*
\vert n_{\bm k}\rangle \langle p_{\bm k}
  \vert \otimes \vert m_{-{\bm k}}\rangle \langle q_{-{\bm k}}\vert .
\end{equation}
The next step, as was discussed around \Eq{eq:projectedState}, is to
consider the following projector which corresponds to the measurement
of an observable of the sub-system ``$-{\bm k}$''
\begin{equation}
\label{eq:defpij}
\hat{\Pi}_j=\widehat{{\rm Id}}_{\bm k}\otimes \vert j_{-{\bm k}}\rangle 
\langle j_{-{\bm k}}\vert \, .
\end{equation}
From this expression, it is easy to show that
\begin{equation}
  \hat{\rho} ({\bm k},-{\bm k})\hat{\Pi}_j =\sum_{n,m}\sum_{p}c_{nm}c_{pj}^*\vert n_{\bm k}
  \rangle \langle p_{\bm k}
  \vert \otimes \vert m_{-{\bm k}}\rangle \langle j_{-{\bm k}}\vert ,
\end{equation}
from which we deduce that
\begin{equation}
  \tr _{-{\bm k}}\left[\hat{\rho} ({\bm k},-{\bm k})\hat{\Pi}_j\right]
  =\sum_i \langle i_{-{\bm k}}\vert \hat{\rho} \hat{\Pi}_j\vert 
  i_{-{\bm k}}\rangle=
  \sum_{n}\sum_p c_{nj}c_{pj}^*
  \vert n_{\bm k}
  \rangle \langle p_{\bm k}
  \vert.
\end{equation}
We also have $\tr \left[\hat{\rho}({\bm k},-{\bm k})
  \hat{\Pi}_j\right]=\sum_nc_{nj}c_{nj}^*$ where, here, the trace is taken
over the full state space. Therefore, the state defined in
\Eq{eq:projectedState} can be expressed as
\begin{equation}
\hat{\rho}({\bm k};\hat{\Pi}_j)=\frac{1}{\sum_nc_{nj}c_{nj}^*}
\sum_{n}\sum_p c_{nj}c_{pj}^*
\vert n_{\bm k}
\rangle \langle p_{\bm k}
\vert\, .
\end{equation}
From this expression, it is straightforward to show that $\hat{\rho}^2({\bm
  k};\hat{\Pi}_j)=\hat{\rho}({\bm k};\hat{\Pi}_j)$. Indeed, using the above explicit
expression, one has
\begin{eqnarray}
  \hat{\rho}^2({\bm k};\hat{\Pi}_j) &=& 
  \frac{1}{\sum _nc_{nj}c_{nj}^*}\frac{1}{\sum _mc_{mj}c_{mj}^*}
  \sum_{pq}c_{pj}c_{qj}^*\vert p_{\bm k}
  \rangle \langle q_{\bm k}
  \vert
  \sum_{rs}c_{rj}c_{sj}^*\vert r_{\bm k}
  \rangle \langle s_{\bm k} \vert \\
  &=& 
  \frac{1}{\sum _nc_{nj}c_{nj}^*}\frac{1}{\sum _mc_{mj}c_{mj}^*}
  \sum_{pq}\sum_{rs}c_{pj}c_{qj}^*
  c_{rj}c_{sj}^*
  \vert p_{\bm k}\rangle
  \langle q_{\bm k}\vert r_{\bm k}
  \rangle \langle s_{\bm k} \vert \\
  &=& \frac{1}{\sum _nc_{nj}c_{nj}^*}\frac{1}{\sum _mc_{mj}c_{mj}^*}
  \sum_{p}\sum_{rs}c_{pj}c_{rj}^*c_{rj}c_{sj}^*
  \vert p_{\bm k}
  \rangle \langle s_{\bm k}
  \vert
  \\
  &=& \frac{1}{\sum _nc_{nj}c_{nj}^*}\frac{1}{\sum _mc_{mj}c_{mj}^*}
  \sum_r c_{rj}c_{rj}^*
  \sum_{p}\sum_{s}c_{pj}c_{sj}^*
  \vert p_{\bm k}
  \rangle \langle s_{\bm k}
  \vert
  \\
  &=& \frac{1}{\sum _nc_{nj}c_{nj}^*}
  \sum_{p}\sum_{s}c_{pj}c_{sj}^*
  \vert p_{\bm k}
  \rangle \langle s_{\bm k}
  \vert
  =\hat{\rho}({\bm k};\hat{\Pi}_j)\, .
\end{eqnarray}
It is therefore a pure state and, consequently, as noticed in the main
text, its entropy vanishes.

\section{Quantum Discord for a General Splitting of the System}
\label{sec:genediscord}

In this appendix, we calculate the discord for a general splitting of
the quantum system as discussed in Sec.~\ref{sec:discord}. The
calculations presented here lead to
\Fig{fig:GeneralizedDiscord}. The state of the system is given by
\Eq{eq:qstate} and can be written as
\begin{align}
\label{eq:defrho12}
\hat{\rho} &=\frac{1}{\cosh^2(r_{\bm k})}\sum_{n,n^\prime=0}^\infty
\ee^{2i(n-n^\prime)\varphi_{\bm k}}
(-1)^{n+n^\prime}
\tanh^{n+n^\prime}(r_{\bm k})
\vert n_{\bm k},n_{-\bm{k}}\rangle\langle n^\prime_{\bm k},n^\prime_{-\bm{k}}\vert 
= \sum_{n,n^\prime=0}^\infty a_{n,n^\prime}\vert n_{\bm k},n_{-\bm{k}}
\rangle\langle n^\prime_{\bm k},n^\prime_{-\bm{k}}\vert\, ,
\end{align}
which defines the coefficients $a_{n,n^\prime}$. Here, the density
matrix is expanded over the states $\vert n_{\bm k}, n_{-\bm k}
\rangle$ which is especially convenient when the system is split
according to ${\cal E}={\cal E}_{\bm k}\otimes {\cal E}_{-{\bm
    k}}$. In this appendix, however, we consider the case where ${\cal
  E}={\cal E}_{1}\otimes {\cal E}_{2}$, the corresponding ladder
operators being given by \Eqs{eq:defa1} and~(\ref{eq:defa2}). It 
is therefore more convenient to use the states
\begin{align}
\label{eq:defstate12}
\vert n_1, m_2\rangle = \frac{\left(\hat{a}_1^\dagger\right)^{n}}
{\sqrt{n!}}\frac{\left(\hat{a}_2^\dagger\right)^{m}}{\sqrt{m!}} 
\vert 0_1, 0_2\rangle,
\end{align}
containing $n$ quantas in the mode $1$ and $m$ quantas in the mode
$2$. As a consequence, one must first write $\vert n_{\bm k}, n_{-\bm
  k} \rangle = \sum_{m,p}\langle m_1,p_2\vert n_{\bm k}, n_{-\bm
  k}\rangle \vert m_1,p_2\rangle $ and calculate $\langle m_1,p_2\vert
n_{\bm k}, n_{-\bm k}\rangle $. This quantity can be written as
\begin{align}
\label{eq:matrixelement}
\langle m_1,p_2 \vert n_{\bm k}, n_{-\bm k} \rangle = 
\frac{1}{\sqrt{m!p!n!n!}}
\langle 0_1, 0_2 \vert
\left(\hat{a}_1\right)^{m}
\left(\hat{a}_2\right)^{p}
\left(\hat{c}_{\bm k}^\dagger\right)^{n}
\left(\hat{c}_{-\bm k}^\dagger\right)^{n}
\vert 0_1,0_2\rangle\, ,
\end{align}
where we used \Eq{eq:defstate12} and that $\vert 0_{\bm
  k},0_{-{\bm k}}\rangle=\vert 0_1,0_2\rangle$. Then, the idea of the
calculation is to permutate the operators such that
$(\hat{a}_1)^{m}$ and $(\hat{a}_2)^{p}$ directly act on $\vert
0_1,0_2\rangle$. In order to do so, we use the fact~\cite{2013JPhA...46c5304P}  that, for two
operators $\hat{X}$ and $\hat{Y}$ such that
$[\hat{X},\hat{Y}]=d\hat{I}$, one has
\begin{align}
\left[\hat{X}^n,\hat{Y}^m\right]=\sum_{\ell =1}^{\mathrm{min}(n,m)}d^\ell 
\ell ! \binom{n}{\ell}\binom{m}{\ell}\hat{Y}^{m-\ell}\hat{X}^{n-\ell}\, ,
\end{align}
where the coefficient $\binom{n}{\ell}=n!/[(n-\ell)!\ell!]$. This
immediately implies that
\begin{align}
\hat{X}^n\hat{Y}^m=\sum_{\ell =0}^{\mathrm{min}(n,m)}d^\ell \ell ! 
\binom{n}{\ell}\binom{m}{\ell}\hat{Y}^{m-\ell}\hat{X}^{n-\ell}\, .
\end{align}
Let us make use of this relation. Since $[\hat{a}_1,\hat{c}_{\bm
  k}^\dagger]=\cos \alpha$, see \Eq{eq:defa1}, one has
\begin{align}
\hat{a}_1^{m}\left(\hat{c}_{\bm k}^\dagger\right)^{n}
=\sum_{\ell _1=0}^{\mathrm{min}(m,n)}(\cos \alpha)^{\ell_1}  \ell_1 ! 
\binom{m}{\ell_1}\binom{n}{\ell_1}
\left(\hat{c}_{\bm k}^\dagger\right)^{n-\ell_1}\hat{a}_1^{m-\ell_1}\, .
\end{align}
Moreover, since $\hat{a}_1$ and $\hat{a}_2$ commute, one can first
permute them and then use the relation we have just established. In
that case \Eq{eq:matrixelement} can be written as
\begin{align}
\langle m_1,p_2 \vert n_{\bm k}, n_{-\bm k} \rangle = 
\frac{1}{\sqrt{m!p!n!n!}}
\sum_{\ell_1 =0}^{\mathrm{min}(m,n)}(\cos \alpha)^{\ell_1} 
\ell_1 ! \binom{m}{\ell_1}\binom{n}{\ell_1}
\langle 0_1, 0_2 \vert
\left(\hat{a}_2\right)^{p}
\left(\hat{c}_{\bm k}^\dagger\right)^{n-\ell_1}\hat{a}_1^{m-\ell_1}
\left(\hat{c}_{-\bm k}^\dagger\right)^{n}
\vert 0_1,0_2\rangle\, .
\end{align}
In order for $\hat{a}_1$ to act on the vacuum, we see that one more
permutation is needed. It can be obtained by using again the same
trick. Since $[\hat{a}_1,\hat{c}_{-\bm k}^\dagger]=\sin \alpha $, see
\Eq{eq:defa1}, one has
\begin{align}
\hat{a}_1^{m-\ell_1}
\left(\hat{c}_{-\bm k}^\dagger\right)^{n} = 
\sum_{\ell_2 =0}^{\mathrm{min}(m-\ell_1,n)}(\sin \alpha)^{\ell_2} \ell_2 ! 
\binom{m-\ell_1}{\ell_2}\binom{n}{\ell_2}
\left(\hat{c}_{-\bm k}^\dagger\right)^{n-\ell_2}\hat{a}_1^{m-\ell_1-\ell_2}\, ,
\end{align}
and, therefore, one can write
\begin{align}
\langle m_1,p_2 \vert n_{\bm k}, n_{-\bm k} \rangle = &
\frac{1}{\sqrt{m!p!n!n!}}
\sum_{\ell_1 =0}^{\mathrm{min}(m,n)}
\sum_{\ell_2 =0}^{\mathrm{min}(m-\ell_1,n)}
(\cos \alpha)^{\ell_1} \ell_1 !
\binom{m}{\ell_1}\binom{n}{\ell_1}
(\sin \alpha)^{\ell_2} \ell_2 ! \binom{m-\ell_1}{\ell_2}\binom{n}{\ell_2}
\nonumber \\ & 
\langle 0_1, 0_2 \vert
\left(\hat{a}_2\right)^{p}
\left(\hat{c}_{\bm k}^\dagger\right)^{n-\ell_1}
\left(\hat{c}_{-\bm k}^\dagger\right)^{n-\ell_2}\hat{a}_1^{m-\ell_1-\ell_2}
\vert 0_1,0_2\rangle\, .
\end{align}
This expression vanishes unless $m-\ell_1-\ell_2=0$ or
$\ell_2=m-\ell_1$. Moreover, clearly, if $n<m-\ell_1$, then
$\mathrm{min}(m-\ell_1, n)=n$, the sum over $\ell_2$ runs from $0$ to
$n$, and $\ell_2$ can never reach $m-\ell_1$. In that case, the above
expression is zero. On the contrary, if $n>m-\ell_1$, then
$\mathrm{min}(m-\ell_1, n)=m-\ell_1$ and the sum over $\ell_2$ runs
from $0$ to $m-\ell_1$. But if $n>m-\ell_1$, it also means that
$\ell_1>m-n$ and, if this quantity is positive, the sum over $\ell_1$
must start from this value rather than from zero. In other words, it
must start from $\mathrm{max}(0,m-n)$. Therefore, one obtains
\begin{align}
\langle m_1,p_2 \vert n_{\bm k}, n_{-\bm k} \rangle = &
\frac{1}{\sqrt{m!p!n!n!}}
\sum_{\ell_1 =\mathrm{max}(0,m-n)}^{\mathrm{min}(m,n)}
(\cos \alpha)^{\ell_1} \ell_1 !
\binom{m}{\ell_1}\binom{n}{\ell_1}
(\sin \alpha)^{m-\ell_1} \left(m-\ell_1\right) !\binom{n}{m-\ell_1}
\nonumber \\ & 
\langle 0_1, 0_2 \vert
\left(\hat{a}_2\right)^{p}
\left(\hat{c}_{\bm k}^\dagger\right)^{n-\ell_1}
\left(\hat{c}_{-\bm k}^\dagger\right)^{n-m+\ell_1}
\vert 0_1,0_2\rangle\, .
\end{align}
The next step is now to bring the operator $\hat{a}_2$ to the left and
the same method we have just described can be applied. After
straightforward manipulations, the final expression reads
\begin{align}
\langle m_1,p_2 \vert n_{\bm k}, n_{-\bm k} \rangle = &
\sqrt{\frac{m!p!}{n!n!}}
\sum_{\ell_1 =\mathrm{max}(0,m-n)}^{\mathrm{min}(m,n)}
(\cos \alpha)^{2\ell_1+n-m} 
(\sin \alpha)^{m+n-2\ell_1} 
(-1)^{n- \ell_1}
\ee^{2ip\alpha} 
\nonumber \\ &
\binom{n}{m-\ell_1}
\binom{n}{\ell_1}
\delta\left(m+p,2n\right)\, .
\end{align}
The appearance of the $\delta$ function makes sense since it means
that the same total number of particles must be contained in both
states. In particular, one can check that, in the specific case
$\alpha=0$, the previous expression gives $\delta(m-n)\delta(p-n)$,
as it should.

\par

Using the expression previously established, one can then write
\begin{align}
\vert n_{\bm k}, n_{-\bm k} \rangle = &
\sum_{m=0}^{+\infty}\sum_{p=0}^{+\infty}\langle m_1,p_2
\vert n_{\bm k}, n_{-\bm k}\rangle \vert m_1,p_2\rangle
=  
\sum_{m=0}^{2n}\langle m_1,(2n-m)_2\vert n_{\bm k}, n_{-\bm k}\rangle 
\vert m_1,(2n-m)_2\rangle
\\ = & 
\left[-(
\cos \alpha) 
(\sin \alpha) \right]^{n}
\sum_{m=0}^{2n}
\frac{\sqrt{m!(2n-m)!}}{n!}
\ee^{2i(2n-m)\alpha} 
(\tan \alpha)^m
\nonumber\\ &
\sum_{\ell =\mathrm{max}(0,m-n)}^{\mathrm{min}(m,n)}  
( \tan \alpha)^{-2\ell} 
(-1)^{-\ell}
\binom{n}{m-\ell}
\binom{n}{\ell}
\vert m_1,(2n-m)_2\rangle\, .
\end{align}
This expression can be further simplified as the second sum can be
calculated in terms of a hypergeometric function. Indeed, the sum
over $m$ can be split into a sum from $m=0$ to $m=n$, in which case
one has $m\leq n$ and a sum from $m=n+1$ to $m=2n$, in which case one
has $m>n$. But in these two cases, the corresponding sum over $\ell$
is known since for $m\leq n$, one has
\begin{align}
\sum_{\ell=0}^m \left[-\frac{1}{\tan^2 \alpha}\right]^\ell\binom{n}{m-\ell}
\binom{n}{\ell}=\binom{n}{m}{}_2F_1
\left(-m,-n,1-m+n,-\frac{1}{\tan^2\alpha }\right)\, ,
\end{align}
while, for $m\geq n$, one can write
\begin{align}
\sum_{\ell=m-n}^n \left[-\frac{1}{\tan^2(\alpha)}\right]^\ell\binom{n}{m-\ell}
\binom{n}{\ell}=\left[-\frac{1}{\tan^2(\alpha)}\right]^{m-n}
\binom{n}{m-n}{}_2F_1\left(m-2n,-n,1+m-n,-\frac{1}{\tan^2 \alpha}\right)\, .
\end{align}
As a consequence, the state $\vert n_{\bm k}, n_{-\bm k} \rangle $ can
be expressed as
\begin{align}
\label{eq:nstate12}
\vert n_{\bm k}, n_{-\bm k} \rangle = &
\frac{\left[-
\cos(\alpha) 
\sin(\alpha) \right]^{n}}{n!}
\ee^{4in\alpha}
\left[
\sum_{m=0}^{n}
\sqrt{(m)!(2n-m)!}
\ee^{-2im\alpha} 
(\tan \alpha )^m
\right. \nonumber\\ & \left.
\binom{n}{m}{}_2F_1\left(-m,-n,1-m+n,-\frac{1}{\tan^2 \alpha}\right) 
\vert m_1,(2n-m)_2\rangle
+
\sum_{m=n+1}^{2n}
\sqrt{(m)!(2n-m)!}
\ee^{-2im\alpha} 
\right. \nonumber\\ & \left.
(-1)^{m-n}
(\tan \alpha)^{2n-m}
\binom{n}{m-n}{}_2F_1\left(m-2n,-n,1+m-n,
-\frac{1}{\tan^2 \alpha}\right)\vert m_1,(2n-m)_2\rangle
\right]
\\ 
\label{eq:nstate122}
= &
\sum_{m=0}^{2n}b_{nm}\vert m_1,(2n-m)_2\rangle
\, ,
\end{align}
which defines the coefficients $b_{nm}$. Combining
\Eqs{eq:defrho12} and~(\ref{eq:nstate122}), one can write the
expression of the density matrix as
\begin{align}
\label{eq:finalrho12}
\hat{\rho}_{12}=\sum_{n,n^\prime=0}^\infty\sum_{m=0}^{2n}\sum_{m^\prime=0}^{2n^\prime}
a_{nn^\prime}b_{nm}b^*_{n^\prime m^\prime}\vert m,2n-m\rangle
\langle m^\prime,2n^\prime-m^\prime\vert \, .
\end{align}

We are now in a position to calculate the discord along the lines of
Sec.~\ref{sec:discord}. The quantity ${\cal I}_{12}$ is defined by
${\cal I}_{12}=S(\hat{\rho}_1)+S(\hat{\rho}_2)-S(\hat{\rho}_{12})$,
where $\hat{\rho}_1$ ($\hat{\rho}_2$) is obtained from
$\hat{\rho}_{12}$ by tracing out the degrees of freedom of the
sub-system $2$ ($1$). Then let us imagine that we make a measurement
on the sub-system $1$ corresponding to an operator $\hat{\Pi}_j$. This
defines
$\hat{\rho}_2(\hat{\Pi}_j)=\tr_1(\hat{\rho}_{12}\hat{\Pi}_j/p_j)$
where $p_j=\tr (\hat{\rho}_{12}\hat{\Pi})$ and ${\cal
  J}_{12}=S(\hat{\rho}_2)
-\sum_jp_jS\left[\hat{\rho}_2(\hat{\Pi}_j)\right]$. However, we have
shown in Appendix~\ref{sec:appendixpure} that
$S\left[\hat{\rho}_2(\hat{\Pi}_j)\right]=0$ since $\hat{\rho}_{12}$ is
still a pure state. As a result, ${\cal J}_{12}=S(\hat{\rho}_2)$ and,
hence, the discord is just given by $\delta_{12}=S(\hat{\rho}_1)$. We
see that we only need to calculate the entropy of $\hat{\rho}_1$. In
order to obtain this reduced density matrix, we trace out the second
set of degrees of freedom. Using \Eq{eq:finalrho12}, this leads
to the following expression:
\begin{align}
  \hat{\rho}_1= & \sum_{p=0}^\infty\langle p_2\vert\hat{\rho}\vert
  p_2\rangle =
  \sum_{p=0}^\infty\sum_{n,n^\prime=0}^\infty\sum_{m=0}^{2n}
  \sum_{m^\prime=0}^{2n^\prime}a_{nn^\prime}b_{nm}b^*_{n^\prime
    m^\prime} \langle p_2\vert m,2n-m\rangle\langle
  m^\prime,2n^\prime-m^\prime\vert p_2\rangle ,\\ = &
  \sum_{n,n^\prime=0}^\infty\sum_{m=0}^{2n}a_{n,n^\prime}b_{n,m}b^*_{n^\prime,
    m+2(n^\prime-n)} \vert m\rangle\langle m+2(n^\prime-n)\vert .
\end{align}

Since $\hat{\rho}_1$ is a priori not a thermal state, one cannot
directly apply \Eq{eq:thermalentropy}. In an analogy with
\Eqs{eq:defq} and~(\ref{eq:defpi}), let us therefore introduce
the quantities $\hat{q}_{1,2}$ and $\hat{\pi}_{1,2}$ defined by
$\hat{q}_{1,2}=(\sqrt{2k})^{-1}(a_{1,2}+a_{1,2}^{\dagger})$ and
$\hat{\pi}_{1,2}=-i\sqrt{k/2}(a_{1,2}-a_{1,2}^{\dagger})$. This
defines the vector
$\hat{P}_j=(k^{1/2}\hat{q}_1,k^{-1/2}\hat{\pi}_1,k^{1/2}\hat{q}_2,
k^{1/2}\hat{\pi_2})$ which is the equivalent of $\hat{R}_j$ defined
after \Eq{eq:WeylOperator:def}. Then the covariance matrix
$\gamma_{jk}$ is defined by $\left\langle \hat{P}_j\hat{P}_k\right
\rangle = \gamma_{jk}/2+iJ_{jk}/2$; see below \Eq{eq:chi:gamma}
in the text and the definition of the matrix $J_{jk}$ given by
\Eq{eq:defJ}. In fact, since we want to calculate the entropy
of the sub-system $1$ only, it is sufficient to consider the
covariance matrix in this sub-system. Straightforward calculations lead
to
\begin{align}
\gamma_{11}& =k\langle
\hat{q}_1^2\rangle =\langle
\hat{a}_1^2+(\hat{a}_1^{\dagger})^2+2\hat{a}_1\hat{a}_1^{\dagger}\rangle-1
=\tr\{\hat{\rho}_1[\hat{a}_1^2+(\hat{a}_1^{\dagger})^2
+2\hat{a}_1\hat{a}_1^{\dagger}]\}-1, \\
\gamma_{12} &=\gamma_{21}=i\left\langle 
(\hat{a}_1^{\dagger})^2-\hat{a}_1^2\right\rangle = i\tr\lbrace\hat{\rho}_1[(\hat{a}_1^{\dagger})^2-\hat{a}_1^2 ] \rbrace, \\
\gamma_{22} &= -\langle
\hat{a}_1^2+(\hat{a}_1^{\dagger})^2-2\hat{a}_1\hat{a}_1^{\dagger}\rangle-1=-\tr\{\hat{\rho}_1[\hat{a}_1^2+(\hat{a}_1^{\dagger})^2-2\hat{a}_1\hat{a}_1^{\dagger}]\}-1,
\end{align}
which is explicitly known since we have determined the density matrix
$\hat{\rho}_1$; see \Eq{eq:finalrho12}. As is evident from the
above expressions, in order to find $\gamma_{jk}$, it is sufficient to
determine the three following quantities, $\langle \hat{a}_1^2\rangle$,
$\langle (\hat{a}_1^{\dagger})^2\rangle$ and $\langle
\hat{a}_1\hat{a}_1^{\dagger}\rangle$. For the sake of illustration,
let us calculate the first one. One has
\begin{align}
\langle
\hat{a}_1^2\rangle &= \mathrm{Tr}(\hat{\rho}_1 \hat{a}_1^2)= 
\sum_{p=0}^\infty 
\sum_{n,n^\prime=0}^\infty\sum_{m=0}^{2n}a_{n,n^\prime}b_{n,m}b^*_{n^\prime, m+2(n^\prime-n)}
\langle p \vert
 m\rangle\langle m+2(n^\prime-n)\vert
 \hat{a}_1^2
\vert p \rangle,
\\ = &
\sum_{p=2}^\infty 
\sum_{n,n^\prime=0}^\infty\sum_{m=0}^{2n}a_{n,n^\prime}b_{n,m}b^*_{n^\prime, m+2(n^\prime-n)}
\sqrt{p(p-1)}
\langle p \vert
 m\rangle\langle m+2(n^\prime-n)
\vert p-2 \rangle,
\\ = &
\sum_{n=1}^\infty
\sum_{m=2}^{2n}
a_{n,n-1}b_{n,m}b^*_{n-1,m-2}
\sqrt{m(m-1)}.
\end{align}
The two other quantities are given by similar calculations. Then it
is sufficient to calculate the eigenvalues $\kappa $ of the matrix
$(\gamma/2) J_1$ to directly obtain the symplectic spectrum, thanks to
Williamson's theorem~\cite{Williamson:1936}. One obtains
\begin{equation}
\kappa=\pm i\sqrt{\left(\langle \hat{a}_1 \hat{a}_1^{\dagger}\rangle-\frac12\right)^2
-\langle \hat{a}_1^2\rangle \langle (\hat{a}_1^{\dagger})^2\rangle }\equiv \pm i\sigma\, ,
\end{equation}
which defines $\sigma$.
Finally, the von Neumann entropy, and therefore the discord
$\delta_{12}$, is given by the following expression:
\begin{align}
\delta_{12}=S(\hat{\rho}_1)=-\mathrm{Tr}\left(\hat{\rho}_1\log\hat{\rho}_1\right)=
\left(\sigma+\frac{1}{2}\right)\log_2\left(\sigma+\frac{1}{2}\right)
-\left(\sigma-\frac{1}{2}\right)\log_2\left(\sigma-\frac{1}{2}\right)\, .
\end{align}
We have evaluated this expression numerically. It is displayed as a
function of the parameter $\alpha$ in
Fig.~\ref{fig:GeneralizedDiscord}. The dependence in $k\eta$ is hidden
in the coefficients $a_{nn'}$ which depend on the squeezing
parameters, these ones being explicit time-dependent quantities.

\section{Covariance Matrix of a Two-Mode Squeezed State}
\label{sec:appendixcm}

In this appendix, we calculate the covariance matrix of a two-mode
squeezed quantum state. The characteristic function $\chi$ has been
introduced in \Eq{eq:chi:def} and the Weyl operator $\hat{W}$ defined
in \Eq{eq:WeylOperator:def}. Using the form~(\ref{eq:qstatevacuum})
for the density matrix, and, as below \Eq{eq:WeylOperator:def},
defining $\hat{R}=\left(k^{1/2}\hat{q}_{\bm k},k^{-1/2}\hat{\pi}_{\bm
    k},k^{1/2}\hat{q}_{-{\bm k}},k^{-1/2}\hat{\pi}_{-{\bm
      k}}\right)^{\rm T}\equiv
\left(\hat{R}_1,\hat{R}_2,\hat{R}_3,\hat{R}_4\right)^{\rm T}$, the
characteristic function $\chi (\xi)$ (where the components of $\xi$
are denoted as $\xi_1$, $\xi_2$, $\xi_3$ and $\xi_4$) can be written as
\begin{eqnarray}
\chi(\xi) &=& \tr\left[
\hat{S}(r_k,\varphi_k)\hat{R}(\theta_k)\vert 0_{\bm k},0_{-{\bm k}}\rangle \langle 
0_{\bm k},0_{-{\bm k}}\vert \hat{R}^{\dagger}(\theta_k)\hat{S}^{\dagger}(r_k,\varphi_k)
{\rm e}^{i\xi_1\hat{R}_1+i\xi_2\hat{R}_2+i\xi_3\hat{R}_3+i\xi_4\hat{R}_4}\right] \\
&=& \tr\left[\vert 0_{\bm k},0_{-{\bm k}}\rangle \langle 
0_{\bm k},0_{-{\bm k}}\vert \hat{R}^{\dagger}(\theta_k)\hat{S}^{\dagger}(r_k,\varphi_k)
{\rm e}^{i\xi_1\hat{R}_1+i\xi_2\hat{R}_2+i\xi_3\hat{R}_3+i\xi_4\hat{R}_4}
\hat{S}(r_k,\varphi_k)\hat{R}(\theta_k)\right] \\
&=& \sum_{n=0}^{\infty}\sum_{n'=0}^{\infty}
\langle n_{\bm k},n'_{-{\bm k}}
\vert 0_{\bm k},0_{-{\bm k}}\rangle \langle 
0_{\bm k},0_{-{\bm k}}\vert \hat{R}^{\dagger}(\theta_k)\hat{S}^{\dagger}(r_k,\varphi_k)
{\rm e}^{i\xi_1\hat{R}_1+i\xi_2\hat{R}_2+i\xi_3\hat{R}_3+i\xi_4\hat{R}_4}
\hat{S}(r_k,\varphi_k)\hat{R}(\theta_k)\vert n_{\bm k},n'_{-{\bm k}}\rangle \\
&=& \langle 
0_{\bm k},0_{-{\bm k}}\vert \hat{R}^{\dagger}(\theta_k)\hat{S}^{\dagger}(r_k,\varphi_k)
{\rm e}^{i\xi_1\hat{R}_1+i\xi_2\hat{R}_2+i\xi_3\hat{R}_3+i\xi_4\hat{R}_4}
\hat{S}(r_k,\varphi_k)\hat{R}(\theta _k)\vert 0_{\bm k},0_{-{\bm k}}\rangle
\\
&=& \langle 
0_{\bm k},0_{-{\bm k}}\vert \hat{R}^{\dagger}(\theta_k)\hat{S}^{\dagger}(r_k,\varphi_k)
{\rm e}^{i\xi_1\hat{R}_1+i\xi_2\hat{R}_2}{\rm e}^{+i\xi_3\hat{R}_3+i\xi_4\hat{R}_4}
\hat{S}(r_k,\varphi_k)\hat{R}(\theta_k)\vert 0_{\bm k},0_{-{\bm k}}\rangle \\
&=& {\rm e}^{i\xi_1\xi_2/2+i\xi_3\xi_4/2}\langle 
0_{\bm k},0_{-{\bm k}}\vert \hat{R}^{\dagger}(\theta_k)\hat{S}^{\dagger}(r_k,\varphi_k)
{\rm e}^{i\xi_1\hat{R}_1}{\rm e}^{i\xi_2\hat{R}_2}{\rm e}^{i\xi_3\hat{R}_3}
{\rm e}^{i\xi_4\hat{R}_4}
\hat{S}(r_k,\varphi_k)\hat{R}(\theta_k)\vert 0_{\bm k},0_{-{\bm k}}\rangle,
\end{eqnarray}
where we have made use of the Baker-Campbell-Hausdorff formula ${\rm
  e}^{\hat{A}+\hat{B}} ={\rm
  e}^{-[\hat{A},\hat{B}]/2}{\rm e}^{\hat{A}}{\rm
  e}^{\hat{B}}$, valid if the operators $\hat{A}$ and $\hat{B}$ commute with their
commutator $[\hat{A},\hat{B}]$, which is the case here; see below. The next step consists
in introducing the operator $\hat{S}\hat{R}(\hat{S}\hat{R})^{\dagger}$
(which is unity) between each exponential factor. This means that we
must now calculate
\begin{eqnarray}
\hat{R}^{\dagger}(\theta_k)\hat{S}^{\dagger}(r_k,\varphi_k)
{\rm e}^{i\xi_1\hat{R}_1}\hat{S}(r_k,\varphi_k)\hat{R}(\theta_k)
&=& \hat{R}^{\dagger}(\theta_k)\hat{S}^{\dagger}(r_k,\varphi_k)
\sum_{n=0}^{\infty}\frac{1}{n!}i^n\xi_1^n\hat{R}_1^n
\hat{S}(r_k,\varphi_k)\hat{R}(\theta_k)\\
&=&
\sum_{n=0}^{\infty}\frac{1}{n!}i^n\xi_1^n
\hat{R}^{\dagger}(\theta_k)\hat{S}^{\dagger}(r_k,\varphi_k)
\hat{R}_1^n
\hat{S}(r_k,\varphi_k)\hat{R}(\theta_k).
\end{eqnarray}
But one has (in order to avoid cumbersome notation, in this equation,
we do not write the dependence in the squeezing parameters)
\begin{equation}
\hat{R}^{\dagger}\hat{S}^{\dagger}\hat{R}_1^n\hat{S}\hat{R}
=\hat{R}^{\dagger}\hat{S}^{\dagger}
\hat{R}_1\hat{S}\hat{R}\hat{R}^{\dagger}\hat{S}^{\dagger}
\hat{R}_1^{n-1}\hat{S}\hat{R}
=\hat{R}^{\dagger}\hat{S}^{\dagger}
\hat{R}_1\hat{S}\hat{R}\hat{R}^{\dagger}\hat{S}^{\dagger}
\hat{R}_1\hat{S}\hat{R}\hat{R}^{\dagger}\hat{S}^{\dagger}
\hat{R}_1^{n-2}\hat{S}\hat{R}
=\left(\hat{R}^{\dagger}\hat{S}^{\dagger}\hat{R}_1\hat{S}\hat{R}\right)^n.
\end{equation}
As a consequence, one can rewrite the series as
\begin{eqnarray}
\hat{R}^{\dagger}(\theta_k)\hat{S}^{\dagger}(r_k,\varphi_k)
{\rm e}^{i\xi_1\hat{R}_1}\hat{S}(r_k,\varphi_k)\hat{R}(\theta_k)
&=&
\sum_{n=0}^{\infty}\frac{1}{n!}i^n\xi_1^n
\left[\hat{R}^{\dagger}(\theta_k)\hat{S}^{\dagger}(r_k,\varphi_k)
\hat{R}_1
\hat{S}(r_k,\varphi_k)\hat{R}(\theta_k)\right]^n\\
&=& {\rm e}^{i\xi_1\hat{R}^{\dagger}(\theta_k)
\hat{S}^{\dagger}(r_k,\varphi_k)\hat{R}_1
\hat{S}(r_k,\varphi_k)\hat{R}(\theta_k)}.
\end{eqnarray}
Using this result in the expression of the characteristic function,
not only for the first exponential term but for the four terms, we
arrive at the following expression:
\begin{eqnarray}
\label{eq:chiinter}
\chi (\xi) &=& 
{\rm e}^{i\xi_1\xi_2/2+i\xi_3\xi_4/2}\langle 
0_{\bm k},0_{-{\bm k}}\vert 
{\rm e}^{i\xi_1\hat{R}^{\dagger}\hat{S}^{\dagger}\hat{R}_1\hat{S}\hat{R}}
{\rm e}^{i\xi_2\hat{R}^{\dagger}\hat{S}^{\dagger}\hat{R}_2\hat{S}\hat{R}}
{\rm e}^{i\xi_3\hat{R}^{\dagger}\hat{S}^{\dagger}\hat{R}_3\hat{S}\hat{R}}
{\rm e}^{i\xi_4\hat{R}^{\dagger}\hat{S}^{\dagger}\hat{R}_4\hat{S}\hat{R}}
\vert 0_{\bm k},0_{-{\bm k}}\rangle.
\end{eqnarray}
To proceed, one has to evaluate the four terms
$\hat{R}^{\dagger}\hat{S}^{\dagger}\hat{R}_i\hat{S}\hat{R}$. Using \Eqs{eq:ck}, and~(\ref{eq:cminusk}), it is easy to show that
\begin{eqnarray}
\hat{\Omega}_1 &\equiv& \hat{R}^{\dagger}(\theta_k)\hat{S}^{\dagger}(r_k,\varphi_k)
\hat{R}_1\hat{S}(r_k,\varphi_k)\hat{R}(\theta_k) = \hat{R}_1 \cosh r_k 
\cos\theta_k
+\hat{R}_2 \cosh r_k \sin \theta_k
\nonumber \\ & &
-\hat{R}_3 \sinh r_k
\cos\left(\theta_k+2\varphi_k\right)
-\hat{R}_4 \sinh r_k \sin \left(2\varphi_k +\theta_k\right),\\
\hat{\Omega}_2 &\equiv& \hat{R}^{\dagger}(\theta_k )\hat{S}^{\dagger}(r_k,\varphi_k)
\hat{R}_2
\hat{S}(r_k,\varphi_k)\hat{R}(\theta_k ) = -\hat{R}_1 \cosh r_k \sin\theta_k
+\hat{R}_2 \cosh r_k \cos \theta_k
\nonumber \\ & &
-\hat{R}_3 \sinh r_k
\sin\left(2\varphi_k+\theta_k\right)
+\hat{R}_4 \sinh r_k \cos \left(2\varphi_k +\theta_k\right),\\
\hat{\Omega}_3 &\equiv& \hat{R}^{\dagger}(\theta_k )
\hat{S}^{\dagger}(r_k,\varphi_k)\hat{R}_3
\hat{S}(r_k,\varphi_k)\hat{R}(\theta _k) = -\hat{R}_1 \sinh r_k 
\cos\left(2\varphi_k+\theta_k\right)
-\hat{R}_2 \sinh r_k \sin \left(2\varphi_k+\theta_k\right)
\nonumber \\ & &
+\hat{R}_3 \cosh r_k\cos \theta_k
+\hat{R}_4 \cosh r_k \sin \theta_k,\\
\hat{\Omega }_4 &\equiv& \hat{R}^{\dagger}(\theta_k)
\hat{S}^{\dagger}(r_k,\varphi_k)\hat{R}_4
\hat{S}(r_k,\varphi_k)\hat{R}(\theta _k) = -\hat{R}_1 \sinh r_k 
\sin\left(2\varphi_k+\theta_k\right)
+\hat{R}_2 \sinh r_k \cos \left(2\varphi_k +\theta_k\right)
\nonumber \\ & &
-\hat{R}_3 \cosh r_k\sin \theta_k
+\hat{R}_4 \cosh r_k \cos \theta_k .
\end{eqnarray}
The final step is to express the product of four exponentials in
\Eq{eq:chiinter} in terms of a single exponential. For this
purpose, it is interesting to calculate the commutators of the
operators $\left[\hat{\Omega}_i,\hat{\Omega}_j\right]$ in order to use
(again) the Baker-Campbell-Hausdorff formula. One finds
\begin{eqnarray}
\left[\hat{\Omega}_1,\hat{\Omega}_3\right] &=& 
\left[\hat{\Omega}_1,\hat{\Omega}_4\right] 
= \left[\hat{\Omega}_2,\hat{\Omega}_3\right]
=\left[\hat{\Omega}_2,\hat{\Omega}_4\right]=0, \quad
\left[\hat{\Omega}_1,\hat{\Omega}_2\right] 
=\left[\hat{\Omega}_3,\hat{\Omega}_4\right]=i.
\end{eqnarray} 
As a consequence, the characteristic function~(\ref{eq:chiinter}) now
takes the form
\begin{eqnarray}
\chi(\xi) &=& 
{\rm e}^{i\xi_1\xi_2/2+i\xi_3\xi_4/2}\langle 
0_{\bm k},0_{-{\bm k}}\vert 
{\rm e}^{-i\xi_1\xi_2/2}
{\rm e}^{i\xi_1\hat{\Omega}_1+i\xi_2\hat{\Omega}_2}
{\rm e}^{-i\xi_3\xi_4/2}
{\rm e}^{i\xi_3\hat{\Omega}_3+i\xi_4\hat{\Omega} _4}
\vert 0_{\bm k},0_{-{\bm k}}\rangle \\
&=& 
\langle 0_{\bm k},0_{-{\bm k}}\vert 
{\rm e}^{i\xi_1\hat{\Omega}_1+i\xi_2\hat{\Omega}_2+i\xi_3\hat{\Omega}_3+i\xi_4\hat{\Omega }_4}
\vert 0_{\bm k},0_{-{\bm k}}\rangle 
 \\
&=& 
\label{eq:chiinter2}
\langle 0_{\bm k},0_{-{\bm k}}\vert 
{\rm e}^{i\eta_1\hat{R}_1+i\eta_2\hat{R}_2+i\eta_3\hat{R}_3+i\eta_4\hat{R}_4}
\vert 0_{\bm k},0_{-{\bm k}}\rangle \equiv
\chi_{\rm vac}(\eta_1,\eta_2,\eta_3,\eta_4),
\end{eqnarray}
where the coefficients $\eta_i$ can be expressed as
\begin{eqnarray}
\eta_1 &=& \xi_1 \cosh r_k \cos\theta_k -\xi_2 \sinh r_k \sin \theta_k 
-\xi_3 \sinh r_k \cos \left(2\varphi_k+\theta_k\right)
-\xi_4 \sinh r_k \sin\left(2\varphi_k + \theta_k\right), \\
\eta_2 &=& \xi_1 \cosh r_k \sin\theta_k+\xi_2 \cosh r_k \cos \theta_k 
-\xi_3 \sinh r_k \sin \left(2\varphi_k+\theta_k\right)
+\xi_4 \sinh r_k \cos\left(2\varphi_k + \theta_k\right), \\
\eta_3 &=& -\xi_1 \sinh r_k \cos\left(2\varphi_k +\theta_k\right)
-\xi_2 \sinh r_k \sin \left(2\varphi_k+ \theta_k \right)
+\xi_3 \cosh r_k \cos \theta_k
-\xi_4 \cosh r_k \sin \theta_k. \\
\eta_4 &=& -\xi_1 \sinh r_k \sin\left(2\varphi_k +\theta_k\right)
+\xi_2 \sinh r_k \cos \left(2\varphi_k+ \theta_k \right)
+\xi_3 \cosh r_k \sin \theta_k
+\xi_4 \cosh r_k \cos \theta_k.
\end{eqnarray}
In \Eq{eq:chiinter2}, $\chi_{\rm vac}(\eta_1,
\eta_2,\eta_3,\eta_4)$ represents the characteristic function of the
vacuum. But this function is well-known and easy to calculate. The
vacuum state is a Gaussian state and one has, see for instance
Eq.~(33) of \Refc{2007PhR...448....1W},
\begin{equation}
\chi_{\rm vac}(\eta_1,\eta_2,\eta_3,\eta_4)
={\rm e}^{-\eta^{\rm T}{\rm Id}_4 \eta/4}\, ,
\end{equation}
where ${\rm Id}_4$ is the identity matrix in four dimensions. Since
the definition of the covariance matrix $\gamma$ is given by the
expression
\begin{equation}
\label{eq:chiGauss}
\chi (\xi_1,\xi_2,\xi_3,\xi_4)
={\rm e}^{-\xi^{\rm T}\gamma \xi/4},
\end{equation}
this implies that
\begin{equation}
\sum_{i=1}^{4}\eta_i^2=\sum_{i=1}^4\sum_{j=1}^4\gamma_{ij}\xi_i\xi_j,
\end{equation}
which allows us to infer the components of the covariance
matrix. Then lengthy but straightforward calculations lead to the
following covariance matrix for the two-mode squeezed state:
\begin{equation}
\gamma=
\begin{pmatrix}
\cosh \left(2r_k\right) & 0 & -\sinh \left(2r_k\right) 
\cos\left(2\varphi_k\right)
& -\sinh\left( 2r_k\right) 
\sin \left(2\varphi_k\right) \\
0 & \cosh \left(2r_k\right) & -\sinh \left(2r_k\right) 
\sin\left(2\varphi_k\right)
& \sinh\left( 2r_k\right) 
\cos \left(2\varphi_k\right) \\
-\sinh \left(2r_k\right) 
\cos\left(2\varphi_k\right)
& -\sinh \left(2r_k\right) 
\sin\left(2\varphi_k\right)
& \cosh\left(2r_k\right) & 0 \\
-\sinh\left( 2r_k\right) 
\sin \left(2\varphi_k\right)
& \sinh\left( 2r_k\right) 
\cos \left(2\varphi_k\right)
& 0 & \cosh\left(2r_k\right)
\end{pmatrix}.
\end{equation}
We notice that $\gamma $ does not depend on the rotation angle $\theta_{k}$. Once
we have the covariance matrix, we can determine the two-point
correlation functions by means of the relation $\langle
R_jR_k\rangle=\gamma_{jk}/2+iJ_{jk}/2$ derived in appendix~\ref{sec:Wigner:meanValue}. As mentioned in the main text, the
matrix $J$ is defined by
\begin{equation}
J=
\begin{pmatrix}
0 & 1 & 0 & 0 \\
-1 & 0 & 0 &0 \\
0 & 0 & 0 & 1 \\
0 & 0 & -1 & 0
\end{pmatrix}.
\end{equation}
Using the above expression of the correlation matrix, the explicit
form of the two-point correlators are given by
\begin{eqnarray}
\label{eq:qq}
\langle \hat{q}_{\bm k}\hat{q}_{\bm k}\rangle &=& 
\langle \hat{q}_{-{\bm k}}\hat{q}_{-{\bm k}}\rangle =
\frac{1}{2k}\cosh(2r_k), 
\quad
\langle \hat{\pi}_{\bm k}\hat{\pi}_{\bm k}\rangle = 
\langle \hat{\pi}_{-{\bm k}}\hat{\pi}_{-{\bm k}}\rangle
=\frac{k}{2}\cosh(2r_k)
\\
\label{eq:qminusq}
\langle \hat{q}_{\bm k}\hat{q}_{-{\bm k}}\rangle &=& 
-\frac{1}{2k}\sinh(2r_k)\cos(2\varphi_k), 
\quad
\langle \hat{\pi}_{\bm k}\hat{\pi}_{-{\bm k}}\rangle
=
\frac{k}{2}\sinh(2r_k)\cos(2\varphi_k), 
\\
\label{eq:qminuspi}
\langle \hat{q}_{\bm k}\hat{\pi}_{-{\bm k}}\rangle &=& 
\langle \hat{\pi}_{\bm k}\hat{q}_{-{\bm k}}\rangle
=
-\frac{1}{2}\sinh(2r_k)\sin(2\varphi_k), 
\quad
\langle \hat{q}_{{\bm k}}\hat{\pi}_{{\bm k}}\rangle 
=\langle \hat{q}_{-{\bm k}}
\hat{\pi}_{-{\bm k}}\rangle=-\langle \hat{\pi}_{{\bm k}}
\hat{q}_{{\bm k}}\rangle =-\langle \hat{\pi}_{-{\bm k}}
\hat{q}_{-{\bm k}}\rangle=\frac{i}{2}.
\end{eqnarray}
These results are used in the main text in order to calculate the two-point correlation functions of $\hat{v}_{\bm k}$ and $\hat{p}_{\bm
  k}$; see \Eqs{eq:vvminus}, (\ref{eq:ppminus})
and~(\ref{eq:vpminus}), related to $\hat{q}_{\bm k}$ and
$\hat{\pi}_{\bm k}$ through \Eqs{eq:vqpi} and~(\ref{eq:pqpi}).

\section{Discord of a Classical State}
\label{sec:appendixdiscord}

In this appendix, we show that classical states given
by \Eq{eq:classicalstate:2} have vanishing
quantum discord. Let us first calculate the reduced density matrix
$\hat{\rho}_{\rm class}({\bm k})$ obtained by tracing
out~\Eq{eq:classicalstate:2} over degrees of freedom belonging to space
``$-{\bm k}$''. One obtains
\begin{equation}
\hat{\rho}_{\rm class}({\bm k})=\sum _m \langle m_{-{\bm k}}\vert 
\sum _{i=0}^{\infty}\sum_{j=0}^{\infty}
p_{ij}\vert i_{\bm k}\rangle \langle i_{\bm k}\vert 
\otimes 
\vert j_{-{\bm k}}\rangle \langle j_{-{\bm k}}\vert
m_{-{\bm k}}\rangle =\sum_{ij}p_{\ij}\vert i_{\bm k}\rangle \langle i_{\bm k}\vert\, . 
\end{equation}
In the same manner, one obtains a similar expression for $\hat{\rho}_{\rm
  class}(-{\bm k})$ where $i_{\bm k}$ is replaced by $j_{-\bm k}$ in
the above formula. It is easy to show that these matrices are diagonal
in the basis $\vert n_{\bm k}\rangle$. Indeed,
\begin{equation}
\langle n_{\bm k}\vert \hat{\rho}_{\rm class}({\bm k})\vert m_{\bm k}\rangle =\delta _{nm}
\sum_jp_{nj}\, .
\end{equation}
Of course, one has a similar relation (but not identical if the
probabilities are not symmetric) in the other half space, namely
\begin{equation}
\langle n_{-{\bm k}}\vert \hat{\rho}_{\rm class}({\bm k})
\vert m_{-{\bm k}}\rangle =\delta _{nm}
\sum_ip_{in}\, .
\end{equation}
Therefore, it is straightforward to calculate the corresponding
entropy. Indeed, the calculation of the entropy requires in general estimating
the logarithm of a matrix, which is a non-trivial
task. But when the matrix is diagonal, its logarithm is just a
diagonal matrix whose entries are given by the logarithm of the
entries of the original matrix. As a consequence, one immediately
deduces that
\begin{equation}
S[\hat{\rho}_{\rm class}({\bm k})]=\sum_i\left[\left(\sum_{\ell}p_{i\ell}\right)
\log\left(\sum_j p_{ij}\right)\right], \quad 
S[\hat{\rho}_{\rm class}(-{\bm k})]=\sum_i\left[\left(\sum_{\ell}p_{\ell i}\right)
\log\left(\sum_j p_{j i}\right)\right]\, .
\end{equation}
On the other hand, since the full density matrix $\hat{\rho}_{\rm
  class}({\bm k},-{\bm k})$ satisfies
\begin{equation}
\langle n_{\bm k}m_{-{\bm k}}\vert \hat{\rho}_{\rm class}({\bm k},-{\bm k})
\vert n_{\bm k}'m_{-{\bm k}}'\rangle =p_{nm}\delta _{nn'}\delta_{mm'}\, ,
\end{equation}
one can write its entropy as
\begin{equation}
S\left[\hat{\rho}_{\rm class}({\bm k},-{\bm k})\right]
=\sum_{ij}p_{ij}\log p_{ij}.
\end{equation}
It follows that the mutual information ${\cal I}({\bm k},-{\bm k})$ is
given by
\begin{equation}
{\cal I}({\bm k},-{\bm k})=\sum_i\left[\left(\sum_{\ell}p_{i\ell}\right)
\log\left(\sum_j p_{ij}\right)\right]+
\sum_i\left[\left(\sum_{\ell}p_{\ell i}\right)
\log\left(\sum_j p_{j i}\right)\right]
-\sum_{ij}p_{ij}\log p_{ij}\, .
\end{equation}
This completes the first step in the calculation of the discord.

\par

Let us now calculate the quantity ${\cal J}({\bm k},-{\bm k})$. We
consider again the projector $\hat{\Pi}_j$ introduced in
\Eq{eq:defpij}. It is easy to show that
\begin{equation}
\label{eq:rhopiappendix}
\hat{\rho}_{\rm class}({\bm k},-{\bm k})\hat{\Pi}_j=\sum_n
p_{nj}\vert n_{\bm k}\rangle \langle n_{\bm k}\vert 
\otimes 
\vert j_{-{\bm k}}\rangle \langle j_{-{\bm k}}\vert\, .
\end{equation}
Moreover, using this expression, one can also establish that
\begin{equation}
\tr \left[\hat{\rho}_{\rm class}({\bm k},-{\bm k})\hat{\Pi}_j\right]
=\sum_n p_{nj}\, .
\end{equation}
Then, in order to derive a formula for $\hat{\rho}_{\rm class}({\bm
  k};\hat{\Pi}_j)$, one has to trace out the degrees of freedom contained in
``$-{\bm k}$'' in \Eq{eq:rhopiappendix}, see also
\Eq{eq:projectedState}, and it comes to
\begin{equation}
\hat{\rho}_{\rm class}({\bm k};\hat{\Pi}_j)=\frac{1}{\sum _\ell p_{\ell j}}
\sum_n p_{nj} \vert n_{\bm k}\rangle \langle n_{\bm k}\vert \, .
\end{equation}
This density matrix is diagonal with elements $p_{nj}/\sum _\ell
p_{\ell j}$ on the diagonal. Consequently, the conditional entropy $S_{\rm cond}$
which, as before, requires the calculation of the logarithm of the
above mentioned matrix, can be easily estimated and reads
\begin{eqnarray}
S_{\rm cond}&\equiv & \sum_j \left(\sum _{\ell}p_{\ell j}\right)
S\left[\hat{\rho}_{\rm class}({\bm k};\hat{\Pi}_j)\right]
=\sum_j \left(\sum _{\ell}p_{\ell j}\right) \sum _n \frac{p_{nj}}{\sum _m p_{mj}}
\log\left (\frac{p_{nj}}{\sum _i p_{ij}}\right) \\
&=& \sum_{nj} p_{nj} \log\left(\frac{p_{nj}}{\sum _i p_{ij}}\right) 
=\sum_{nj} p_{nj} \log\left(p_{nj}\right)
-\sum_{nj} p_{nj} \log\left(\sum _i p_{ij}\right) \\
&=& \sum_{nj} p_{nj} \log\left(p_{nj}\right)
-\sum_{j} \left[\left(\sum _n p_{nj}\right) 
\log\left(\sum _i p_{ij}\right)\right] \, .
\end{eqnarray}
It follows that the quantity ${\cal J}({\bm k},-{\bm k})$ can be
expressed as
\begin{eqnarray}
{\cal J}({\bm k},-{\bm k})\equiv S[\hat{\rho}_{\rm class}({\bm k})]
-S_{\rm cond}
&=&\sum_i\left[\left(\sum_{\ell}p_{i\ell}\right)
\log\left(\sum_j p_{ij}\right)\right]
-\sum_{nj} p_{nj} \log\left(p_{nj}\right)
\nonumber \\ & &
+\sum_{j} \left[\left(\sum _n p_{nj}\right) 
\log\left(\sum _i p_{ij}\right)\right]={\cal I}({\bm k},-{\bm k}) \, .
\end{eqnarray}
Therefore, we see that, for the choice of $\hat{\Pi}_j$ made above, the
difference between ${\cal J}({\bm k},-{\bm k})$ and ${\cal I}({\bm
  k},-{\bm k})$ vanishes. In order to obtain the discord, one must
minimize this quantity over the set of projectors, see
\Eq{eq:defdiscord}. However, the discord is a positive definite
quantity~\cite{Datta:2010} and, as a consequence, if for one specific choice of $\hat{\Pi}_j$
one proves it is zero, then it obviously means that the minimum is
also zero. Therefore, as announced in the main text, we conclude that
the quantum discord of the ``classical'' state vanishes.

\section{Characteristic Function of a Classical State}
\label{sec:appendixcmclassical}

In this appendix, we calculate the characteristic function of the
classical state~(\ref{eq:classicalstate:2}). Let us recall that the
characteristic function $\chi$ is defined in \Eq{eq:chi:def}
[and the Weyl operator in
\Eq{eq:WeylOperator:def}]. As a consequence, we have
\begin{align}
\chi(\xi)&= \tr\left[\hat{\rho}_{\rm class}({\bm k},-{\bm k})
    {\rm e}^{i\xi_1\hat{R}_1+i\xi_2\hat{R}_2+i\xi_3\hat{R}_3+i\xi_4\hat{R}_4}\right]\\
  &= \sum _{n,m}\langle n_{\bm k},m_{-{\bm k}}\vert
  \sum _{i,j}
  p_{ij}\vert i_{\bm k}\rangle \langle i_{\bm k}\vert 
  \otimes 
  \vert j_{-{\bm k}}\rangle \langle j_{-{\bm k}}\vert
  {\rm e}^{i\xi_1\hat{R}_1+i\xi_2\hat{R}_2+i\xi_3\hat{R}_3+i\xi_4\hat{R}_4}
  \vert n_{\bm k},m_{-{\bm k}}\rangle \\
  &= {\rm e}^{i\xi_1\xi_2/2+i\xi_3\xi_4/2}
  \sum_{n,m}p_{nm}
  \langle n_{\bm k},m_{-{\bm k}}\vert
  {\rm e}^{i\xi_1\hat{R}_1}{\rm e}^{i\xi_2\hat{R}_2}
{\rm e}^{i\xi_3\hat{R}_3}{\rm e}^{i\xi_4\hat{R}_4}
  \vert n_{\bm k},m_{-{\bm k}}\rangle \\
\label{eq:interchiclassical}
  &= {\rm e}^{i\xi_1\xi_2/2+i\xi_3\xi_4/2}
  \sum_{n,m}p_{nm}\int  \dd \left(k^{1/2}x_{\bm k}\right) \,
\dd \left(k^{1/2}x_{-{\bm k}}\right)
  \dd \left(k^{1/2}y_{\bm k}\right) \, \dd \left(k^{1/2}y_{-{\bm k}}\right)
  \langle n_{\bm k},m_{-{\bm k}}\vert
  {\rm e}^{i\xi_1\hat{R}_1}{\rm e}^{i\xi_2\hat{R}_2}
  \vert x_{\bm k},x_{-{\bm k}}\rangle 
\nonumber \\ & \times 
\langle x_{\bm k},x_{-{\bm k}}\vert
  {\rm e}^{i\xi_3\hat{R}_3}{\rm e}^{i\xi_4\hat{R}_4}
  \vert y_{\bm k},y_{-{\bm k}}\rangle 
  \langle y_{\bm k},y_{-{\bm k}}
  \vert n_{\bm k},m_{-{\bm k}}\rangle,
\end{align}
where, in the last equality, we have introduced the closure relation
twice. In this expression, $\vert x_{\bm k}\rangle $ represents the
eigenstate of the ``position operator'' $\hat{q}_{\bm k}$. This is
especially convenient because the action of an operator of the form
${\rm e}^{i\xi_i\hat{R}_i}$ on this state is particular
simple. Indeed, one has
\begin{eqnarray}
{\rm e}^{i\xi_3\hat{R}_3}{\rm e}^{i\xi_4\hat{R}_4}
\vert y_{\bm k},y_{-{\bm k}}\rangle 
&=&{\rm e}^{i\xi_3k^{1/2}\hat{q}_{-{\bm k}}}
{\rm e}^{i\xi_4k^{-1/2}\hat{\pi}_{-{\bm k}}}
\vert y_{\bm k},y_{-{\bm k}}\rangle 
={\rm e}^{i\xi_3k^{1/2}\hat{q}_{-{\bm k}}}
\vert y_{\bm k},y_{-{\bm k}}-\xi_4k^{-1/2}\rangle
\\ 
&=& {\rm e}^{i\xi_3k^{1/2}(y_{-{\bm k}}-\xi_4k^{-1/2})}
\vert y_{\bm k},y_{-{\bm k}}-\xi_4k^{-1/2}\rangle \, .
\end{eqnarray}
In the same way, one can also write
\begin{eqnarray}
{\rm e}^{i\xi_1\hat{R}_1}{\rm e}^{i\xi_2\hat{R}_2}
\vert x_{\bm k},x_{-{\bm k}}\rangle 
={\rm e}^{i\xi_1k^{1/2}(x_{{\bm k}}-\xi_2k^{-1/2})}
\vert x_{\bm k}-\xi_2k^{-1/2},x_{-{\bm k}}\rangle \, .
\end{eqnarray}
As a consequence, using these two last relations,
\Eq{eq:interchiclassical} now takes the following form
\begin{eqnarray}
\chi(\xi ) &=& {\rm e}^{i\xi_1\xi_2/2+i\xi_3\xi_4/2}
\sum_{n,m}k^2p_{nm}\int  \dd x_{\bm k} \, \dd x_{-{\bm k}}
 \dd y_{\bm k} \, \dd y_{-{\bm k}}
{\rm e}^{i\xi_1k^{1/2}x_{\bm k}}
{\rm e}^{-i\xi_1\xi_2}
{\rm e}^{i\xi_3k^{1/2}y_{-{\bm k}}}
{\rm e}^{-i\xi_3\xi_4}
\langle n_{\bm k},m_{-{\bm k}}\vert
x_{\bm k}-\xi_2k^{-1/2},x_{-{\bm k}}\rangle 
\nonumber \\ & & \times
\langle x_{\bm k},x_{-{\bm k}}
\vert y_{\bm k},y_{-{\bm k}}-\xi_4k^{-1/2}\rangle 
\langle y_{\bm k},y_{-{\bm k}}
\vert n_{\bm k},m_{-{\bm k}}\rangle
\\
&=& {\rm e}^{-i\xi_1\xi_2/2+i\xi_3\xi_4/2}
\sum_{n,m}kp_{nm}\int  \dd x_{\bm k} \, \dd x_{-{\bm k}}\, 
{\rm e}^{i\xi_1k^{1/2}x_{\bm k}}
{\rm e}^{i\xi_3k^{1/2}x_{-{\bm k}}}
\langle n_{\bm k}\vert
x_{\bm k}-\xi_2k^{-1/2}\rangle
\langle m_{-{\bm k}}\vert
x_{-{\bm k}}\rangle
\nonumber \\ & & \times
\langle x_{\bm k}\vert n_{\bm k}\rangle
\langle 
x_{-{\bm k}}+\xi_4k^{-1/2}\vert 
m_{-{\bm k}}\rangle .
\end{eqnarray}
Then, in order to proceed, one has to calculate scalar products of the
form $\langle x_{\bm k}\vert n_{\bm k}\rangle$. But this is nothing
but the wavefunction of a state containing $n$ particles and, as is
well known, it can be expressed in terms of Hermite polynomials ${\rm
  H}_n$. As a consequence, one obtains
\begin{eqnarray}
\chi (\xi)&=& 
 {\rm e}^{-i\xi_1\xi_2/2+i\xi_3\xi_4/2}
\sum_{n,m}kp_{nm}\int  \dd x_{\bm k} \, \dd x_{-{\bm k}}\, 
{\rm e}^{i\xi_1k^{1/2}x_{\bm k}}
{\rm e}^{i\xi_3k^{1/2}x_{-{\bm k}}}
\frac{1}{\sqrt{\sqrt{\pi}2^nn!}}
{\rm e}^{-\left(k^{1/2}x_{\bm k}-\xi_2\right)^2/2}
{\rm H}_n\left(k^{1/2}x_{\bm k}-\xi_2\right)
\nonumber \\ & & \times
\frac{1}{\sqrt{\sqrt{\pi}2^mm!}}
{\rm e}^{-k\left(x_{-{\bm k}}\right)^2/2}
{\rm H}_m\left(k^{1/2}x_{-{\bm k}}\right)
\frac{1}{\sqrt{\sqrt{\pi}2^nn!}}
{\rm e}^{-k\left(x_{{\bm k}}\right)^2/2}
{\rm H}_n\left(k^{1/2}x_{{\bm k}}\right)
\nonumber \\ & & \times 
\frac{1}{\sqrt{\sqrt{\pi}2^mm!}}
{\rm e}^{-\left(k^{1/2}x_{-{\bm k}}+\xi_4\right)^2/2}
{\rm H}_m\left(k^{1/2}x_{-{\bm k}}+\xi_4\right)
\\
&=& \frac{1}{\pi}{\rm e}^{-i\xi_1\xi_2/2+i\xi_3\xi_4/2}
{\rm e}^{-\xi_1^2/4+i\xi_1\xi_2/2-\xi_2^2/2}
{\rm e}^{-\xi_3^2/4-i\xi_3\xi_4/2-\xi_4^2/4}
\nonumber \\ & & \times 
\sum_{n,m}\frac{p_{nm}}{2^{n+m}n!m!}
\int  \dd u \, {\rm e}^{-u^2}
{\rm H}_n\left(u-\frac{\xi_2}{2}+\frac{i\xi_1}{2}\right)
{\rm H}_n\left(u+\frac{\xi_2}{2}+\frac{i\xi_1}{2}\right)
\nonumber \\ & & \times
\int  \dd v \, {\rm e}^{-v^2}
{\rm H}_m\left(v-\frac{\xi_4}{2}+\frac{i\xi_3}{2}\right)
{\rm H}_m\left(v+\frac{\xi_4}{2}+\frac{i\xi_3}{2}\right).
\end{eqnarray}
Thanks to Eq.~(7.377) of \Refc{Gradshteyn:1965aa}, the two integrals
involving an exponential function and the product of two Hermite
polynomials can be performed exactly. As a result, one obtains
\begin{eqnarray}
  \chi(\xi) &=& {\rm e}^{-\left(\xi_1^2+\xi_2^2+\xi_3^2+\xi_4^2\right)/4}
  \sum_{n,m}p_{nm}{\rm L}_n\left(\frac{\xi_1^2}{2}+\frac{\xi_2^2}{2}\right)
  {\rm L}_m\left(\frac{\xi_3^2}{2}+\frac{\xi_4^2}{2}\right),
  \label{eq:characteristicfunction:class:Laguerre}
\end{eqnarray}
where ${\rm L}_n^0\equiv {\rm L}_n$ is a Laguerre polynomial; see
Eq.~(8.970.2) of \Refc{Gradshteyn:1965aa}. Of course, to go further,
one needs to specify the distribution $p_{nm}$. But the last equation
has the advantage of showing that, in general, the characteristic
function of the classical state has no reason to be Gaussian. If one
makes the choice $p_{nm}=\left(1-{\rm e}^{-\beta_k}\right){\rm
  e}^{-\beta_kn}\delta(n-m)$, see the discussion around \Eq{eq:betak}, then the characteristic function reduces to
\begin{eqnarray}
  \chi(\xi) &=& {\rm e}^{-\left(\xi_1^2+\xi_2^2+\xi_3^2+\xi_4^2\right)/4}
\left(1-{\rm e}^{-\beta_k}\right)
  \sum_{n=0}^{\infty}{\rm
  e}^{-\beta_kn}{\rm L}_n\left(\frac{\xi_1^2}{2}+\frac{\xi_2^2}{2}\right)
  {\rm L}_n\left(\frac{\xi_3^2}{2}+\frac{\xi_4^2}{2}\right).
\end{eqnarray}
This series can be calculated explicitly by means of Eq.~(8.976.1) of
\Refc{Gradshteyn:1965aa}. The result reads
\begin{align}
\label{eq:chiclassical}
\chi(\xi) & =
{\rm e}^{-\tanh^{-1}\left(\frac{\beta_k}{2}\right)\frac{\xi_1^2+\xi_2^2+\xi_3^2+\xi_4^2}{4}}
{\rm I}_0\left[
\frac{\sqrt{\left(\xi_1^2+\xi_2^2\right)
\left(\xi_3^2+\xi_4^2\right)}}{2\sinh(\beta_k/2)}
\right]\, ,
\end{align}
where ${\rm I}_0$ is a modified Bessel function of order
zero~\cite{Gradshteyn:1965aa}. The virtue of this expression is
twofold. First, since the argument of the Bessel function is always
positive and ${\rm I}_0$ is also always positive, the
characteristic function is positive everywhere. Second,
even with the simple choice
of $p_{nm}$ made before, one can see that the characteristic function is not
Gaussian (note that in the sub-Hubble limit where $\beta_k\rightarrow \infty$, the argument of the Bessel function vanishes and the Bessel function becomes 1; hence the characteristic function is Gaussian in this limit).
\section{Quantum Correlators and Weyl Transforms}
\label{sec:Wigner:meanValue}
In this appendix we show how mean values of quantum operators can be
calculated by means of the Wigner function~\cite{Case:2008}. Using the definition of
the Weyl transform given in \Eq{eq:Weyltransform:def} and that of the
Wigner function, see \Eq{eq:Wigner:def}, let us first calculate the
following integral $\int \tilde{A} W$. One has
\begin{eqnarray} \int \tilde{A} W
\dd q_{\bm k} \dd \pi_{\bm k} \dd q_{-\bm k} \dd \pi_{-\bm k}
&=&\frac{1}{\left(2\pi\right)^2} \int \left\langle q_{\bm
    k}+\frac{x}{2},q_{-\bm k}+\frac{y}{2} \vert \hat{A} \vert q_{\bm
    k}-\frac{x}{2},q_{-\bm k}-\frac{y}{2} \right\rangle \ee^{-i\pi_{\bm
    k}\left(x+x^\prime\right)-i\pi_{-\bm k}\left(y+y^\prime\right)}
\nonumber\\ & & \times \left\langle q_{\bm k}+\frac{x^\prime}{2},q_{-\bm
    k}+\frac{y^\prime}{2} \vert \hat{\rho} \vert q_{\bm
    k}-\frac{x^\prime}{2},q_{-\bm k}-\frac{y^\prime}{2} \right\rangle
\dd q_{\bm k} \dd \pi_{\bm k} \dd q_{-\bm k} \dd \pi_{-\bm k}\dd x\, \dd
y\, \dd x^\prime \dd y^\prime\, .
\end{eqnarray} 
The integration over $\pi_{\bm k}$ and $\pi_{-\bm k}$ gives rise to
Dirac delta functions, which can then be used to perform the integrals
over $x^\prime$ and $y^\prime$. As a consequence, one obtains
\begin{eqnarray} \int \tilde{A} W \dd
q_{\bm k} \dd \pi_{\bm k} \dd q_{-\bm k} \dd \pi_{-\bm k} &=& \int
\left\langle q_{\bm k}+\frac{x}{2},q_{-\bm k}+\frac{y}{2} \vert
  \hat{A} \vert q_{\bm k}-\frac{x}{2},q_{-\bm k}-\frac{y}{2}
\right\rangle \nonumber \\ & & \times \left\langle q_{\bm
    k}-\frac{x}{2},q_{-\bm k}-\frac{y}{2} \vert \hat{\rho} \vert
  q_{\bm k}+\frac{x}{2},q_{-\bm k}+\frac{y}{2} \right\rangle \dd
q_{\bm k} \dd q_{-\bm k} \dd x\, \dd y \, .
\end{eqnarray} 
The next step consists in performing a change of variables. Instead of
working in terms of $q_{\bm k}$, $x$ and $q_{-{\bm k}}$, $y$, we
define the quantities $u_{\pm {\bm k}}$ and $w_{\pm {\bm k}}$ such
that $u_{\bm k}=q_{\bm k}-x/2$, $w_{\bm k}=q_{\bm k}+x/2$, $u_{-\bm
  k}=q_{-\bm k}-y/2$ and $w_{-\bm k}=q_{-\bm k}+y/2$. The determinant
of the Jacobian of these transformations being one, one has $\dd
q_{\bm k}\dd x = \dd u_{\bm k} \dd v_{\bm k} $ and $\dd q_{-\bm k}\dd
y = \dd u_{-\bm k} \dd v_{-\bm k} $; hence
\begin{eqnarray}
\int \tilde{A} W \dd q_{\bm k}
\dd \pi_{\bm k} \dd q_{-\bm k} \dd \pi_{-\bm k} &=& \int \left\langle
  w_{\bm k} , w_{-\bm k} \vert \hat{A} \vert u_{\bm k} , u_{-\bm k}
\right\rangle
\left\langle u_{\bm k} , u_{-\bm k} \vert \hat{\rho} 
\vert w_{\bm k} , w_{-\bm k} \right\rangle
\dd u_{\bm k} \dd w_{\bm k} \dd u_{-\bm k} \dd w_{-\bm k} \, .
\nonumber\\
&=& \int \left\langle w_{\bm k} , w_{-\bm k}   
\vert \hat{A} \hat{\rho} \vert w_{\bm k} , w_{-\bm k} \right\rangle
\dd w_{\bm k}  \dd w_{-\bm k} \\
&=&  \mathrm{Tr}\left( \hat{A} \hat{\rho} \right)
=\langle \hat{A}\rangle \, .
\end{eqnarray}
As a consequence, the expectation value of $\hat{A}$ can be obtained
by something which is the average of the physical quantity represented
by $\tilde{A}$ over phase space with probability density $W$
characterizing the state. Therefore, as explained in the main text, the
quantum problem can be replaced by a stochastic approach in
$\tilde{A}$. It is thus interesting to study under which conditions
$A$ and $\tilde{A}$ can be identified. A general expression for any analytic
operator $\hat{A}$ is provided by its Taylor series in terms of the
operators $\hat{q}_{\bm k}$, $\hat{\pi}_{\bm k}$, $\hat{q}_{-\bm k}$
and $\hat{\pi}_{-\bm k}$. Making use of the canonical commutation
relations, this can always be written as 
\begin{eqnarray} 
\hat{A}&=&
\sum_{n_1,n_2}a_{n_1n_2} \hat{q}_{\bm k}^{n_1} \hat{\pi}_{\bm k}^{n_2}
+\sum_{n_1,n_2} b_{n_1n_2} \hat{q}_{-\bm k}^{n_1}  \hat{\pi}_{-\bm k}^{n_2}
+\sum_{n_1,n_2} c_{n_1n_2} \hat{q}_{\bm k}^{n_1}  \hat{q}_{-\bm k}^{n_2}
+\sum_{n_1,n_2} d_{n_1n_2} \hat{\pi}_{\bm k}^{n_1}  \hat{\pi}_{-\bm k}^{n_2}
\nonumber\\ & &
+\sum_{n_1,n_2} e_{n_1n_2} \hat{q}_{\bm k}^{n_1}  \hat{\pi}_{-\bm k}^{n_2}
+\sum_{n_1,n_2} f_{n_1n_2} \hat{\pi}_{\bm k}^{n_1}  \hat{q}_{-\bm k}^{n_2}\, ,
\label{eq:Weyl:Aexpansion}
\end{eqnarray}
where $a_{n_1n_2}$, \dots, $f_{n_1n_2}$ are numerical
coefficients. Let us calculate the Weyl
transform of each of these terms. From the
definition~(\ref{eq:Weyltransform:def}), one can write that
\begin{eqnarray}
\widetilde{q_{\bm k}^{n_1} \pi_{\bm k}^{n_2}}&=&\int\dd x\, \dd y\, 
\ee^{-i\pi_{\bm k} x - i \pi_{-\bm k} y}
\left\langle q_{\bm k}+\frac{x}{2} , 
q_{-\bm k}+\frac{y}{2}\vert \hat{q}_{\bm k}^{n_1} 
\hat{\pi}_{\bm k}^{n_2} \vert q_{\bm k}-\frac{x}{2} , 
q_{-\bm k}-\frac{y}{2}\right\rangle\\
&=&\int\dd x\, \dd y \, 
\ee^{-i\pi_{\bm k} x - i \pi_{-\bm k} y}\left(q_{\bm k}
+\frac{x}{2}\right)^{n_1}\left\langle q_{\bm k}
+\frac{x}{2} , q_{-\bm k}+\frac{y}{2}\vert 
\hat{\pi}_{\bm k}^{n_2} \vert q_{\bm k}-\frac{x}{2} , 
q_{-\bm k}-\frac{y}{2}\right\rangle\, \\
&=&\int\dd x\, \dd y \, \dd \pi_{\bm k}^\prime \dd \pi_{\bm
  k}^{\prime\prime} \dd \pi_{-\bm k}^\prime \dd \pi_{-\bm
  k}^{\prime\prime} \ee^{-i\pi_{\bm k} x - i \pi_{-\bm k} y}\left(q_{\bm
    k}+\frac{x}{2}\right)^{n_1} \left\langle q_{\bm k}+\frac{x}{2} ,
  q_{-\bm k}+\frac{y}{2}\vert\pi_{\bm k}^\prime,\pi_{-\bm
    k}^\prime\right\rangle \nonumber\\ & & \times \left\langle \pi_{\bm
    k}^\prime,\pi_{-\bm k}^\prime \vert \hat{\pi}_{\bm k}^{n_2}\vert
  \pi_{\bm k}^{\prime\prime},\pi_{-\bm
    k}^{\prime\prime}\right\rangle\left\langle \pi_{\bm
    k}^{\prime\prime},\pi_{-\bm k}^{\prime\prime}\vert q_{\bm
    k}-\frac{x}{2} , q_{-\bm k}-\frac{y}{2}\right\rangle\, , 
\end{eqnarray}
where we have inserted the identity operator twice. Making use of the
relation $\langle q_{\bm k} \vert \pi_{\bm k} \rangle = \exp(i q_{\bm
  k}\pi_{\bm k})/\sqrt{2\pi}$, the previous expression can be
simplified and one obtains
\begin{eqnarray} 
\widetilde{q_{\bm
    k}^{n_1} \pi_{\bm k}^{n_2}}&=&\frac{1}{\left(2\pi\right)^2}
\int\dd x\, \dd y \, \dd \pi_{\bm k}^\prime \dd \pi_{\bm k}^{\prime\prime} \dd
\pi_{-\bm k}^\prime \dd \pi_{-\bm k}^{\prime\prime} \ee^{-i\pi_{\bm k} x
  - i \pi_{-\bm k} y} \left(q_{\bm k}+\frac{x}{2}\right)^{n_1}
\left\langle \pi_{\bm k}^\prime,\pi_{-\bm k}^\prime \vert
  \hat{\pi}_{\bm k}^{n_2}\vert \pi_{\bm k}^{\prime\prime},\pi_{-\bm
    k}^{\prime\prime}\right\rangle \nonumber\\ & &
\times \exp\left[i\left(q_{\bm k}+\frac{x}{2}\right)\pi_{\bm k}^\prime
  +i\left(q_{-\bm k}+\frac{y}{2}\right)\pi_{-\bm k}^\prime
  -i\left(q_{\bm k}-\frac{x}{2}\right)\pi_{\bm k}^{\prime\prime}
  -i\left(q_{-\bm k}-\frac{y}{2}\right)\pi_{-\bm k}^{\prime\prime}\right]\\
&=&\frac{1}{\left(2\pi\right)^2}\int\dd x\, \dd y \, \dd \pi_{\bm k}^\prime 
\dd \pi_{\bm k}^{\prime\prime}  \dd \pi_{-\bm k}^\prime 
\dd \pi_{-\bm k}^{\prime\prime} \ee^{-i\pi_{\bm k} x - i \pi_{-\bm k} y}
\left(q_{\bm k}+\frac{x}{2}\right)^{n_1} 
\left(\pi_{\bm k}^{\prime\prime}\right)^{n_2}
\delta(\pi_{\bm k}^\prime-\pi_{\bm k}^{\prime\prime})\nonumber\\
& & \times \delta(\pi_{-\bm k}^\prime-\pi_{-\bm k}^{\prime\prime})
\exp\left[i\left(q_{\bm k}+\frac{x}{2}\right)\pi_{\bm k}^\prime
  +i\left(q_{-\bm k}+\frac{y}{2}\right)\pi_{-\bm k}^\prime
  -i\left(q_{\bm k}-\frac{x}{2}\right)\pi_{\bm k}^{\prime\prime}
  -i\left(q_{-\bm k}-\frac{y}{2}\right)\pi_{-\bm k}^{\prime\prime}\right]
\\
&=&\frac{1}{\left(2\pi\right)^2}\int\dd x\, \dd y \, \dd \pi_{\bm k}^\prime
\dd \pi_{-\bm k}^\prime \ee^{i x \left(\pi_{\bm k}^\prime - \pi_{\bm
      k}\right)+i y\left( \pi_{-\bm k}^\prime-\pi_{-\bm
      k}\right)}\left(q_{\bm k}+\frac{x}{2}\right)^{n_1}
\left(\pi_{\bm k}^{\prime}\right)^{n_2}
\label{eq:qn1pin2tilde:interm}\\
&=&\frac{1}{2\pi}\int\dd x\, \dd \pi_{\bm k}^\prime 
\ee^{i x \left(\pi_{\bm k}^\prime - \pi_{\bm k}\right)}
\left(q_{\bm k}+\frac{x}{2}\right)^{n_1}\left(\pi_{\bm k}^{\prime}\right)^{n_2}\\
&=&\frac{1}{2\pi}\int\dd x 
\, \ee^{-i x \pi_{\bm k}}\left(q_{\bm k}+\frac{x}{2}\right)^{n_1}
\int \dd \pi_{\bm k}^\prime \, \ee^{i x \pi_{\bm k}^\prime}
\left(\pi_{\bm k}^{\prime}\right)^{n_2}\, .
\end{eqnarray}
This expression can be further simplified since the second integral is
the Fourier transform of a monomial function. This leads to
\begin{eqnarray} 
\widetilde{q_{\bm k}^{n_1} \pi_{\bm
    k}^{n_2}}&=& \left(-i\right)^{n_2} \int\dd x \ee^{-i x \pi_{\bm k}}
\left(q_{\bm k}+\frac{x}{2}\right)^{n_1}
\delta^{(n_2)}\left(x\right)\, , 
\end{eqnarray} 
where $\delta^{(n_2)}$ stands for the $n_2{}^\mathrm{th}$ derivative
of the delta function. After integrating by parts $n_2$ times, one
obtains the following formula:
\begin{eqnarray} 
\widetilde{q_{\bm
    k}^{n_1} \pi_{\bm k}^{n_2}}&=& i^{n_2} \int\dd x
\frac{\partial^{n_2}}{\partial x^{n_2}}\left[ \ee^{-i x \pi_{\bm k}}
  \left(q_{\bm k}+\frac{x}{2}\right)^{n_1}\right]
\delta\left(x\right)\, .
\label{eq:pqtilde:interm}
\end{eqnarray}
The derivative can be calculated by making use of the binomial formula
and this leads to
\begin{eqnarray}
\frac{\partial^{n_2}}{\partial x^{n_2}}
\left[\ee^{-i x \pi_{\bm k}}
\left(q_{\bm k}+\frac{x}{2}\right)^{n_1}\right]
& = &\sum_{j=0}^{n_2}\binom{n_2}{j}\frac{\partial^{j}}{\partial x^{j}}
\left(\ee^{-i x \pi_{\bm k}}\right)\frac{\partial^{n_2-j}}
{\partial x^{n_2-j}}\left[\left(q_{\bm k}+\frac{x}{2}\right)^{n_1}\right]\\
& = &\sum_{j=0}^{n_2}\binom{n_2}{j}\left(-i\pi_{\bm k}\right)^j
\ee^{-i x \pi_{\bm k}}
\frac{n_1!}{2^{n_2-j}\left(n_1-n_2+j\right)!}
\nonumber\\ & & \times 
\left(q_{\bm k}+\frac{x}{2}\right)^{n_1-n_2+j}\theta\left(j+n_1-n_2\right)\, ,
\end{eqnarray}
where $\theta\left(j+n_1-n_2\right)=1$ if $j\geq n_2-n_1$ and $0$
otherwise. Inserting this expression into \Eq{eq:pqtilde:interm}, one
obtains, after integrating out the $\delta(x)$ function, 
\begin{eqnarray}
\widetilde{q_{\bm k}^{n_1} \pi_{\bm k}^{n_2}}&=& \frac{q_{\bm
    k}^{n_1-n_2}}{\left(-2i\right)^{n_2}}
\sum_{j=0}^{n_2}\binom{n_2}{j} \frac{n_1!\left(-2i\pi_{\bm k}q_{\bm
      k}\right)^j}{\left(n_1-n_2+j\right)!}
\theta\left(j+n_1-n_2\right)
 \label{eq:WeylTransform:qnpinprime}
\, .
\end{eqnarray}
Obviously, one finds a similar expression for $\widetilde{q_{-\bm
    k}^{n_1} \pi_{-\bm k}^{n_2}}$. For $n_1=0$, this immediately leads
to $\widetilde{\pi_{\bm k}^{n}}= \pi_{\bm k}^{n}$ while for $n_2=0$,
one obtains $\widetilde{q_{\bm k}^{n}}=q_{\bm k}^{n}$. This means that
any analytic function of $\hat{q}_{\bm k}$ only or of $\hat{\pi}_{\bm
  k}$ only has a trivial Weyl transform, namely $\widetilde{f(q_{\bm
    k})}=f({q}_{\bm k})$ and $\widetilde{f(\pi_{\bm k})}=f({\pi}_{\bm
  k})$. However, this is not true for mixed terms containing both
$\hat{q}_{\bm k}$ and $\hat{\pi}_{\bm k}$. For example, when
$n_1=n_2=1$, one finds $\widetilde{q_{\bm k} \pi_{\bm k}}=q_{\bm k}
\pi_{\bm k}+i/2$. This is actually due to the fact that $\hat{q}_{\bm
  k}$ and $\hat{\pi}_{\bm k}$ do not commute. When $n_1=n_2=2$, one finds $\widetilde{q_{\bm k}^2 \pi_{\bm k}^2}=q_{\bm k}^2
\pi_{\bm k}^2+2iq_{\bm k}\pi_{\bm k}-1/2$ and $\widetilde{\pi_{\bm k}^2 q_{\bm k}^2}=\pi_{\bm k}^2
q_{\bm k}^2-2i\pi_{\bm k}q_{\bm k}-1/2$, and so forth.
For all other terms in
\Eq{eq:Weyl:Aexpansion}, the operators commute and the Weyl transform
is trivial. For example, let us work out $\widetilde{q_{\bm k}^{n_1}
  \pi_{-\bm k}^{n_2}}$ (the other terms proceed in exactly the same
way). One has
\begin{eqnarray}
\widetilde{q_{\bm k}^{n_1} \pi_{-\bm k}^{n_2}}&=&\int\dd x\,\dd y\,
\ee^{-i\pi_{\bm k} x - i \pi_{-\bm k} y}
\left\langle q_{\bm k}+\frac{x}{2} , 
q_{-\bm k}+\frac{y}{2}\vert \hat{q}_{\bm k}^{n_1} 
\hat{\pi}_{-\bm k}^{n_2} \vert q_{\bm k}-\frac{x}{2} , 
q_{-\bm k}-\frac{y}{2}\right\rangle\\
&=&\int\dd x\, \dd y \, \ee^{-i\pi_{\bm k} x - i \pi_{-\bm k} y}
\left(q_{\bm k}+\frac{x}{2}\right)^{n_1}\left\langle 
q_{\bm k}+\frac{x}{2} , q_{-\bm k}+\frac{y}{2}\vert 
\hat{\pi}_{-\bm k}^{n_2} \vert q_{\bm k}-\frac{x}{2} , 
q_{-\bm k}-\frac{y}{2}\right\rangle\, .
\end{eqnarray}
At this stage, one can proceed exactly as before (namely, again
introduce twice the identity operator and integrate out the delta
Dirac functions) and obtain an equation resembling
\Eq{eq:qn1pin2tilde:interm}. It reads
\begin{eqnarray}
\widetilde{q_{\bm k}^{n_1} \pi_{-\bm k}^{n_2}}
&=&\frac{1}{\left(2\pi\right)^2}\int\dd x\, \dd y\, 
\dd \pi_{\bm k}^\prime   \dd \pi_{-\bm k}^\prime 
\ee^{i x \left(\pi_{\bm k}^\prime - \pi_{\bm k}\right)
+i y\left( \pi_{-\bm k}^\prime-\pi_{-\bm k}\right)}
\left(q_{\bm k}+\frac{x}{2}\right)^{n_1}
\left(\pi_{-\bm k}^{\prime}\right)^{n_2}\\
&=&\frac{1}{2\pi}\int\dd x\, \dd \pi_{\bm k}^\prime 
\ee^{i x \left(\pi_{\bm k}^\prime - \pi_{\bm k}\right)}
\left(q_{\bm k}+\frac{x}{2}\right)^{n_1}
\left(\pi_{-\bm k}\right)^{n_2}\\
&=&q_{\bm k}^{n_1} \pi_{-\bm k}^{n_2}\, ,
\end{eqnarray} 
that is to say the announced result. For similar reasons, one also has
$\widetilde{q_{\bm k}^{n_1} q_{-\bm k}^{n_2}}=q_{\bm k}^{n_1} q_{-\bm
  k}^{n_2}$, $\widetilde{\pi_{\bm k}^{n_1} \pi_{-\bm
    k}^{n_2}}=\pi_{\bm k}^{n_1} \pi_{-\bm k}^{n_2}$ and
$\widetilde{\pi_{\bm k}^{n_1} q_{-\bm k}^{n_2}}=\pi_{\bm k}^{n_1}
q_{-\bm k}^{n_2}$. In fact, combining these results, the Weyl
transform of terms for which $n_1+n_2=1$ can be written in a compact
manner, namely
\begin{equation}
\label{eq:RjRktilde}
\widetilde{R_j R_k}=R_j R_k+\frac{1}{2}\left[\hat{R}_j,\hat{R}_k\right]\, ,
\end{equation}
where the commutator $\left[\hat{R}_j,\hat{R}_k\right]$ is given by the matrix $J$; see
\Eqs{eq:defJ} and~(\ref{eq:weylR}).

\section{Wigner Function of a Gaussian State}
\label{appendix:Wigner:Gaussian}
In this section, we want to write the Wigner function in terms of the
characteristic function and the correlation matrix. In order to do so,
let us first derive an explicit expression of the characteristic
function in terms of the density matrix elements. As already noticed
in the text, a first remark is that, if two operators $\hat{A}$ and
$\hat{B}$ commute with their commutator
($[\hat{A},[\hat{A},\hat{B}]]=[\hat{B},[\hat{A},\hat{B}]]=0$), the
Baker-Campbell-Haussdorf formula reads
$\ee^{\hat{A}+\hat{B}}=\ee^{-[\hat{A},\hat{B}]/2}\ee^{\hat{A}}\ee^{\hat{B}}$. Recalling
once more that $\hat{R}\equiv \left(k^{1/2}\hat{q}_{\bm
    k},k^{-1/2}\hat{\pi}_{\bm k},k^{1/2}\hat{q}_{-{\bm
      k}},k^{-1/2}\hat{\pi}_{-{\bm k}}\right)^{\rm T}\equiv
\left(\hat{R}_1,\hat{R}_2,\hat{R}_3,\hat{R}_4\right)^{\rm T}$, we see
that each $\hat{R}_i$ satisfies this property. As a consequence, the
Weyl operator defined in \Eq{eq:WeylOperator:def} can be
written as
\begin{align}
\hat{\mathcal{W}}\left(\xi\right)\equiv &
\exp\left(i\xi_1k^{1/2}\hat{q}_{\bm{k}}+i\xi_2k^{-1/2}
\hat{\pi}_{\bm{k}}+i\xi_3k^{1/2}\hat{q}_{-\bm{k}}
+i\xi_4k^{-1/2}\hat{q}_{-\bm{k}}\right)
\\ =& \ee^{i\left(\xi_1\xi_2+\xi_3\xi_4\right)/2}
\ee^{i\xi_1k^{1/2}\hat{q}_{\bm{k}}}\ee^{i\xi_2k^{-1/2}
\hat{\pi}_{\bm{k}}}\ee^{i\xi_3k^{1/2}\hat{q}_{-\bm{k}}}
\ee^{i\xi_4k^{-1/2}\hat{q}_{-\bm{k}}}\,.  
\end{align}
Using the above expression, one can express the characteristic function
$\chi (\xi)$ [defined in \Eq{eq:chi:def}] as
\begin{eqnarray}
\chi\left(\xi\right)&=&
\mathrm{Tr}\left[\hat{\rho}\hat{\mathcal{W}}\left(\xi\right)\right] \\&=&
\int\dd q_{\bm {k}}\, \dd q_{-\bm {k}}\left\langle q_{\bm {k}}, q_{-\bm
    {k}} \left\vert \hat{\rho}\hat{\mathcal{W}}\left(\xi\right) \right\vert
  q_{\bm {k}}, q_{-\bm {k}} \right\rangle \\&=&
\ee^{i\left(\xi_1\xi_2+\xi_3\xi_4\right)/2} \int\dd q_{\bm
  {k}}\, \dd q_{-\bm {k}}\left\langle q_{\bm {k}}, q_{-\bm {k}}
  \left\vert
    \hat{\rho}\ee^{i\xi_1k^{1/2}\hat{q}_{\bm{k}}}
\ee^{i\xi_2k^{-1/2}\hat{\pi}_{\bm{k}}}\ee^{i\xi_3k^{1/2}\hat{q}_{-\bm{k}}}
\ee^{i\xi_4k^{-1/2}\hat{q}_{-\bm{k}}}
  \right\vert q_{\bm {k}}, q_{-\bm {k}} \right\rangle \nonumber 
\\&=&
\ee^{i\left(\xi_1\xi_2+\xi_3\xi_4\right)/2} \int\dd q_{\bm
  {k}}\, \dd q_{-\bm {k}}\dd q^\prime_{\bm {k}}\, \dd q\prime_{-\bm
  {k}}\left\langle q_{\bm {k}}, q_{-\bm {k}} \vert \hat{\rho} \vert
  q^\prime_{\bm {k}}, q^\prime_{-\bm {k}}\right\rangle \nonumber\\ & &
\times \left\langle q^\prime_{\bm {k}}, q^\prime_{-\bm {k}}\left\vert
    \ee^{i\xi_1k^{1/2}\hat{q}_{\bm{k}}}\ee^{i\xi_2k^{-1/2}\hat{\pi}_{\bm{k}}}
\ee^{i\xi_3k^{1/2}\hat{q}_{-\bm{k}}}\ee^{i\xi_4k^{-1/2}\hat{q}_{-\bm{k}}}
  \right\vert q_{\bm {k}}, q_{-\bm {k}} \right\rangle 
\\&=&
\label{eq:chiinterwig}
\ee^{i\left(\xi_1\xi_2+\xi_3\xi_4\right)/2} \int\dd q_{\bm
  {k}}\, \dd q_{-\bm {k}}\dd q^\prime_{\bm {k}}\, \dd q^\prime_{-\bm
  {k}}\left\langle q_{\bm {k}}, q_{-\bm {k}} \vert \hat{\rho} \vert
  q^\prime_{\bm {k}}, q^\prime_{-\bm {k}}\right\rangle 
\left\langle q^\prime_{\bm {k}} \left\vert
    \ee^{i\xi_1k^{1/2}\hat{q}_{\bm{k}}}\ee^{i\xi_2k^{-1/2}\hat{\pi}_{\bm{k}}}
  \right\vert q_{\bm {k}} \right\rangle 
\nonumber \\ & & \times
\left\langle q^\prime_{-\bm
    {k}}\left\vert
    \ee^{i\xi_3k^{1/2}\hat{q}_{-\bm{k}}}\ee^{i\xi_4k^{-1/2}\hat{q}_{-\bm{k}}}
  \right\vert q_{-\bm {k}} \right\rangle\, , 
\end{eqnarray} 
where we have utilized the closure relation in order to simplify the
above expression. We see that the characteristic function has indeed
been written in terms of the density matrix elements. However, we
still need to calculate the two other terms which remain in the
integral~(\ref{eq:chiinterwig}). Recalling that $\langle q_{\bm k}
\vert \pi_{\bm k} \rangle = \exp(i q_{\bm k}\pi_{\bm k})/\sqrt{2\pi}$,
one arrives at
\begin{eqnarray} 
\left\langle q^\prime_{\bm {k}}
  \left\vert
    \ee^{i\xi_1k^{1/2}\hat{q}_{\bm{k}}}\ee^{i\xi_2k^{-1/2}\hat{\pi}_{\bm{k}}}
  \right\vert q_{\bm {k}} \right\rangle &=& \int\dd \bar{q}_{\bm k}\, \dd
\bar{\pi}_{\bm k} \left\langle q^\prime_{\bm {k}} \left\vert
    \ee^{i\xi_1k^{1/2}\hat{q}_{\bm{k}}} \right\vert \bar{q}_{\bm
    k}\right\rangle \left\langle \bar{q}_{\bm k} \left\vert
    \ee^{i\xi_2k^{-1/2}\hat{\pi}_{\bm{k}}} \right\vert \bar{\pi}_{\bm
    k}\right\rangle\left\langle \bar{\pi}_{\bm k} \vert q_{\bm {k}}
\right\rangle \nonumber\\ &=& \frac{1}{2\pi} \ee^{i\xi_1k^{1/2}
  q^\prime_{\bm{k}}} \int\dd \bar{q}_{\bm k}\, \dd \bar{\pi}_{\bm k}
\delta\left(q^\prime_{\bm{k}}-\bar{q}_{\bm k}\right)
\ee^{i\xi_2k^{-1/2} \bar{\pi}_{\bm{k}}}
\ee^{i\bar{q}_{\bm{k}}\bar{\pi}_{\bm{k}}} \ee^{-i\bar{\pi}_{\bm{k}}
  q_{\bm{k}}} \nonumber\\ &=& \frac{1}{2\pi} \ee^{i\xi_1k^{1/2}
  q^\prime_{\bm{k}}} \int \dd \bar{\pi}_{\bm k}
\ee^{i\bar{\pi}_{\bm{k}}\left(k^{-1/2}\xi_2+q^\prime_{\bm k}-q_{\bm k}
  \right) } \nonumber\\ &=& \ee^{i\xi_1k^{1/2}q^\prime_{\bm{k}}}
\delta\left(k^{-1/2}\xi_2+q^\prime_{\bm k}-q_{\bm k} \right)\, .  
\end{eqnarray}
Of course, one has a similar expression for the last term in
\Eq{eq:chiinterwig}. As a consequence, the expression of the
characteristic function takes the following form:
\begin{eqnarray} 
\chi\left(\xi\right)&=&
\ee^{i\left(\xi_1\xi_2+\xi_3\xi_4\right)/2} \int\dd q_{\bm
  {k}}\, \dd q_{-\bm {k}}\dd q^\prime_{\bm {k}}\, \dd q^\prime_{-\bm
  {k}}\left\langle q_{\bm {k}}, q_{-\bm {k}} \vert \hat{\rho} \vert
  q^\prime_{\bm {k}}, q^\prime_{-\bm {k}}\right\rangle \nonumber\\ & &
\ee^{i\xi_1k^{1/2}q^\prime_{\bm{k}}}
\delta\left(k^{-1/2}\xi_2+q^\prime_{\bm k}-q_{\bm k} \right)
\ee^{i\xi_3k^{1/2}q^\prime_{-\bm{k}}}
\delta\left(k^{-1/2}\xi_4+q^\prime_{-\bm k}-q_{-\bm k} \right) \\& =&
\ee^{-i\left(\xi_1\xi_2+\xi_3\xi_4\right)/2} \int\dd q_{\bm
  {k}}\, \dd q_{-\bm {k}} \left\langle q_{\bm {k}}, q_{-\bm {k}} \left \vert
  \hat{\rho} \right \vert q_{\bm {k}}-k^{-1/2}\xi_2, q_{-\bm
    {k}}-k^{-1/2}\xi_4\right\rangle \ee^{i\left(\xi_1
    k^{1/2}q_{\bm{k}} +\xi_3 k^{1/2}q_{-\bm{k}}\right)}\, .
\nonumber\\
\end{eqnarray} 

Let us now introduce the quantity $\mathcal{A}\left(\xi\right)$
defined by
\begin{eqnarray}
  \mathcal{A}\left(\xi\right)\equiv\frac{1}{\left(2\pi\right)^4}
  \int\dd^4\eta\, \ee^{-i\xi^\mathrm{T}\eta}\chi\left(\eta\right)\dd\eta\,
  , \end{eqnarray} 
and express this phase-space function in terms of the density
matrix elements. Using the expression of the characteristic function 
derived above, one has 
\begin{eqnarray}
\mathcal{A}\left(\xi\right)&=&
\frac{1}{\left(2\pi\right)^4}\int\dd^4\eta\, \dd
q_{\bm {k}}\dd q_{-\bm {k}}\, \ee^{-i\xi^\mathrm{T}\eta}
\ee^{-i\left(\eta_1\eta_2+\eta_3\eta_4\right)/2}
\ee^{i\left(\eta_1 k^{1/2}q_{\bm{k}} +\eta_3 k^{1/2}q_{-\bm{k}}\right)}
\nonumber \\& & \left\langle q_{\bm {k}}, q_{-\bm {k}} \vert
  \hat{\rho} \vert q_{\bm {k}}-k^{-1/2}\eta_2, q_{-\bm
    {k}}-k^{-1/2}\eta_4\right\rangle\, .  
\end{eqnarray} 
The integral over $\eta_1$ (respectively $\eta_3$) can be easily
performed and gives rise to a $\delta(k^{1/2}q_{\bm
  k}-\eta_2/2-\xi_1)$ [respectively $\delta(k^{1/2}q_{-\bm
  k}-\eta_4/2-\xi_3)$] function. Then this allows us to integrate
over $\eta_2$ and $\eta_4$ and it follows that 
\begin{eqnarray}
\mathcal{A}\left(\xi\right)&=&\frac{4}{\left(2\pi\right)^2}\int \dd
q_{\bm {k}}\dd q_{-\bm {k}}
\exp\left[2i\left(\xi_1\xi_2+\xi_3\xi_4-k^{1/2}q_{\bm
      k}\xi_2-k^{1/2}q_{-\bm k}\xi_4\right)\right] \nonumber \\& &
\left\langle q_{\bm {k}}, q_{-\bm {k}} \left \vert \hat{\rho} \right \vert
  2k^{-1/2}\xi_1-q_{\bm k} , 2k^{-1/2}\xi_3-q_{-\bm k}\right\rangle\, .  
\end{eqnarray}
Let us now perform the change of integration variable $q_{\bm
  k}=k^{-1/2}\xi_1+x/2$ and $q_{-{\bm k}}=k^{-1/2}\xi_3+y/2$. Then
one obtains that $\mathcal{A}\left(\xi\right)$ is given by
\begin{align} 
\mathcal{A}\left(\xi\right)=\frac{1}{\left(2\pi\right)^2}\int
\dd x\, \dd y \exp\left[-ik^{1/2}\left(\xi_2 x+\xi_4 y\right)\right]
\left\langle k^{-1/2}\xi_1+\frac{x}{2},
  k^{-1/2}\xi_3+\frac{y}{2} \right\vert \hat{\rho} \left\vert
  k^{-1/2}\xi_1-\frac{x}{2} , k^{-1/2}\xi_3-\frac{y}{2}\right\rangle\, .  
\end{align} 
Finally, inserting
$\hat{\rho}=\vert\Psi\rangle\langle\Psi\vert$ in the above expression
yields 
\begin{align}
\mathcal{A}\left(\xi\right) &=\frac{1}{\left(2\pi\right)^2}\int \dd
x\dd y \exp\left[-ik^{1/2}\left(\xi_2 x+\xi_4 y\right)\right]
\Psi\left( k^{-1/2}\xi_1+\frac{x}{2},
  k^{-1/2}\xi_3+\frac{y}{2} \right) \Psi^*\left(
  k^{-1/2}\xi_1-\frac{x}{2} , k^{-1/2}\xi_3-\frac{y}{2}\right)\, ,
\end{align} 
which exactly coincides with \Eq{eq:Wigner:def}. Therefore, we have
shown that
\begin{align}
\label{eq:wigchi}
W\left(\xi\right)=\frac{1}{\left(2\pi\right)^4}
\int\dd^4\eta \, \ee^{-i\xi^\mathrm{T}\eta}\chi\left(\eta\right)\, ,  
\end{align}
which is the standard expression of the Wigner function in terms of the
characteristic one.

\par

From here, an expression of $W$ in terms of $\gamma$ can be
obtained. Indeed, as written in \Eq{eq:chiGauss}, for a Gaussian
state, one has
$\chi\left(\eta\right)=\exp\left(-\eta^\mathrm{T}\gamma\eta/4\right)$,
where $\gamma$ is, by definition, the covariance matrix. Inserting
this last expression in \Eq{eq:wigchi}, one obtains
\begin{eqnarray}
W\left(\xi\right)&=&\frac{1}{\left(2\pi\right)^4}
\int\dd^4\eta\exp\left(-i\xi^\mathrm{T}\eta-\frac{1}{4}
\eta^\mathrm{T}\gamma\eta\right)\dd\eta
\\ &=&
\frac{1}{\left(2\pi\right)^4}
\exp\left(-\xi^\mathrm{T}\gamma^{-1}\xi\right)
\int\dd^4\eta\exp\left[-\frac{1}{4}
\left(\eta+2i\gamma^{-1}\xi\right)^\mathrm{T}
\gamma\left(\eta+2i\gamma^{-1}\xi\right)\right]
\, , 
\end{eqnarray} 
where we have used the fact that the covariance matrix is symmetric,
$\gamma^\mathrm{T}=\gamma$, and the fact that $\xi$ and $\eta$ are
real vectors; hence $\eta^\mathrm{T}\xi=\xi^\mathrm{T}\eta$. After
performing the change of integration variable $\eta\rightarrow
\eta+2i\gamma^{-1}\xi$, the Gaussian integration finally leads to 
\begin{eqnarray}
W\left(\xi\right) &=&
\frac{1}{\pi^2\sqrt{\mathrm{det}\gamma}}\ee^{-\xi^T\gamma^{-1}\xi}\, ,
\label{eq:Wigner:gamma}
\end{eqnarray}
which is exactly the formula used in the text, in particular, in
Sec.~\ref{sec:stocha}.

\section{Evolution Equation for the Wigner Function}
\label{appendix:Wigner:EOM}
In this appendix, we derive the equation controlling the time
evolution of the Wigner function. The time derivative of $W$ can be
obtained from \Eq{eq:Wigner:def}. Explicitly, one has
\begin{align}
  \frac{\dd}{\dd\eta}W&=\frac{1}{\left(2\pi\right)^2}
\int\dd x\, \dd y \left[\frac{\dd}{\dd\eta}\Psi^*
\left(q_{\bm k}-\frac{x}{2},q_{-\bm k}-\frac{y}{2}\right)\right]
\ee^{-i \pi_{\bm k} x -i\pi_{-\bm k} y}\Psi
\left(q_{\bm k}+\frac{x}{2},q_{-\bm k}+\frac{y}{2}\right)\nonumber\\
  & + \frac{1}{\left(2\pi\right)^2}\int\dd x\, \dd y \Psi^*
\left(q_{\bm k}-\frac{x}{2},q_{-\bm k}-\frac{y}{2}\right)
\ee^{-i \pi_{\bm k} x -i\pi_{-\bm k} y}\frac{\dd}{\dd\eta}\Psi
\left(q_{\bm k}+\frac{x}{2},q_{-\bm k}+\frac{y}{2}\right)\nonumber\\
  &=\frac{i}{\left(2\pi\right)^2}\int\dd x\, \dd y 
\left\langle q_{\bm k}+\frac{x}{2},q_{-{\bm k}}+\frac{y}{2}
\left \vert \hat{H}\right \vert \Psi \right\rangle ^*
\ee^{-i \pi_{\bm k} x -i\pi_{-\bm k} y}\Psi\left(q_{\bm k}
+\frac{x}{2},q_{-\bm k}+\frac{y}{2}\right)\nonumber\\
  &  -\frac{i}{\left(2\pi\right)^2}\int\dd x\, \dd y \,
\Psi^*\left(q_{\bm k}-\frac{x}{2},q_{-\bm k}-\frac{y}{2}\right)
\ee^{-i \pi_{\bm k} x -i\pi_{-\bm k} y}
\left\langle q_{\bm k}+\frac{x}{2},q_{-{\bm k}}+\frac{y}{2}
\left \vert \hat{H}\right \vert \Psi \right\rangle\, ,
\end{align}
where in the second equality we have used the Schr\"odinger equation
$\partial\Psi/\partial\eta=-i\hat{H}\Psi$ and the fact that the
Hamiltonian is Hermitian. In order to proceed, one needs to express
the Hamiltonian density in terms of the Gaussian operators
$\hat{q}_{\bm k}$, $\hat{\pi}_{\bm k}$, $\hat{q}_{-\bm k}$ and
$\hat{\pi}_{-\bm k}$. Inserting \Eqs{eq:vqpi} and~(\ref{eq:pqpi}) into
\Eq{eq:Hamiltonian:vandp}, one obtains the following expression:
\begin{equation}
\label{eq:hamiltonian:qpi}
\hat{H}_{\bm k} = 
\frac{k^2}{2}\left(\hat{q}_{\bm k}^2+\hat{q}_{-\bm k}^2\right)
+\frac{\hat{\pi}_{\bm k}^2+\hat{\pi}_{-\bm k}^2}{2}
+\frac{z^\prime}{z}\left(\hat{q}_{\bm k}\hat{\pi}_{-\bm k}+\hat{q}_{-\bm k}
\hat{\pi}_{\bm k}\right)\, .
\end{equation}
The three terms in the above Hamiltonian give rise to three terms for
the time evolution of $W$ which, in the following, we denote by
$\dd W_{q^2}/\dd\eta$, $\dd W_{\pi^2}/\dd\eta$ and $\dd
W_{q\pi}/\dd\eta$. Let us work them out one by one. The first term can
be expressed as
\begin{eqnarray} \frac{\dd W_{q^2}}{\dd
    \eta}&=&-i\left(\frac{k}{2\pi}\right)^2 \int\dd x\dd y
  \Psi^*\left(q_{\bm k}-\frac{x}{2},q_{-\bm k}-
    \frac{y}{2}\right)\ee^{-i \pi_{\bm k} x -i\pi_{-\bm k} y}
  \left(xq_{\bm k}+yq_{-\bm k}\right)\Psi\left(q_{\bm k}
    +\frac{x}{2},q_{-\bm k}+\frac{y}{2}\right)\, .\nonumber\\
\end{eqnarray}
Noticing that differentiating \Eq{eq:Wigner:def} with respect to
$\pi_{\bm k}$ and $\pi_{-\bm k}$ precisely produces factors $x$ and
$y$, it is easy to check that the above expression can also be written
as
\begin{equation} 
\label{eq:q2}
\frac{\dd W_{q^2}}{\dd
  \eta}=k^2\left(q_{\bm k}\frac{\dd}{\dd\pi_{\bm k}}+q_{-\bm
    k}\frac{\dd}{\dd\pi_{-\bm k}}\right)W\, .  
\end{equation} 
Then recalling that the momentum is represented by $\hat{\pi}_{\bm
  k}=-i\partial/\partial q_{\bm k}$, the second term reads
\begin{eqnarray}
\frac{\dd W_{\pi^2}}{\dd \eta}&=&\frac{i/2}{\left(2\pi\right)^2}
\int\dd x\dd y \left(\frac{\partial^2}{\partial q_{\bm k}^2}
+\frac{\partial^2}{\partial q_{-\bm k}^2}\right)\Psi^*
\left(q_{\bm k}-\frac{x}{2},q_{-\bm k}-\frac{y}{2}\right)
\ee^{-i \pi_{\bm k} x -i\pi_{-\bm k} y}
\Psi\left(q_{\bm k}+\frac{x}{2},q_{-\bm k}+\frac{y}{2}\right)\nonumber\\
& -& \frac{i/2}{\left(2\pi\right)^2}\int\dd x\dd y 
\Psi^*\left(q_{\bm k}-\frac{x}{2},q_{-\bm k}-\frac{y}{2}\right)
\ee^{-i \pi_{\bm k} x -i\pi_{-\bm k} y}\left(\frac{\partial^2}{\partial q_{\bm k}^2}
+\frac{\partial^2}{\partial q_{-\bm k}^2}\right) 
\Psi\left(q_{\bm k}+\frac{x}{2},q_{-\bm k}+\frac{y}{2}\right)\, .\nonumber\\
\end{eqnarray} 
In this expression, derivatives with respect to $q_{\bm k}$ and
$q_{-\bm k}$ can be written as derivatives with respect to $x$ and $y$
by noticing that, for example, 
\begin{equation} 
\frac{\partial^2}{\partial q_{\bm
    k}^2}\Psi^*\left(q_{\bm k}-\frac{x}{2},q_{-\bm
    k}-\frac{y}{2}\right)=-2\frac{\partial^2}{\partial q_{\bm
    k}\partial x}\Psi^*\left(q_{\bm k}-\frac{x}{2},q_{-\bm
    k}-\frac{y}{2}\right)\, , 
\end{equation} 
and similar expressions for the three other double derivatives. Then
integrating by part the obtained expression yields the formula
\begin{eqnarray}
\label{eq:pi2}
\frac{\dd W_{\pi^2}}{\dd \eta} &= & -\left(\pi_{\bm
    k}\frac{\partial}{\partial q_{\bm k}}+\pi_{-\bm
    k}\frac{\partial}{\partial q_{-\bm k}}\right) W\, .  
\end{eqnarray} 
Finally, the third term remains to be calculated. It is given by
\begin{eqnarray}
\frac{\dd W_{q\pi}}{\dd \eta}&=&-\frac{z^\prime/z}
{\left(2\pi\right)^2}\int\dd x\dd y 
\, \ee^{-i \pi_{\bm k} x -i\pi_{-\bm k} y}
\Psi\left[\left(q_{\bm k}-\frac{x}{2}\right)
\frac{\partial}{\partial q_{-\bm k}}
+\left(q_{-\bm k}-\frac{y}{2}\right)
\frac{\partial}{\partial q_{\bm k}}\right]\Psi^*\nonumber\\
&-& \frac{z^\prime/z}{\left(2\pi\right)^2}
\int\dd x\dd y \, \ee^{-i \pi_{\bm k} x -i\pi_{-\bm k} y}
\Psi^*\left[\left(q_{\bm k}+\frac{x}{2}\right)
\frac{\partial}{\partial q_{-\bm k}}
+\left(q_{-\bm k}+\frac{y}{2}\right)
\frac{\partial}{\partial q_{\bm k}}\right] \Psi\, ,
\end{eqnarray} 
where, for simplicity, we have omitted the wavefunction
arguments. Using the same manipulations as for the other terms, one
obtains the following expression:
\begin{align} 
\label{eq:qpi}
\frac{\dd
  W_{q\pi}}{\dd \eta} &= \frac{z^\prime}{z}\left(\pi_{-\bm
    k}\frac{\partial}{\partial\pi_{\bm k}}+\pi_{\bm
    k}\frac{\partial}{\partial\pi_{-\bm k}}-q_{\bm
    k}\frac{\partial}{\partial q_{-\bm k}}-q_{-\bm
    k}\frac{\partial}{\partial q_{\bm k}}\right)W\, .  
\end{align} 
Combining \Eqs{eq:q2}, (\ref{eq:pi2}) and~(\ref{eq:qpi}), one
finally obtains 
\begin{align} \frac{\dd W}{\dd
    \eta}&=\left[k^2\left(q_{\bm k}\frac{\partial}{\partial\pi_{\bm
          k}}+q_{-\bm k}\frac{\partial}{\partial\pi_{-\bm
          k}}\right)-\pi_{\bm k}\frac{\partial}{\partial q_{\bm
        k}}-\pi_{-\bm k}\frac{\partial}{\partial q_{-\bm k}}
  +\frac{z^\prime}{z}\left(\pi_{-\bm
        k}\frac{\partial}{\partial\pi_{\bm k}}+\pi_{\bm
        k}\frac{\partial}{\partial\pi_{-\bm k}}-q_{\bm
        k}\frac{\partial}{\partial q_{-\bm k}}-q_{-\bm
        k}\frac{\partial}{\partial q_{\bm k}}\right)\right]W\, .
\label{eq:dWdeta}
\end{align}
It turns out that this last equation is nothing but the classical
Liouville equation for the distribution $W$. Indeed, if one introduces
the Poisson bracket $\lbrace \rbrace_{\mathrm{PB}}$ defined by
\begin{equation}
\left\lbrace f,g\right\rbrace _{\mathrm{PB}}= \frac{\partial
  f}{\partial q_{\bm k}} \frac{\partial g}{\partial \pi_{\bm k}} -
\frac{\partial f}{\partial \pi_{\bm k}} \frac{\partial g}{\partial
  q_{\bm k}} + \left(\bm{k} \leftrightarrow -\bm{k}\right)\, ,  
\end{equation}
then it is straightforward to write \Eq{eq:dWdeta} as 
\begin{equation} \frac{\dd W}{\dd \eta} = \left\lbrace H_{\bm
    k},W \right\rbrace_{\mathrm{PB}}\, ,
\end{equation} 
where we have used \Eq{eq:hamiltonian:qpi}. Let us mention that
although we have established this result for the Hamiltonian of
cosmological perturbations, it can in fact be generalized to any
quadratic Hamiltonian.
%
\twocolumngrid
\bibliography{discord}
\end{document}